\documentclass[secnumarabic,aps,prd,floats,floatfix,nofootinbib,showkeys,showpacs,10pt]{revtex4-1}

\usepackage{times}
\usepackage{amsfonts}
\usepackage{amsmath}
\usepackage{amssymb}
\usepackage{natbib}
\usepackage{graphicx}
\usepackage{enumitem}
\usepackage{xcolor}
\usepackage[colorlinks]{hyperref}
\usepackage[outdir=./fig/]{epstopdf}

\def\apj{ApJ}
\def\apjl{ApJL}
\def\apss{Astrophysics and Space Science}
\def\jcap{JCAP}
\def\mnras{MNRAS}
\def\physrep{Physics Reports}

\graphicspath{{./fig/}}

\begin{document}
\title{Approximation of the potential in scalar field dark energy models}
\author{Richard~A.~\surname{Battye}}
\email{richard.battye@manchester.ac.uk}
\author{Francesco~\surname{Pace}}
\email[Corresponding author: ]{francesco.pace@manchester.ac.uk}

\affiliation{Jodrell Bank Centre for Astrophysics, School of Physics and Astronomy, The University of Manchester, 
Manchester, M13 9PL, United Kingdom}

\label{firstpage}

\date{\today}

\begin{abstract}
We study the nature of potentials in scalar field based models for dark energy - with both canonical and noncanonical 
kinetic terms. We calculate numerically, and using an analytic approximation around $a\approx 1$, potentials for 
models with constant equation-of-state parameter, $w_{\phi}$. We find that for a wide range of models with canonical 
and noncanonical kinetic terms there is a simple approximation for the potential that holds when the scale factor is 
in the range $0.6\lesssim a\lesssim 1.4$. We discuss how this form of the potential can also be used to represent 
models with nonconstant $w_{\phi}$ and, hence, how it could be used in reconstruction from cosmological data.
\end{abstract}

\pacs{98.80.-k, 95.36.+x}

\keywords{Cosmology; scalar field; k-essence; dark energy; equation of state}

\maketitle

\section{Introduction}
The origin of the cosmic acceleration is one of the most significant open questions in cosmology and fundamental 
physics. A cosmological constant is still very much consistent with the data \citep{Planck2015_XIII,Planck2015_XIV}, 
but in order to either refute or confirm this simple hypothesis one needs to consider alternative models to explain 
the observations. One very simple idea is to postulate a dark energy component dominated by a scalar field either with 
a canonical or noncanonical kinetic term. Such models are known as quintessence models 
\citep{Ford1987,Peebles1988,Ratra1988a,Wetterich1988,Caldwell1998,Copeland1998,Steinhardt1999,Barreiro2000} and 
$k$-essence models \citep{ArmendarizPicon1999,Chiba2000,Mukhanov2006}, respectively.

The standard approach when constraining cosmological models with a dark energy component that is not the cosmological 
constant is to define an equation-of-state parameter $w_{\phi}=P_{\phi}/\rho_{\phi}\neq -1$, where $P_{\phi}$ is the 
pressure of dark energy and $\rho_{\phi}$ is its density, making no assumption as to the origin of the dark energy. 
In principle this is a general function of time, but it is often considered to be either constant, or to be 
represented by a specific functional form, for example \cite{Chevallier2001,Linder2003}. At the moment the data barely 
constrain anything beyond a constant $w_{\phi}$, but this is likely to change in the near future as more observations 
probing the equation of state become available, such as 
Euclid\footnote{\url{http://www.euclid-ec.org/}} \citep{Laureijs2011,Amendola2013}, 
LSST\footnote{\url{http://www.lsst.org}} \citep{LSST2012}
and SKA\footnote{\url{https://www.skatelescope.org/}} \citep{Bull2015,Camera2015,Raccanelli2015,Santos2015}. 
Various ideas have been put forward to extend to time varying situations. These include various limited functional 
forms \citep{Chevallier2001,Linder2003,Hannestad2004,Jassal2005,Barboza2008}, 
the \textit{Om} diagnostic \citep{Sahni2008,Zunckel2008}, the state-finder approach \citep{Alam2003,Sahni2003} and 
even using principal component analysis on general piecewise linear parametrizations of $w_{\phi}$ 
\citep{Crittenden2009}. For a review of the parametric and nonparametric methods to reconstruct the dark energy 
equation-of-state parameter, we refer to \cite{Sahni2006}. Since many of the observations are sensitive to 
perturbations in the dark energy it is also necessary to make some assumptions about the perturbations, but we will 
not consider this here.

An alternative is to presume that the origin of the dark energy is a model based on a scalar field. 
However, such models usually involve one or more arbitrary functions which would need to be specified before any model 
prediction could be made. One of these is the potential $V(\phi)$ of the scalar field which one might try to 
reconstruct from observations. 
One obvious suggestion \citep{Sahlen2005}, which extends the approach of \cite{Grivell2000} for inflation, is to 
represent the potential as a Taylor series expanded around the present-day value of the field $\phi_0$
\begin{equation}\label{eqn:Vphi_series}
 V(\phi)=V_0+V_1(\phi-\phi_0)+V_2(\phi-\phi_0)^2+\dots\;,
\end{equation}
and attempt to fit for the coefficients $V_i$. However, it is not clear where to truncate this series in a controlled 
way. 
Similar and complementary methods have been proposed by \cite{Copeland1993,Daly2003,Daly2004,Simon2005a}. 
Other reconstruction methods are valid in the slow-roll regime, that is, when $1+w_{\phi}\approx 0$. For quintessence 
models, a one-parameter \citep{Slepian2014} or two-parameter \citep{Crittenden2007,Chiba2009a,Chiba2009b} formula 
has been used and for $k$-essence models we refer to works by \cite{ArmendarizPicon1999,Chiba2009c}.

In this paper we first calculate potentials for a range of minimally coupled scalar field models with canonical 
(\autoref{sect:mcsf}) and noncanonical (\autoref{sect:kessence}) kinetic terms assuming initially that 
$w_{\phi}$ is constant. It is possible to derive an analytic solution for the potential in Quintessence models, but 
this is not possible in general for the case of $k$-essence models and therefore we resort to numerical calculations 
and an analytic approximation around the present day which is valid for $0.6\lesssim a\lesssim 1.4$. Based on this 
analytic approximation we suggest a form of a potential with just four parameters which we demonstrate can lead to a 
wide range of behaviour for $w_{\phi}$ as a function of time (\autoref{sect:wa}) and, by design, includes models with 
constant $w_{\phi}$. Of course, this functional form will not include every possible behaviour in a general model, but 
it does provide more physical insights and it is useful for models which are not significantly different from a 
linearly evolving equation-of-state parameter. We conclude and discuss our results in \autoref{sect:conclusions}.

In the following, we will use natural units with $c=\hbar=1$, the Planck mass is $M_{\rm pl}=G^{-1/2}$ and we assume 
a metric with signature $(-,+,+,+)$.

\section{Minimally coupled scalar fields with constant \texorpdfstring{$w_{\phi}$}{w}}\label{sect:mcsf}
The Lagrangian for minimally coupled scalar fields is
\begin{equation}
 {\cal L}=-\frac{1}{2}\eta g^{\mu\nu}\nabla_{\mu}\phi\nabla_{\nu}\phi-V(\phi)\;,
\end{equation}
and its corresponding stress-energy tensor
\begin{equation}
 T_{\mu\nu} = g_{\mu\nu}{\cal L}+\eta\nabla_{\mu}\phi\nabla_{\nu}\phi
 = \eta\nabla_{\mu}\phi\nabla_{\nu}\phi-
 g_{\mu\nu}\left[\frac{1}{2}\eta g^{\alpha\beta}\nabla_{\alpha}\phi\nabla_{\beta}\phi+V(\phi)\right]\;.
\end{equation}
The constant $\eta$ distinguishes between the Quintessence case ($\eta=+1$, $-1<w_{\phi}<1$) and the phantom case 
($\eta=-1$, $w_{\phi}<-1$) \citep{Caldwell2002}.

Density and pressure are given by
\begin{equation}
 \rho_{\phi}={T^0}_0=\frac{1}{2}\eta{\dot\phi}^2+V(\phi)\;,\quad 
 P_{\phi}=\frac{1}{3} {T^i}_i=\frac{1}{2}\eta{\dot\phi}^2-V(\phi)\;,
\end{equation}
and the conservation equation $\dot{\rho}_{\phi}+3H(\rho_\phi+P_\phi)=0$ gives rise to the Klein-Gordon equation, 
which describes the time evolution of the scalar field
\begin{equation}\label{eqn:eom}
 \ddot{\phi}+3H\dot{\phi}+\eta\frac{dV}{d\phi}=0\,.
\end{equation}

To achieve an accelerated expansion, we require $w_{\phi}<-1/3$. In fact, observations require $w_{\phi}\simeq-1$ 
(due to the cosmological constant case) \citep{Planck2015_XIII}, hence we can evaluate deviations of 
$w_{\phi}$ from $-1$ with the help of (\ref{eqn:eom})
\begin{equation}\label{eqn:deviations}
 1+w_{\phi}=\frac{V_{\phi}^2}{9H^2(\xi_s+1)^2\rho_{\phi}}\;,
\end{equation}
with $\xi_s=\ddot{\phi}/(3H\dot{\phi})$ \citep{Amendola2013}. Note that in a pure slow-roll approximation, $\xi_s=0$.

By using Friedmann equations, we can determine the time evolution of the scalar field and its potential for a given 
$w_{\phi}(a)$ \citep{Ellis1991}
\begin{align}
 \frac{\phi(a)-\phi_0}{M_{\rm pl}} & = \pm\sqrt{\frac{3\Omega_{\rm de}}{8\pi}}
 \int_{1}^{a}\frac{\sqrt{\eta[1+w_{\phi}(x)]g(x)}}{xE(x)}~dx\;,\label{eqn:phia}\\
 V(a) & = \frac{3H_0^2M_{\rm pl}^2\Omega_{\rm de}[1-w_{\phi}(a)]g(a)}{16\pi}\;,\label{eqn:Va}
\end{align}
where $\Omega_{\rm de}$ is the dark energy density parameter today, $H_0$ the Hubble constant and $\phi_0$ the value of 
the scalar field at $a=1$. Finally, $g(a)$ represents the time evolution of the dark energy component
\begin{equation}\label{eqn:ga}
 g(a)=\exp{\left(-3\int_1^{a}\frac{1+w_{\phi}(x)}{x}dx\right)}\;.
\end{equation}
Assuming a flat geometry, the Hubble parameter is given by
\begin{equation}
 H=H_0E(a)=H_0\left[\frac{\Omega_{\rm m}}{a^3}+\Omega_{\rm de}g(a)\right]^{\frac{1}{2}}\;,
\end{equation}
with $\Omega_{\rm m}$ the matter density parameter today.

For a constant equation of state $w_{\phi}$, integral (\ref{eqn:phia}) can be evaluated as
\begin{equation}\label{eqn:phiaw}
 \frac{\phi-\phi_0}{M_{\rm pl}}=\mp\frac{2}{3w_{\phi}}\sqrt{\frac{3\eta(1+w_{\phi})}{8\pi}}
 \left[\sinh^{-1}\left(\sqrt{\frac{\Omega_{\rm de}}{\Omega_{\rm m}}}a^{-\frac{3w_{\phi}}{2}}\right)-
 \sinh^{-1}\sqrt{\frac{\Omega_{\rm de}}{\Omega_{\rm m}}}\right]\;,
\end{equation}
and its inverse gives an expression for the scale factor in terms of the scalar field
\begin{equation}\label{eqn:aphiw}
 a(\phi)=\left(\frac{\Omega_{\rm m}}{\Omega_{\rm de}}\right)^{-\frac{1}{3w_{\phi}}}
 \left[\sinh\left(\mp\frac{3}{2}w_{\phi}\sqrt{\frac{8\eta\pi}{3(1+w_{\phi})}}
 \left(\frac{\phi-\phi_0}{M_{\rm pl}}\right)
 +\sinh^{-1}\sqrt{\frac{\Omega_{\rm de}}{\Omega_{\rm m}}}\right)\right]^{-\frac{2}{3w_{\phi}}}\,.
\end{equation}

With these relations in hand, we can deduce the full expression for the potential $V(\phi)$
\begin{equation}\label{eqn:Vphi0}
 V(\phi) = \frac{3H_0^2M_{\rm pl}^2\Omega_{\rm de}(1-w_{\phi})}{16\pi}
 \left(\frac{\Omega_{\rm m}}{\Omega_{\rm de}}\right)^{\frac{1+w_{\phi}}{w_{\phi}}}
 \left[\sinh\left(\mp\frac{3}{2}w_{\phi}\sqrt{\frac{8\eta\pi}{3(1+w_{\phi})}}
 \left(\frac{\phi-\phi_0}{M_{\rm pl}}\right)
 +\sinh^{-1}\sqrt{\frac{\Omega_{\rm de}}{\Omega_{\rm m}}}\right)\right]^{\frac{2(1+w_{\phi})}{w_{\phi}}}\,.
\end{equation}

Since $\phi_0$ just shifts the potential in the $\phi$-direction, we can make the choice
\begin{equation}
 \phi_0=\pm\frac{2}{3w_{\phi}}M_{\rm pl}\sqrt{\frac{3\eta(1+w_{\phi})}{8\pi}}\sinh^{-1}
 \sqrt{\frac{\Omega_{\rm de}}{\Omega_{\rm m}}}\,,
\end{equation}
that simplifies the form of the potential
\begin{equation}
 V(\phi)=\frac{3H_0^2M_{\rm pl}^2\Omega_{\rm de}(1-w_{\phi})}{16\pi}
 \left(\frac{\Omega_{\rm m}}{\Omega_{\rm de}}\right)^{\frac{1+w_{\phi}}{w_{\phi}}}
 \sinh^{\frac{2(1+w_{\phi})}{w_{\phi}}}\left[\mp\frac{3w_{\phi}}{2}\sqrt{\frac{8\eta\pi}{3(1+w_{\phi})}}
 \frac{\phi}{M_{\rm pl}}\right]\;,
 \label{eqn:pot}
\end{equation}
in agreement with \cite{Sahni2000,Sahni2006}. 
An analytic solution of (\ref{eqn:phia}) can be found also when dark matter has a constant equation-of-state 
parameter $w_{\rm m}\neq 0$, as shown in \cite{Paliathanasis2015}.


While it is possible to find exact solutions for the potential of Quintessence models with a constant equation of 
state, this is not the case when $w_{\phi}$ is a function of time or for more general scalar field models, such 
$k$-essence models. 
Moreover, the expressions for the potential at early and late times are not very useful from an observational point of 
view, since they assume one of the component to be dominant and are not relevant for modelling late-time observations. 
It is, therefore, worthwhile to find approximate solutions valid for $a\approx 1$ that can be probed with data. 
To do this, we expand in series $\frac{d\phi}{da}$ for $a\approx 1$, but a priori it is not clear where to 
truncate the series. We have checked that a first-order expansion is a very good approximation, leading to a 
scalar field evolving quadratically with respect to the scale factor.

The differential equation describing the approximate evolution of the scalar field for $a\approx 1$ is
\begin{equation}
 \frac{d\phi}{da}=\pm\sqrt{\frac{3\eta(1+w_{\phi})M_{\rm pl}^2\Omega_{\rm de}}{8\pi}}
                  \left\{1-\frac{1}{2}\left(2+3\Omega_{\rm m}w_{\phi}\right)(a-1)\right\}\;,
\end{equation}
which implies the following approximate evolution for the scalar field
\begin{equation}
 \frac{\phi-\phi_0}{M_{\rm pl}} = \pm\sqrt{\frac{3\eta(1+w_{\phi})\Omega_{\rm de}}{8\pi}}
 \left[a-1-\frac{1}{4}\left(2+3\Omega_{\rm m}w_{\phi}\right)(a-1)^2\right]\;.
 \label{eqn:phia1}
\end{equation}
By inverting this, we can find a relation between the scale factor and the scalar field
\begin{equation}\label{eqn:aphi}
 a = 1+\frac{2}{2+3\Omega_{\rm m}w_{\phi}}
       \left\{1-\sqrt{1\mp[2+3\Omega_{\rm m}w_{\phi}]
       \sqrt{\frac{8\eta\pi}{3\Omega_{\rm de}[1+w_{\phi}]}}\frac{\phi-\phi_0}{M_{\rm pl}}}\right\}\;,
\end{equation}
which leads to the following functional form for the approximate potential
\begin{equation}\label{eqn:Vphi_ABCD}
 V(\phi)=AH_0^2M_{\rm pl}^2\left(1-\sqrt{B+C\frac{\phi-\phi_0}{M_{\rm pl}}}\right)^D\;,
\end{equation}
where the coefficients $A$, $B$, $C$ and $D$ are dimensionless constants depending on the cosmological parameters 
characterising the model. For minimally coupled models with constant $w_{\phi}$, the four coefficients assume the 
following values
\begin{equation}\label{eqn:Qcoef}
 \begin{split}
  A & = \frac{3\Omega_{\rm de}(1-w_{\phi})}{16\pi}
        \left(\frac{4+3\Omega_{\rm m}w_{\phi}}{2+3\Omega_{\rm m}w_{\phi}}\right)^{-3(1+w_{\phi})}\;, \quad 
  B = \frac{4}{(4+3\Omega_{\rm m}w_{\phi})^2}\;,\\
  C & = \mp 4\frac{2+3\Omega_{\rm m}w_{\phi}}{(4+3\Omega_{\rm m}w_{\phi})^2}
          \sqrt{\frac{8\eta\pi}{3\Omega_{\rm de}(1+w_{\phi})}}\;, \quad \qquad
  D = -3(1+w_{\phi})\;.
 \end{split}
\end{equation}
(\ref{eqn:Vphi_ABCD}) is  an interesting result, showing that for constant equations of state, the potential can be 
represented by a very simple form.

The four parameters in (\ref{eqn:Qcoef}) depend on two quantities, $\Omega_{\rm de}$ and $w_{\phi}$, therefore, we can 
express two of them ($A$ and $C$) in terms of $B$ and $D$:
\begin{equation}
 A = \frac{(D+6)[\sqrt{B}(D+1)-2]}{16\pi\sqrt{B}(D-3)(1-\sqrt{B})^D}\;, \qquad
 C = \mp 2\sqrt{\frac{8\eta\pi\sqrt{B}(3-D)}{D[\sqrt{B}(D+1)-2]}}\sqrt{B}(1-\sqrt{B}) \;.
\end{equation}

\begin{figure}
 \centering
 \includegraphics[width=0.3\textwidth,angle=-90]{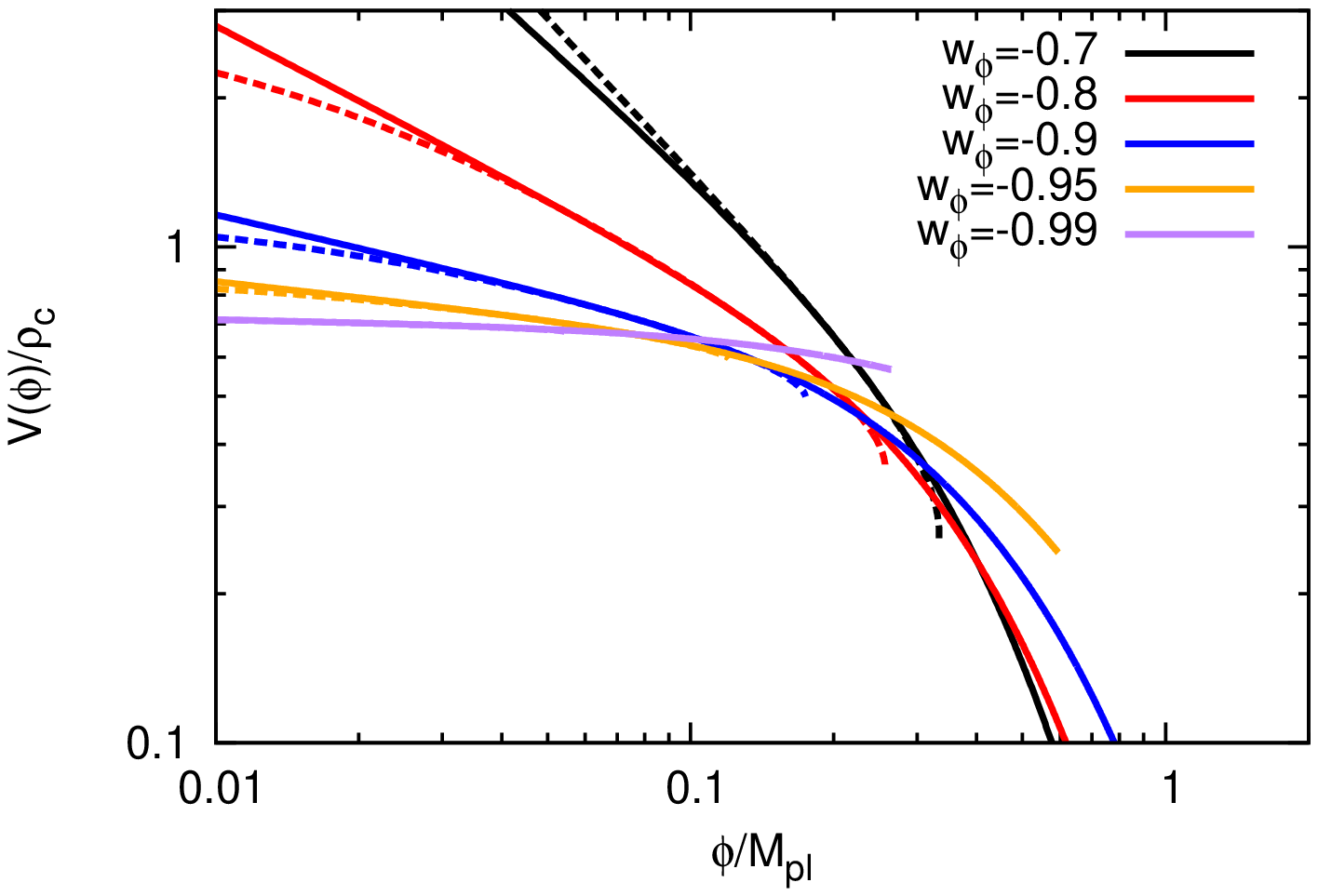}
 \includegraphics[width=0.3\textwidth,angle=-90]{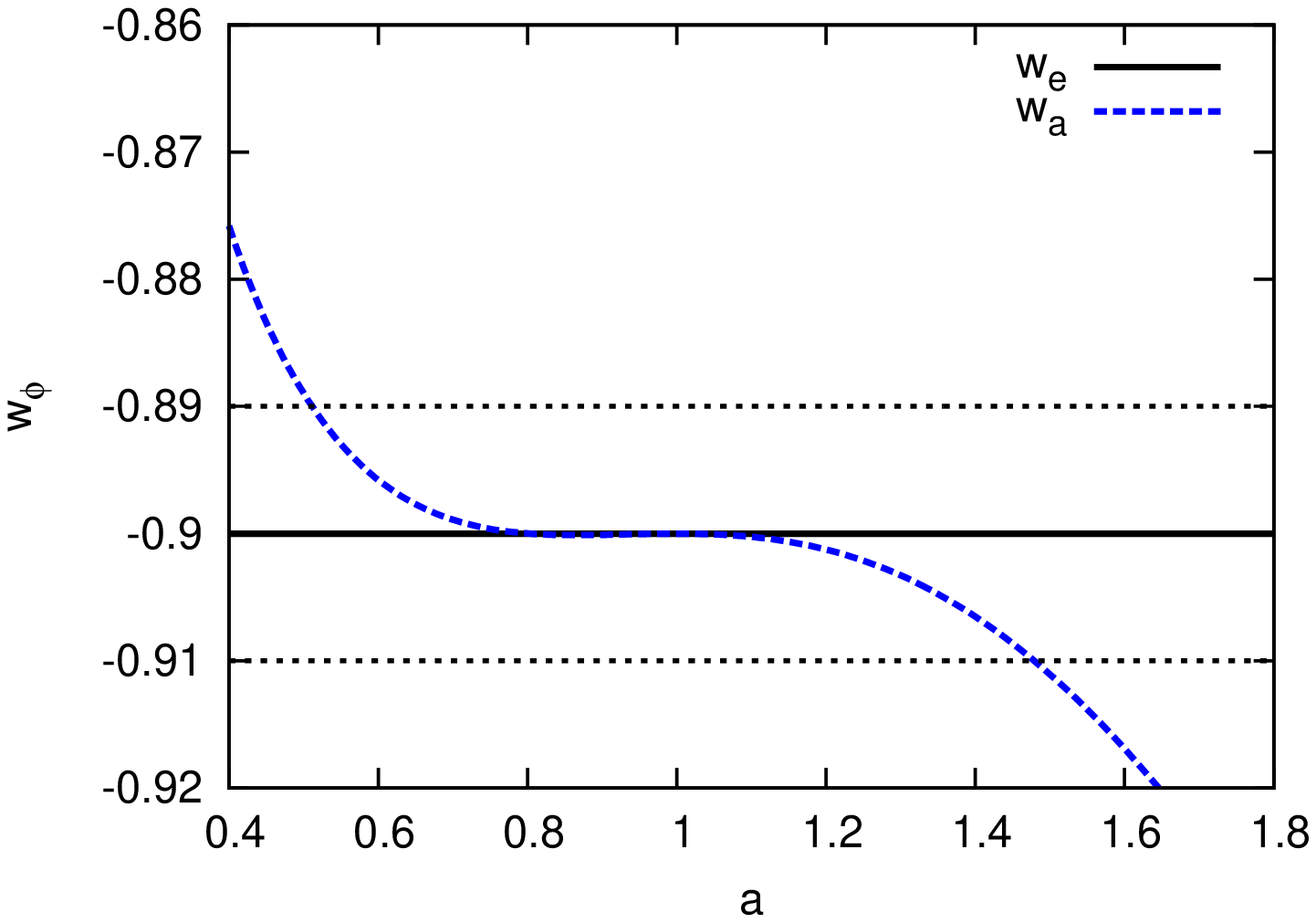}\\
 \includegraphics[width=0.3\textwidth,angle=-90]{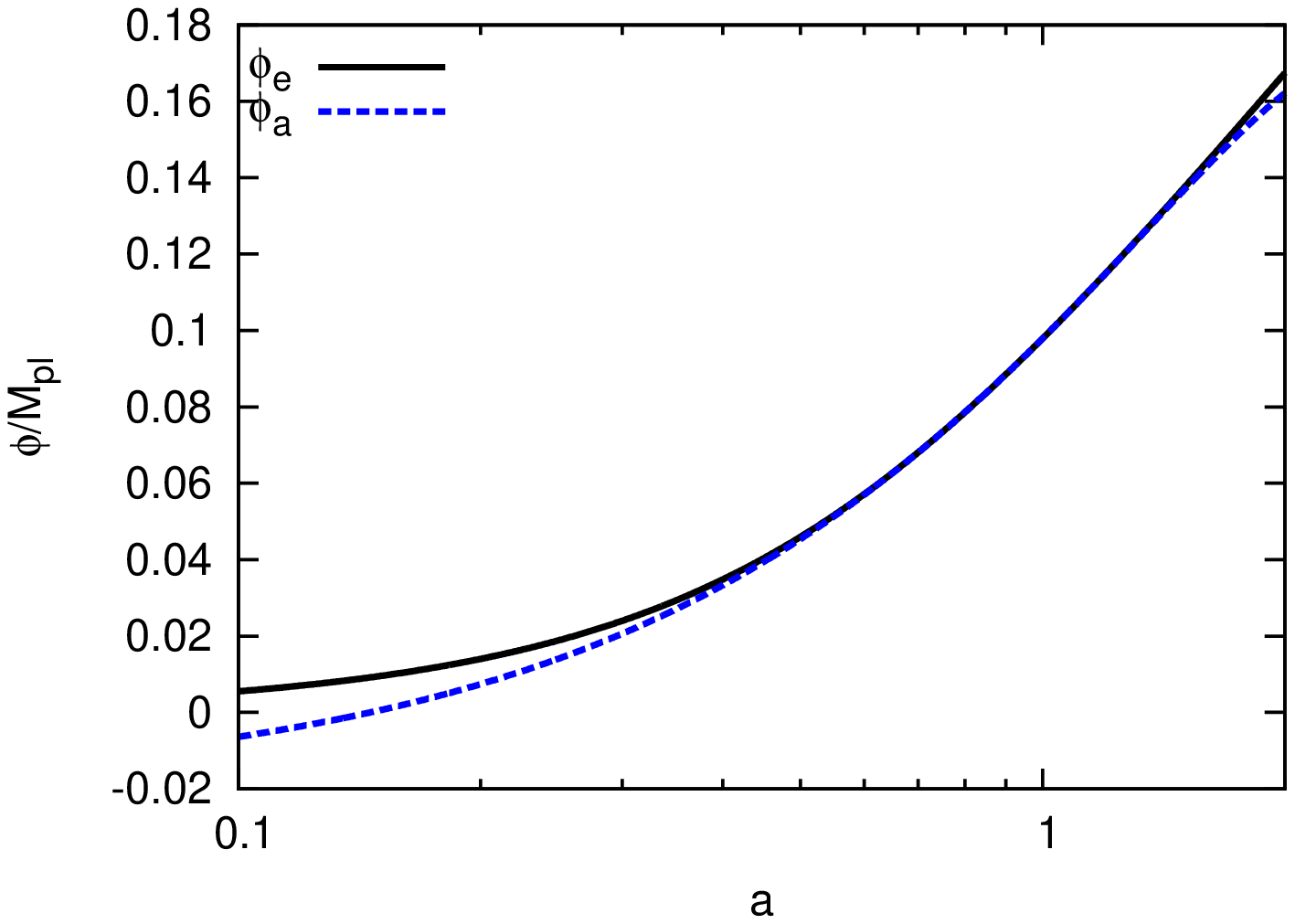}
 \caption{\textit{Top left panel}: Comparison between the exact solution for a constant equation of state for 
 the potential (solid line) and its approximate expression (dashed line), for $a\approx 1$. Different colours refer 
 to different values of $w_{\phi}$. 
 From top to bottom: 
 the black, red, blue, yellow and violet lines correspond to $w_{\phi}=-0.7, -0.8, -0.9, -0.95$ and $-0.99$, 
 respectively. 
 \textit{Top right panel}: Equation of state for the approximated potential of (\ref{eqn:Vphi_ABCD}) for 
 $w_{\phi}=-0.9$. 
 The subscripts {\it a} and {\it e} represent the approximated (blue dashed line) and the exact (black solid line) 
 solutions, respectively. Black horizontal dashed lines show differences of $1\%$ with respect to the exact value. 
 \textit{Bottom panel}: Comparison between the exact solution for the scalar field with a constant equation of state 
 $w_{\phi}=-0.9$ (solid line) and the approximate expression (dashed line), for $a\approx 1$.}
 \label{fig:VphiQn}
\end{figure}

To see how good our approximation is for $a\approx 1$, we compare the approximate expression for the potential to the 
exact solution for different values of the equation-of-state parameter $w_{\phi}$ in the top left panel of 
\autoref{fig:VphiQn}. We assumed the following cosmological parameters: $\Omega_{\rm m}=0.3$, $\Omega_{\rm de}=0.7$ 
and $H_0=70$~km~s$^{-1}$Mpc$^{-1}$. 
The value of $\phi$ at $a=1$ ranges from $\phi/M_{\rm pl}\approx 0.02$ for $w_{\phi}=-0.99$ to 
$\phi/M_{\rm  pl}\approx 0.2$ for $w_{\phi}=-0.7$. 
The approximate solution agrees very well with the analytic one over a range of values centred on $a=1$ (by 
construction) and it deviates from it at both low and high values of the scale factor (corresponding to low and high 
values of the scalar field, respectively). 
In particular, by inspecting the top right panel of \autoref{fig:VphiQn} we 
find an excellent agreement for $0.7\lesssim a \lesssim1.2$. We also note that a better agreement occurs when the 
equation-of-state parameter is not substantially different from $w_{\phi}=-1$: 
this is due to the fact that for the cosmological constant the scalar field and the potential are constant in time. If 
we require a tolerance of $1\%$ in the equation of state derived from the approximate potential, then the confidence 
interval is $0.5\lesssim a \lesssim 1.5$.

In the bottom panel of \autoref{fig:VphiQn} we show the evolution of the scalar field with respect to the scale factor 
for $w_{\phi}=-0.9$. 
We show the time evolution of the scalar field rather than that of the potential because by construction, the latter 
evolves as $a^{-3(1+w_{\phi})}$. Note how the two expressions for the scalar field agree remarkably well over a range 
$0.5\lesssim a\lesssim 1.7$. For values outside this range the approximate solution underestimates the exact one and 
it becomes negative for $a\lesssim 0.2$. This range is largely in agreement with what we found for the reconstructed 
equation-of-state parameter.

\section{\textit{k}-essence with constant \texorpdfstring{$w_{\phi}$}{w}}\label{sect:kessence}
A straightforward extension of minimally coupled scalar fields is given by models with a noncanonical kinetic term. 
These models are described by a Lagrangian of the form ${\cal L}={\cal L}(\phi,\chi)$ \citep{Chiba2000} where 
$\chi=-\frac{1}{2}g_{\mu\nu}\nabla^{\mu}\phi\nabla^{\nu}\phi$ is the canonical kinetic energy term. These models have 
been extensively used to describe dark energy scenario 
\citep{ArmendarizPicon2000,Chiba2000,ArmendarizPicon2001,Padmanabhan2002,Chimento2004,Mukhanov2006} and several works 
studied their dynamics and stability \citep{Garriga1999,Copeland2005,Rendall2006,DeSantiago2013,DeSantiago2014}. 
These models are dubbed ``$k$-essence'' models because the kinetic term $\chi$ can be responsible for the cosmic 
acceleration. A wide variety of models have been proposed and studied in different contests, such as low-energy 
effective string theory \citep{Gasperini2003}, tachyon models \citep{Padmanabhan2002,Abramo2003}, ghost condensates 
\citep{Aguirregabiria2004,Hamed2004,Piazza2004,Copeland2005}, Dirac-Born-Infeld (DBI) theories 
\citep{Alishahiha2004,Silverstein2004,Guo2008}.

The density and pressure are given by $\rho=2\chi{\cal L}_{\chi}-{\cal L}$ and $P={\cal L}$, respectively, where 
${\cal L}_{\chi}=\frac{\partial{\cal L}}{\partial\chi}$ and we will also use 
${\cal L}_{\chi\chi}=\frac{\partial^2{\cal L}}{\partial\chi^2}$ and 
${\cal L}_{\chi\phi}=\frac{\partial^2{\cal L}}{\partial\chi\partial\phi}$.
The sound speed for sub-horizon modes is
\begin{equation}\label{eqn:alphakess}
 \alpha=\frac{P_{\chi}}{\rho_{\chi}}=\left(1+\frac{2\chi{\cal L}_{\chi\chi}}{{\cal L}_{\chi}}\right)^{-1}\,.
\end{equation}

Using $P_{\phi}=w_{\phi}(a)\rho_{\phi}$, we can deduce that
\begin{equation}\label{eqn:gencons1}
 2\chi{\cal L}_{\chi}=\frac{1+w_{\phi}(a)}{w_{\phi}(a)}{\cal L}\;.
\end{equation}
From (\ref{eqn:gencons1}) we see that $k$-essence models can achieve $w_{\phi}\approx -1$ without $\chi\approx 0$. 
This means that such models need not be in the slow-roll regime to act as a dark energy component.

The energy-momentum tensor of $k$-essence is that of a perfect fluid
\begin{equation}
 T_{\mu\nu}={\cal L}_{\chi}\nabla_{\mu}\phi\nabla_{\nu}\phi+{\cal L}g_{\mu\nu}=(\rho+P)u_{\mu}u_{\nu}+Pg_{\mu\nu}\;,
\end{equation}
and velocity $u_{\mu}=\nabla_{\mu}\phi/\sqrt{2\chi}$.
The equation of motion for the scalar field is
\begin{equation}\label{eqn:Keom}
 \dot{\chi}\left({\cal L}_{\chi}+2\chi{\cal L}_{\chi\chi}\right)+
 \sqrt{2\chi}\left(2\chi{\cal L}_{\chi\phi}-{\cal L}_{\phi}\right)+
 6H\chi{\cal L}_{\chi}=0\;,
\end{equation}
and by rearranging the terms in (\ref{eqn:Keom}), the equation of motion reads \citep{Amendola2013}
\begin{equation}\label{eqn:Keom1}
 H^{\mu\nu}\nabla_{\mu}\nabla_{\nu}\phi+2\chi{\cal L}_{\chi\phi}-{\cal L}_{\phi}=
 H^{\mu\nu}\nabla_{\mu}\nabla_{\nu}\phi-{\cal L}_{\chi\phi}g^{\mu\nu}\nabla_{\mu}\phi\nabla_{\nu}\phi
 -{\cal L}_{\phi}=0\;,
\end{equation}
where
\begin{equation}
 H^{\mu\nu}={\cal L}_{\chi\chi}\nabla^{\mu}\phi\nabla^{\nu}\phi-{\cal L}_{\chi}g^{\mu\nu}\;.
\end{equation}
By inspecting (\ref{eqn:Keom1}), we notice that the equation of motion can be written in a very compact form as 
$\nabla_{\mu}J^{\mu}=-{\cal L}_{\phi}$, with $J^{\mu}={\cal L}_{\chi}\nabla^{\mu}\phi$.

Many of the $k$-essence models proposed in literature fall into one of the following types:
\begin{itemize}
 \item[(A)] Models of type A are given by \citep{Mukhanov2006,Bose2009,DeSantiago2011,Sharif2012}
\end{itemize}
\begin{equation}
 {\cal L}=M^4F(\chi)-V(\phi)\;,
\end{equation}
where $M$ has dimensions of mass and $F$ is a dimensionless function. In the following, it is helpful to consider 
$F(\chi)$ to be a power-law $F(\chi)=\left(\frac{\chi}{M^4}\right)^n$, for $n$ constant. Setting $n=1$ implies 
${\cal L}=\chi-V(\phi)$ which corresponds to the Quintessence case discussed in \autoref{sect:mcsf}.

\begin{itemize}
 \item[(B)] Models of type B are given by \citep{ArmendarizPicon1999,Chiba2000,Hamed2004,Scherrer2004}
\end{itemize}
\begin{equation}
 {\cal L}=G(\chi)V(\phi)\;,
\end{equation}
where $G$ is a dimensionless function.

Common Lagrangians proposed in literature, which are mainly for purely kinetic $k$-essence model 
(i.e. $V(\phi)=M^4=$constant), are \citep{Sen2002,Sen2002a,Chimento2004,Sen2005,Yang2008,Yang2012,Yang2015}
\begin{enumerate}
 \item $G(\chi) = -\sqrt{1+2\eta\frac{\chi}{M^4}}$,
 \item $G(\chi) = \left[2\left(\frac{\chi}{M^4}\right)^n-1\right]^{\frac{1}{2n}}$,
 \item $G(\chi) = -\left[1+2\eta\left(\frac{\chi}{M^4}\right)^n\right]^{\frac{1}{2n}}$,
 \item $G(\chi) = A_1\sqrt{\frac{\chi}{M^4}}-A_2\left(\frac{\chi}{M^4}\right)^{\alpha}$,
 \item $G(\chi) = -\left(1-2\frac{\chi}{M^4}\right)^{\beta}$,
 \item $G(\chi) = \frac{\chi}{M^4}-\sqrt{\frac{\chi}{M^4}}$,
\end{enumerate}
where $n$, $\alpha$, $\beta$, $A_1$ and $A_2$ are constant and $\eta=\pm 1$. Typically it is not possible to transform 
between $\eta=+1$ and $\eta=-1$ via a simple redefinition of the scalar field.

An interesting Lagrangian to consider is the ghost condensate model \citep{Hamed2004}
\begin{equation}\label{eqn:ggcm}
 {\cal L}=K(\phi)\chi+L(\phi)\frac{\chi^2}{M^4}\;,
\end{equation}
where $K(\phi)<0$ and $L(\phi)$ are dimensionless potentials and $M$, again, has dimensions of mass. 
If one defines the scalar field $\psi$ by
\begin{equation}
 \left(\frac{d\psi}{d\phi}\right)^2=\frac{L}{|K|}\;,
\end{equation}
and write $X=-\frac{1}{2}g_{\mu\nu}\nabla^{\mu}\psi\nabla^{\nu}\psi,$ then
\begin{equation}
 {\cal L}=V(\phi)\left(-\frac{X}{M^4}+\frac{X^2}{M^8}\right)\;,
\end{equation}
where $V(\phi)=[K(\phi)]^2/L(\phi)$ if $K<0$. Hence, this can be considered as a model of type B with 
$G(\chi)=-\chi/M^4+\chi^2/M^8$.

\begin{itemize}
 \item[(C)] Models of type C are given by \citep{Piazza2004}
\end{itemize}
\begin{equation}
 {\cal L}=-\chi-N(\chi)V(\phi)\;,
\end{equation}
where $N=\left(\frac{\chi}{M^4}\right)^n$ is a dimensionless function. The model represents a generalization of the 
dilatonic ghost condensate model and it is a special case of (\ref{eqn:ggcm}), where $K(\phi)=-1$, 
$L(\phi)=\frac{V(\phi)}{M^4}$ and $N(\chi)=\chi^2$.

\subsection{Type A models with constant \texorpdfstring{$w_{\phi}$}{w}}\label{sect:keA}
For models of type A, with $F(\chi)$ being a power-law, we have $\frac{\chi F_{\chi}}{F}=n$ and 
$\frac{\chi F_{\chi\chi}}{F_{\chi}}=n-1$ and therefore $\alpha=(2n-1)^{-1}$, constant. 
For a general $w_{\phi}(a)$ we have
\begin{align}
 V(\phi) & = \frac{3H_0^2M_{\rm pl}^2\Omega_{\rm de}[1-(2n-1)w_{\phi}(a)]g(a)}{16\pi n}\;,\label{eqn:VaA}\\
 \frac{d\phi}{da} & = \sqrt{2}\frac{M^2}{H_0}
 \left(\frac{3H_0^2M_{\rm pl}^2\Omega_{\rm de}}{16\pi nM^4}\right)^{\frac{1}{2n}}
 \frac{\{[1+w_{\phi}(a)]g(a)\}^{\frac{1}{2n}}}{aE(a)}\;,\label{eqn:phipA}\\
 \frac{\phi-\phi_0}{M_{\rm pl}} & = \sqrt{2}\frac{M^2}{H_0M_{\rm pl}}
 \left(\frac{3H_0^2M_{\rm pl}^2\Omega_{\rm de}}{16\pi nM^4}\right)^{\frac{1}{2n}}
 \int_1^a\frac{\{[1+w_{\phi}(x)]g(x)\}^{\frac{1}{2n}}}{xE(x)}~dx\;.\label{eqn:phiaA}
\end{align}
It is, therefore, possible in principle to find $\phi(a)$, at least numerically. Note that $n\neq 0$, otherwise the 
potential and the scalar field diverge. Our general results are consistent with \cite{Mamon2015} if we set $M^4=1$ 
and $F(\chi)=\chi^2$ and with \cite{Ossoulian2016}.

If $w_{\phi}$ is constant, we can recover analogous results to the Quintessence case. 
In this case, (\ref{eqn:phiaA}) becomes
\begin{equation}
 \frac{\phi-\phi_0}{M_{\rm pl}}=-\frac{2\sqrt{2}M^2}{3w_{\phi}H_0M_{\rm pl}\Omega_{\rm m}^{1/2}}
 \left(\frac{3H_0^2M_{\rm pl}^2\Omega_{\rm de}(1+w_{\phi})}{16\pi nM^4}\right)^{\frac{1}{2n}}
 \left(\frac{\Omega_{\rm m}}{\Omega_{\rm de}}\right)^{\frac{1+w_{\phi}-n}{2nw_{\phi}}}
 \int_{\sinh^{-1}\sqrt{\frac{\Omega_{\rm de}}{\Omega_{\rm m}}}}^{\sinh^{-1}\left(
 \sqrt{\frac{\Omega_{\rm de}}{\Omega_{\rm m}}}a^{-\frac{3w_{\phi}}{2}}\right)}dx\,
 \sinh^{\frac{(1+w_{\phi})(1-n)}{nw_{\phi}}}x\,.
 \label{eqn:phiawA}
\end{equation}
It is not possible to compute this integral analytically for general $w_{\phi}$ and $n$, but it at least illustrates 
that a solution exists and the solution can be computed numerically. It is also important to notice that for a given 
equation-of-state parameter, the scalar field and its potential are not uniquely determined since for a given 
$w_{\phi}$, these two quantities depend also on $n$. 
Note also that equations~(\ref{eqn:VaA}), (\ref{eqn:phiaA}) and~(\ref{eqn:phiawA}) reduce to the Quintessence case for 
$n=1$. 
At early and late times, the potential is given by 
$V_{\rm E}(\phi)\propto\phi^{-\frac{2n(1+w_{\phi})}{[n-(1+w_{\phi})]}}$ and 
$V_{\rm L}(\phi)\propto\phi^{-\frac{2n}{(n-1)}}$, respectively, which are in agreement with \cite{Ossoulian2016}.

Since analytical solutions are not possible, we find it useful to derive approximated expressions also for 
$a\approx 1$. In this case, (\ref{eqn:phipA}) is approximated by
\begin{equation}
 \frac{d\phi}{da}=\sqrt{2}\frac{M^2}{H_0}\left[\frac{3H_0^2M_{\rm pl}^2\Omega_{\rm de}(1+w_{\phi})}
 {16\pi nM^4}\right]^{\frac{1}{2n}}
 \left[1+\frac{n-3+3(\Omega_{\rm de}n-1)w_{\phi}}{2n}(a-1)\right]\;,
\end{equation}
which leads to
\begin{equation}\label{eqn:phiAa1}
 \frac{\phi-\phi_0}{M_{\rm pl}}=\sqrt{2}\frac{M^2}{H_0M_{\rm pl}}
 \left[\frac{3H_0^2M_{\rm pl}^2\Omega_{\rm de}(1+w_{\phi})}{16\pi nM^4}\right]^{\frac{1}{2n}}
 \left[a-1+\frac{n-3+3(\Omega_{\rm de}n-1)w_{\phi}}{4n}(a-1)^2\right]\;.
\end{equation}
By inverting this expression to find $a(\phi)$, the potential can be written with the same functional form as 
(\ref{eqn:Vphi_ABCD}), with the following coefficients:
\begin{equation}\label{eqn:kAcoeff}
 \begin{split}
  A & = \frac{3\Omega_{\rm de}[1-(2n-1)w_{\phi}]}{16\pi n}
        \left(\frac{3+n-3(\Omega_{\rm de}n-1)w_{\phi}}{3-n-3(\Omega_{\rm de}n-1)w_{\phi}}\right)^{-3(1+w_{\phi})}\;, 
        \qquad \qquad \quad 
  B = \frac{4n^2}{[3+n-3(\Omega_{\rm de}n-1)w_{\phi}]^2}\;,\\
  C & = -2\sqrt{2}n\frac{H_0M_{\rm pl}}{M^2}
         \frac{3-n-3(\Omega_{\rm de}n-1)w_{\phi}}{[3+n-3(\Omega_{\rm de}n-1)w_{\phi}]^2}
         \left[\frac{16\pi nM^4}{3H_0^2M_{\rm pl}^2\Omega_{\rm de}(1+w_{\phi})}\right]^{\frac{1}{2n}}\;, \quad 
  D = -3(1+w_{\phi})\;.
 \end{split}
\end{equation}
When $n=1$, this reverts to the coefficients presented in (\ref{eqn:Qcoef}).

\begin{figure}
 \includegraphics[width=0.3\textwidth,angle=-90]{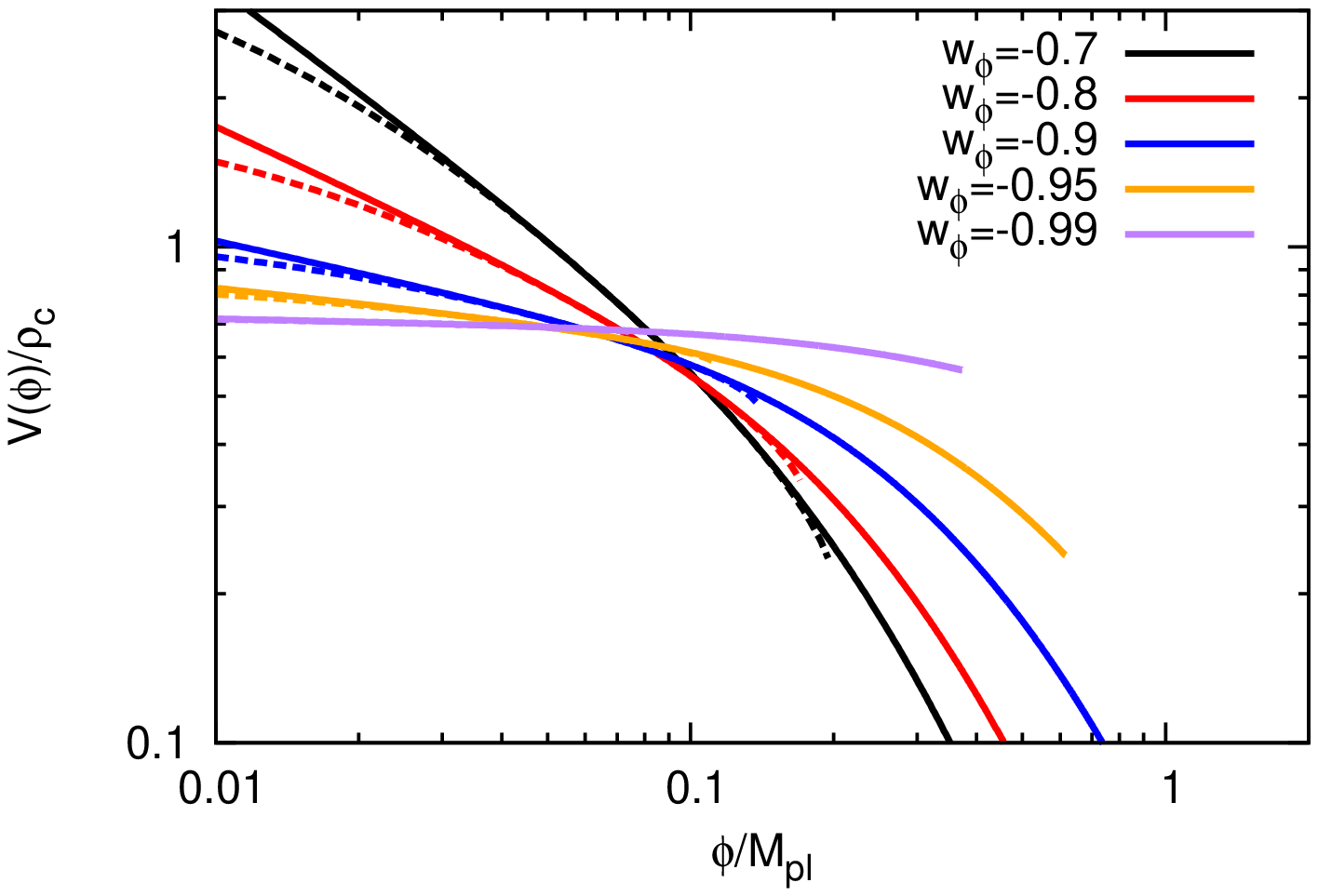}
 \includegraphics[width=0.3\textwidth,angle=-90]{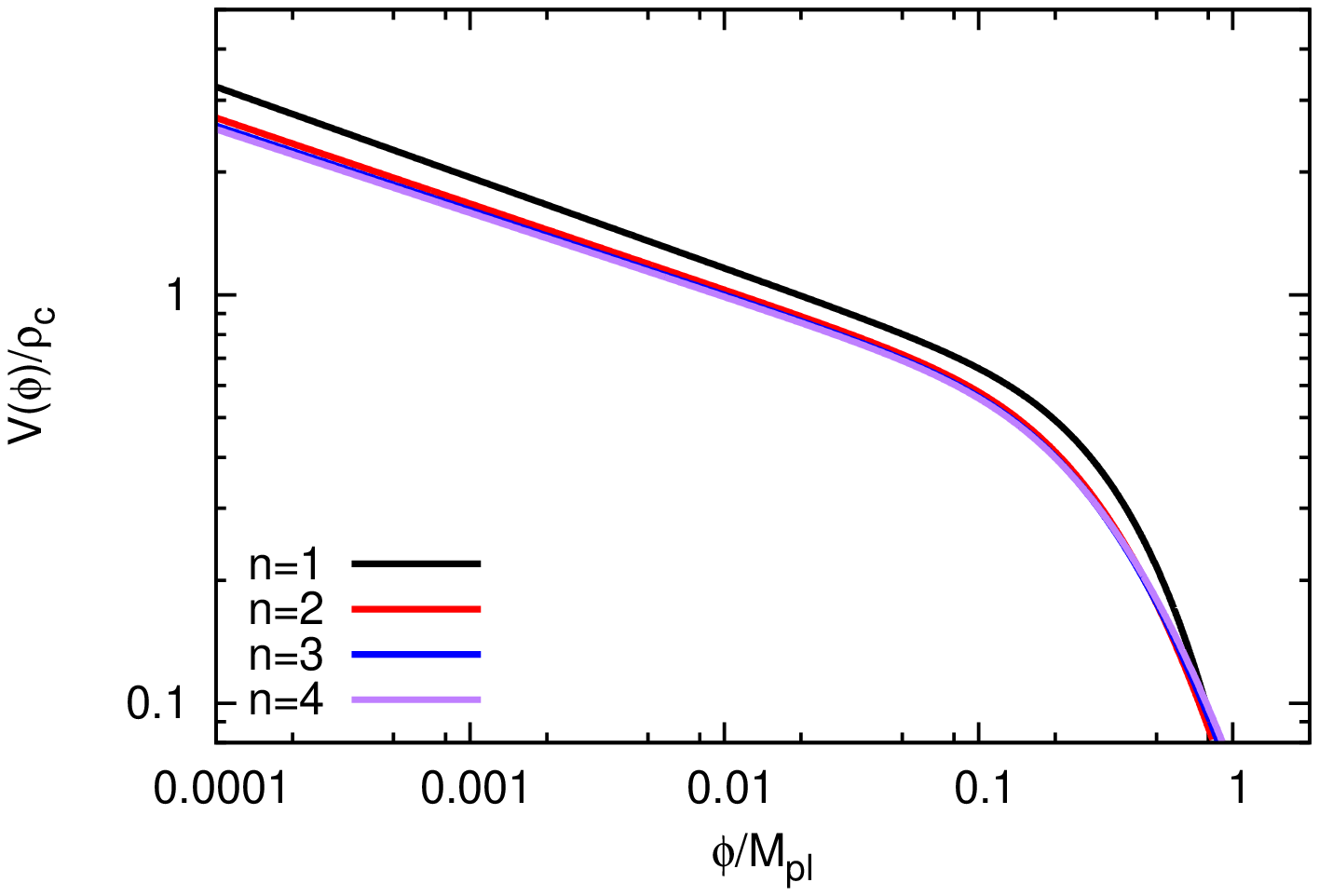}\\
 \includegraphics[width=0.3\textwidth,angle=-90]{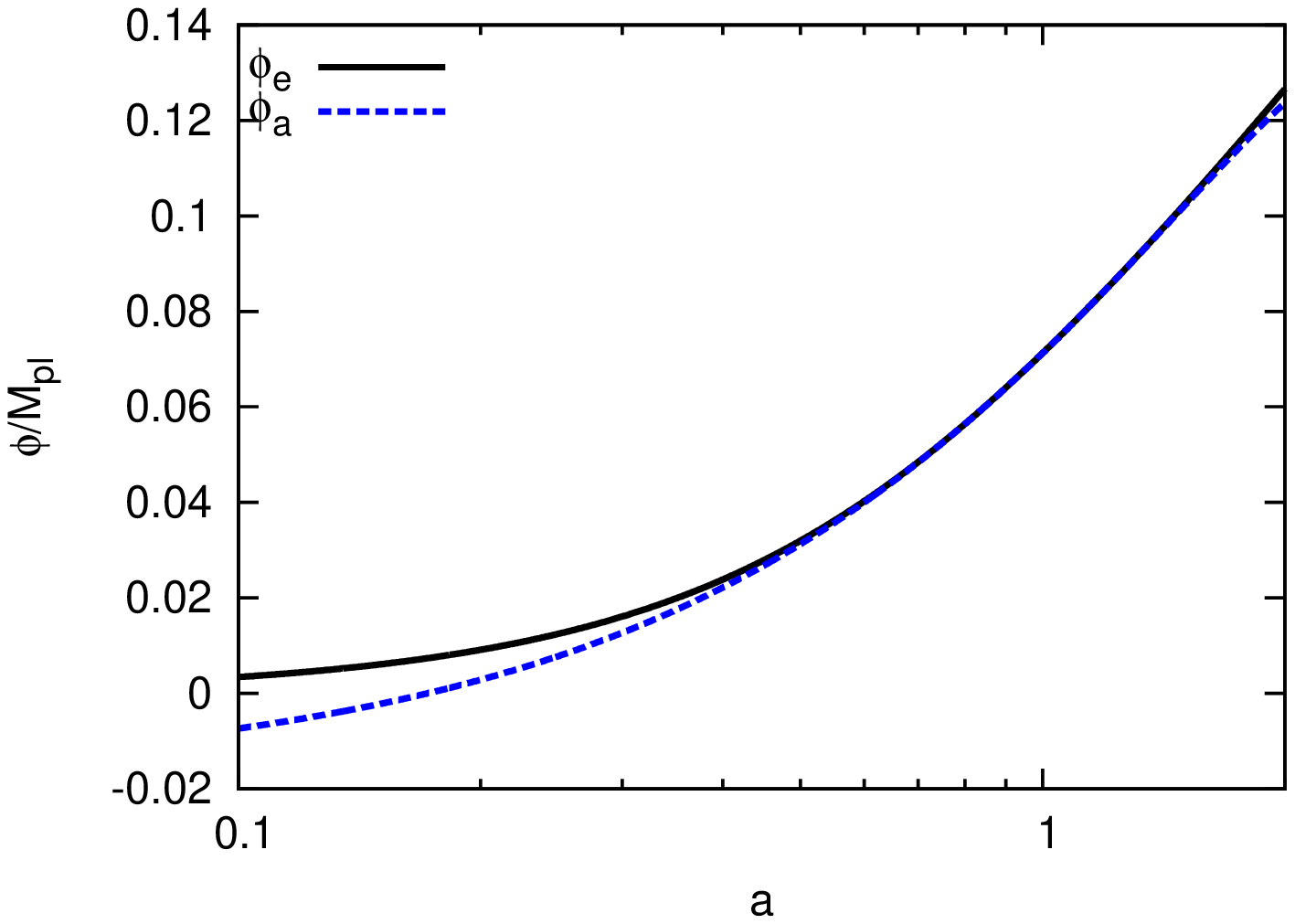}
 \includegraphics[width=0.3\textwidth,angle=-90]{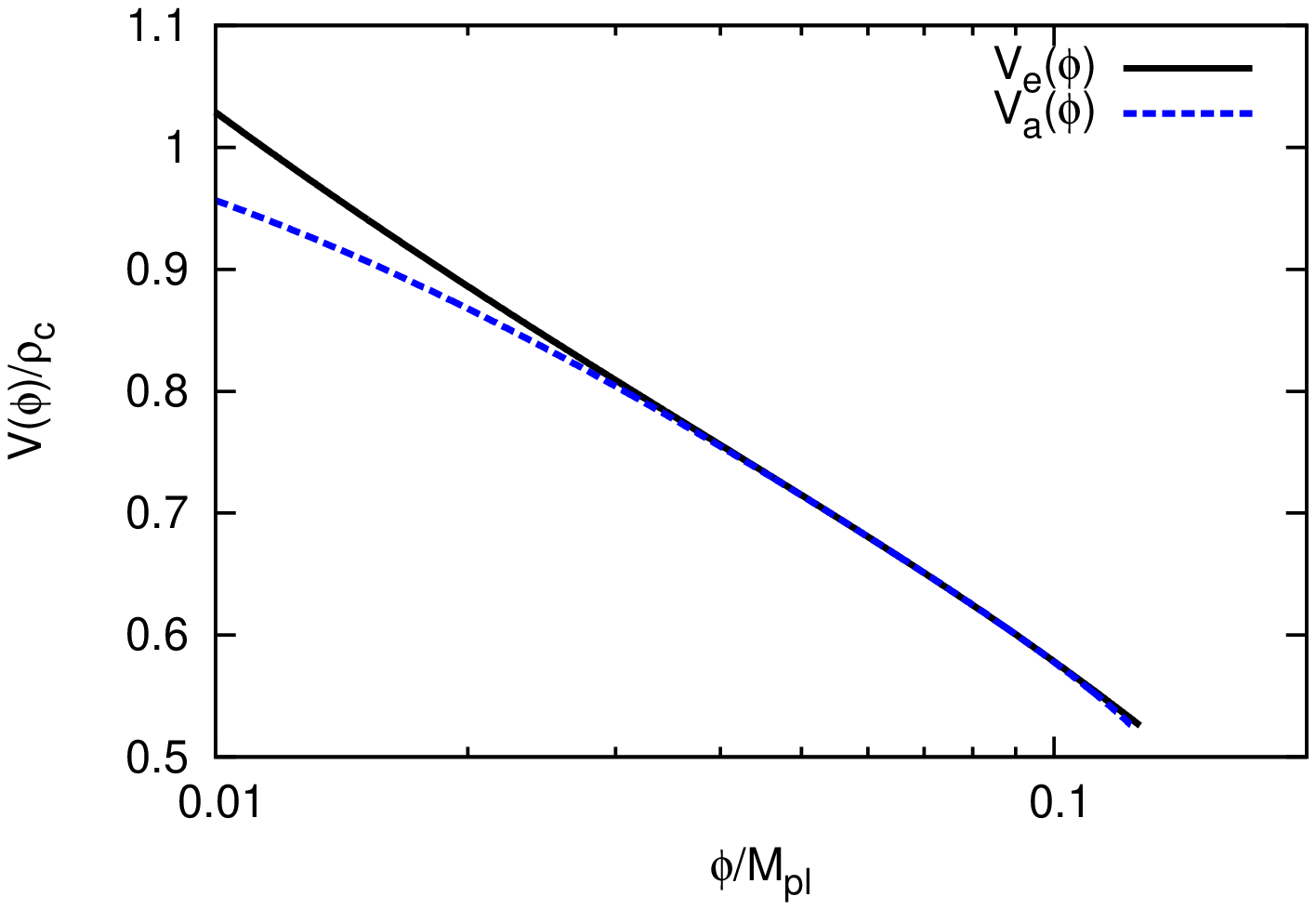}
 \caption{\textit{Top left panel}: Comparison between the full solution, generated numerically, and the approximation 
 for $a\approx 1$ for the scalar field potential for type A model with $n=2$. 
 Solid lines represent the full solution, dashed lines the approximate solution. Different colours show different 
 equations of state, labelled as in \autoref{fig:VphiQn}. 
 \textit{Top right panel}: scalar field potential for $w_{\phi}=-0.9$ for different values of $n$. 
 From top to bottom we show $n=1, 2, 3, 4$. We see that as $n$ increases, the shape of the potential quickly asymptotes 
 to that of $n\rightarrow\infty$. 
 \textit{Bottom left} (\textit{right}) \textit{panel}: 
 Scalar field (potential) for the approximated solution compared with the exact expression (\autoref{eqn:VaA} together 
 with \autoref{eqn:phiaA}) for $w_{\phi}=-0.9$.}
 \label{fig:VphiA}
\end{figure}

In the top left panel of \autoref{fig:VphiA} we make a comparison between exact numerically generated solutions and 
the approximation around $a\approx 1$. As in the case of Quintessence there is a good agreement between the two. 
We also show in the right panel that the potential for $n>2$ quickly asymptotes to the $n\rightarrow\infty$ solution. 
This should be expected from the form of (\ref{eqn:phiawA}). 
In the lower panels we show the range of validity of the approximated scalar field (left) and potential (right) for 
the approximated expression found and described by the four coefficients listed above. 
As for Quintessence models, the approximate potential recovers the exact one only for a limited range in the scale 
factor, hence also the reconstructed equation of state will be limited to the range of validity of the approximate 
potential.

We found that the behaviour of the reconstructed equation of state is very similar to the Quintessence case. More 
quantitatively, we match the true $w_{\phi}$ with $\Delta w_{\phi}=0.01$ for $0.6\lesssim a\lesssim 1.5$; 
a range that is largely in agreement with Quintessence.

\subsection{Type B models with constant \texorpdfstring{$w_{\phi}$}{w}}\label{sect:keB}
Type B models behave quite differently from type A models and it is not possible to make a direct comparison with 
Quintessence or phantom models. These models are commonly studied in literature because the kinetic term is completely 
factorized from the potential term, making the calculations relatively easy.

The evolution of the potential, the kinetic term and the sound speed are given by
\begin{align}
 V(a) & = \frac{3H_0^2M_{\rm pl}^2\Omega_{\rm de}w_{\phi}(a)g(a)}{8\pi G(\chi)}\;,\label{eqn:VaB}\\
 \frac{\chi G_{\chi}}{G} & = \frac{1+w_{\phi}(a)}{2w_{\phi}(a)}\;,\label{eqn:typeb_cons}\\ 
 \alpha & = \left(1+2\frac{\chi G_{\chi\chi}}{G_{\chi}}\right)^{-1}\;.
\end{align}
One approach would be to solve for $G(\chi)$ from (\ref{eqn:typeb_cons}). 
When $w_{\phi}$ is constant it is given by 
\begin{equation}\label{eqn:Gchi1}
 G(\chi)=\left(\frac{\chi}{M^4}\right)^{\frac{1+w_{\phi}}{2w_{\phi}}}\,.
\end{equation}
However, if we do this then we find that $\alpha=w_{\phi}$ which would mean that perturbations would be unstable if 
$w_{\phi}<0$. This is, therefore, not the correct approach for deducing a potential from constant $w_{\phi}$.

The alternative is to specify $G(\chi)$ and consider (\ref{eqn:typeb_cons}) as a constraint on $\chi$ which will be 
constant. Let $\hat\chi$ be the constant value which solves (\ref{eqn:typeb_cons}) for a specific choice of $G(\chi)$, 
then
\begin{equation}
 \frac{d\phi}{da}=\frac{1}{H_0}\left(\frac{2{\hat\chi}a}{\Omega_{\rm m}+\Omega_{\rm de}a^{-3w_{\phi}}}\right)^{1/2}\;,
\end{equation}
whose solution is
\begin{equation}
 \frac{\phi-\phi_0}{M_{\rm pl}}=-\frac{2}{3w\Omega_{\rm m}^{1/2}}\frac{\sqrt{2\hat{\chi}}}{H_0M_{\rm pl}}
      \left(\frac{\Omega_{\rm de}}{\Omega_{\rm m}}\right)^{\frac{1}{2w_{\phi}}}
      \int_{{\rm sinh}^{-1}\sqrt{\frac{\Omega_{\rm de}}{\Omega_{\rm m}}}}^
      {{\rm sinh}^{-1}\left(\sqrt{\frac{\Omega_{\rm de}}{\Omega_{\rm m}}}a^{-\frac{3w_{\phi}}{2}}\right)}dx\,
      \sinh^{-\frac{1+w_{\phi}}{w_{\phi}}}x \;.\label{eqn:typebsoln}
\end{equation}
Again this at least proves the existence of an $\phi(a)$, and hence a $V(\phi)$, which gives rise to constant 
$w_{\phi}$, although there is no analytic solution for general $w_{\phi}$. Note that this expression is equivalent to 
(\ref{eqn:phiawA}) in the limit $n\rightarrow\infty$ and with $\sqrt{\hat{\chi}}=M^2$. 
We are able to find useful approximations at early and late times. In particular we find 
$V_{\rm E}(\phi)\propto\phi^{-2(1+w_{\phi})}$ at early times and $V_{\rm L}(\phi)\propto\phi^{-2}$ at late times, 
respectively.

Let us now consider the specific case of
\begin{equation}\label{eqn:gchi}
 G(\chi)=-\frac{\chi}{M^4}+\frac{\chi^2}{M^8}\;.
\end{equation}
from which we can deduce that $\frac{\hat{\chi}}{M^4}=\frac{1-w_{\phi}}{1-3w_{\phi}}$. Hence, we find that
$\alpha=\frac{1-2\frac{\hat{\chi}}{M^4}}{1-6\frac{\hat{\chi}}{M^4}}=\frac{1+w_{\phi}}{5-3w_{\phi}}$. 
Note that $0\le\alpha\le 1$ implies that $-1\le w_{\phi} \le 1$. From (\ref{eqn:ggcm}), one finds that
\begin{equation}
 \frac{\chi L(\phi)}{M^4K(\phi)}=\frac{w_{\phi}-1}{1-3w_{\phi}}\;,
\end{equation}
and the corresponding sound speed is
\begin{equation}
 \alpha=\frac{1+2\frac{L}{K}\frac{\chi}{M^4}}{1+6\frac{L}{K}\frac{\chi}{M^4}}=\frac{1+w_{\phi}}{5-3w_{\phi}}\;,
\end{equation}
which is what should be expected from the general discussion about the generalised ghost condensate model. 
Note that in type B models with constant $w_{\phi}$, the sound speed becomes a function the of equation of state 
$\alpha=\alpha(w_{\phi})$; such models have been also studied in \cite{Battye2015} where the authors used shear and 
CMB lensing data to constrain dark energy perturbations.

\begin{figure}
 \centering
 \includegraphics[width=0.3\textwidth,angle=-90]{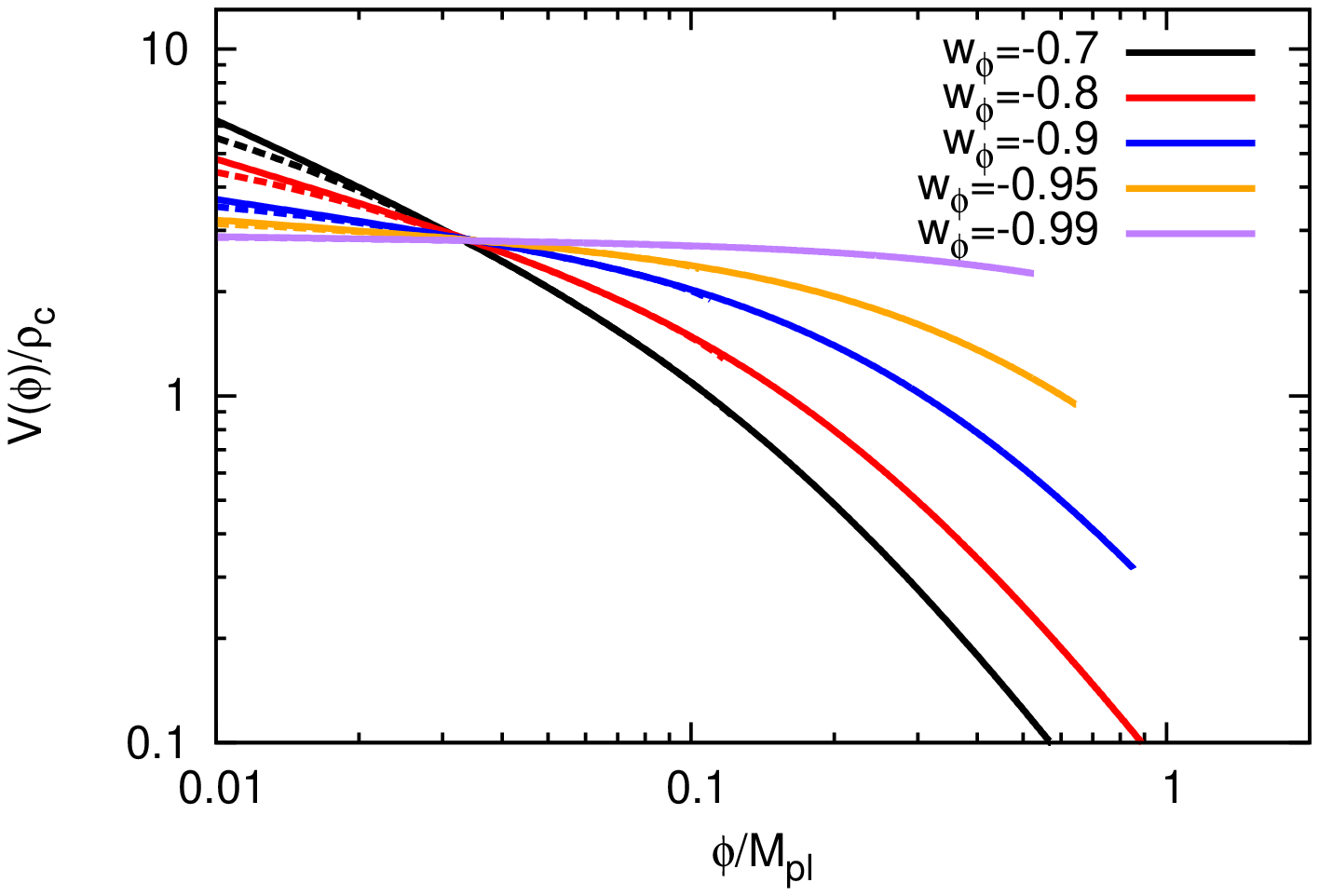}
 \includegraphics[width=0.3\textwidth,angle=-90]{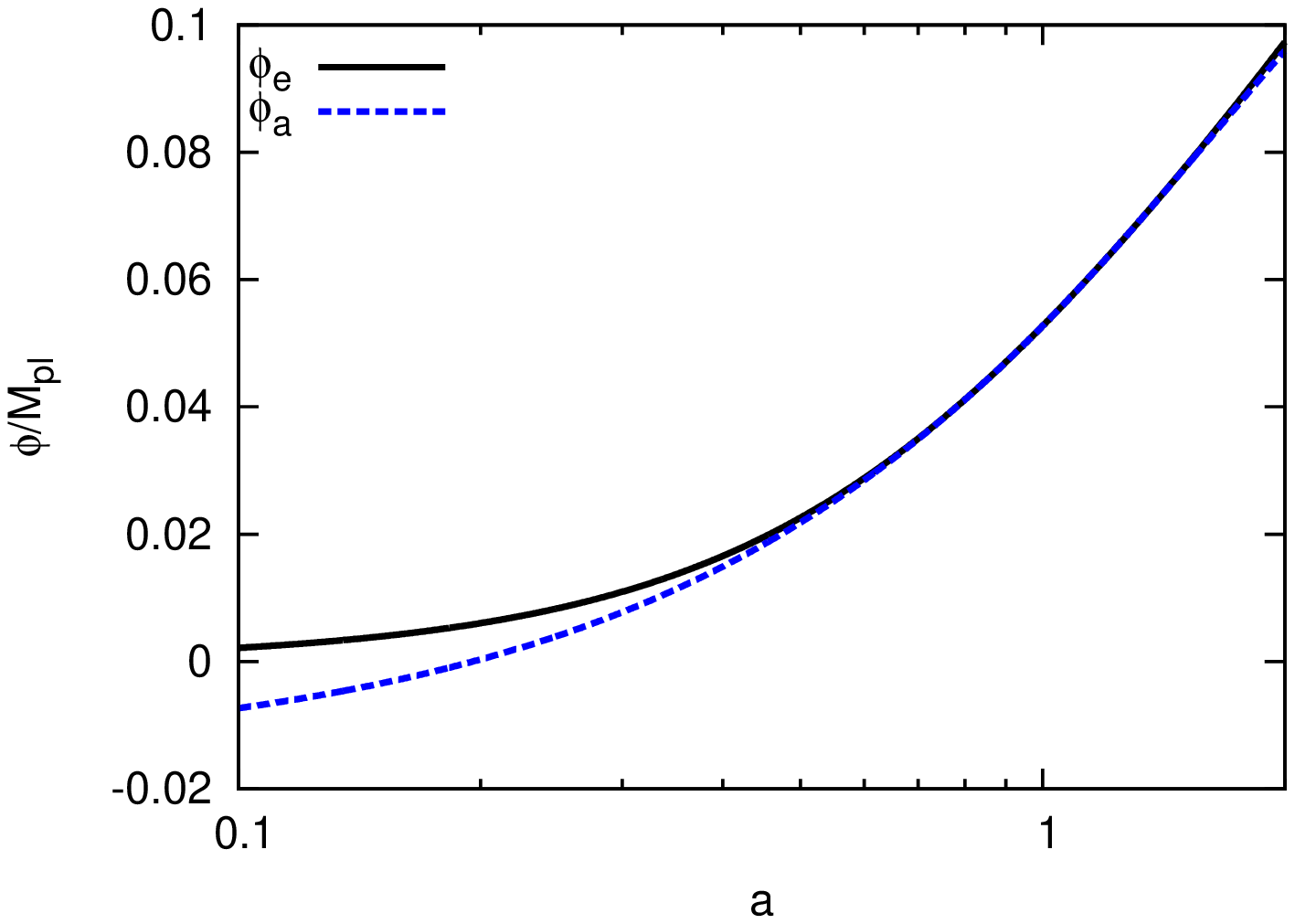}\\
 \includegraphics[width=0.3\textwidth,angle=-90]{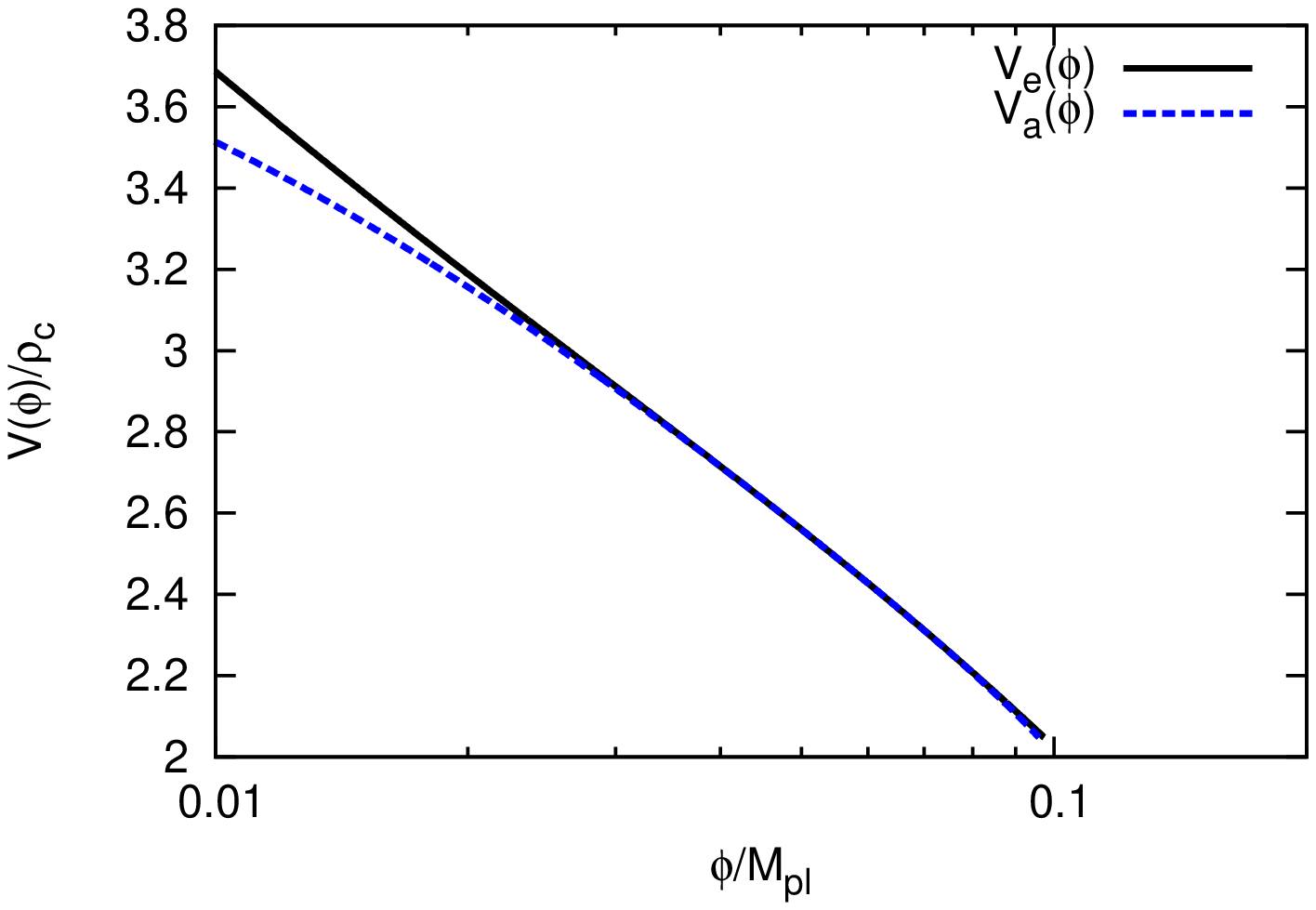}
 \caption{\textit{Left panel}: 
 Approximation of the scalar field potential for $a\approx 1$ for different constant equation-of-state 
 parameters $w_{\phi}$. Solid lines represent the exact numerical solution, while the dashed line show the 
 approximated solution. Colours are as in \autoref{fig:VphiQn}. 
 \textit{Middle} (\textit{right}) \textit{panel}: scalar field (potential) for the approximated solution compared with 
 the exact expression (\ref{eqn:VaB}) together with (\ref{eqn:typebsoln}) for $w_{\phi}=-0.9$. In all the panels we 
 assume $G(\chi)=-\frac{\chi}{M^4}+\frac{\chi^2}{M^8}$.}
 \label{fig:VphiB}
\end{figure}

For $a\approx 1$, the differential equation governing the evolution of the scalar field $\phi$ as a function of the 
scale factor $a$ is
\begin{equation}
 \frac{d\phi}{da}=\frac{\sqrt{2\hat{\chi}}}{H_0}\left[1+\frac{1}{2}(1+3\Omega_{\rm de}w_{\phi})(a-1)\right]\;,
\end{equation}
which leads to
\begin{equation}
 \frac{\phi-\phi_0}{M_{\rm pl}}=\frac{\sqrt{2\hat{\chi}}}{H_0M_{\rm pl}}
 \left[a-1+\frac{1}{4}\left(1+3\Omega_{\rm de}w_{\phi}\right)(a-1)^2\right]\;.
\end{equation}
The relation between the scale factor and the scalar field is then
\begin{equation}
 a=1-\frac{2}{1+3\Omega_{\rm de}w_{\phi}}\left\{1-\sqrt{1+(1+3\Omega_{\rm de}w_{\phi})
 \frac{H_0}{\sqrt{2\hat{\chi}}}(\phi-\phi_0)}\right\}\;.
\end{equation}
Hence, also for type B models we will have an approximate potential of the form of (\ref{eqn:Vphi_ABCD}) with 
dimensionless coefficients
\begin{equation}\label{eqn:kBcoeff}
 \begin{split}
  A & = \frac{3\Omega_{\rm de}w_{\phi}}{8\pi G(\hat{\chi})}
        \left(\frac{3\Omega_{\rm de}w_{\phi}-1}{3\Omega_{\rm de}w_{\phi}+1}\right)^{-3(1+w_{\phi})}\;, \quad 
  B = \frac{4}{(1-3\Omega_{\rm de}w_{\phi})^2}\;,\\
  C & = 4\frac{1+3\Omega_{\rm de}w_{\phi}}{(1-3\Omega_{\rm de}w_{\phi})^2}
        \frac{H_0M_{\rm pl}}{\sqrt{2\hat{\chi}}}\;, \qquad \qquad \quad
  D = -3(1+w_{\phi})\;.
 \end{split}
\end{equation}

In the left panel of \autoref{fig:VphiB} we show the validity of the approximation for the scalar field potential for 
$a\approx 1$ for different constant equation-of-state parameters, as described in the caption, for a model with 
$G(\chi)=-\frac{\chi}{M^4}+\frac{\chi^2}{M^8}$. We obtain a similar level of agreement as for Quintessence and type A 
models. The accuracy increases with the decrease of $w_{\phi}$ and is limited to an epoch centred on $a=1$. 
In the right and middle panels of \autoref{fig:VphiB} we compare the approximated expression for the potential and the 
corresponding scalar field evolution with the exact solution. Note that since (\ref{eqn:typebsoln}) is a limiting 
case of (\ref{eqn:phiawA}), the range of agreement of the equation of state for type B models is similar to what 
found for type A models.

In the general discussion of type B models, we showed that $\alpha=\alpha(w_{\phi})$ and 
$\hat{\chi}=\hat{\chi}(w_{\phi})$; each model will have, therefore, its own particular functional form and a range 
of values for the parameters ensuring their stability. 
In \autoref{tab:typeB} we show the specific functional form of $\hat{\chi}$ and $\alpha$ for several forms of 
$G(\chi)$ proposed in literature and determine when their perturbations are stable ($\alpha\geq 0$) and subluminal 
($\alpha\leq 1$).

\begin{table}[!ht]
 \caption{Dependence of $\hat{\chi}$ and $\alpha$ on the constant equation-of-state parameter $w_{\phi}$ and 
 stability conditions for the model.}
 \begin{center}
  \begin{tabular}{|c|c|c|c|}
   \hline
   \hline
   $G(\chi)$ & $\frac{\hat{\chi}(w_{\phi})}{M^4}$ & $\alpha(w_{\phi})$ & Stability \\
   \hline
   $\left(\frac{\chi}{M^4}\right)^{\frac{1+w_{\phi}}{2w_{\phi}}}$ & - & $w_{\phi}$ & $w_{\phi}\geq 0$ \\
   \hline
   $-\frac{\chi}{M^4}+\frac{\chi^2}{M^8}$ & $\frac{1-w_{\phi}}{1-3w_{\phi}}$ & $\frac{1+w_{\phi}}{5-3w_{\phi}}$ & 
   $-1\leq w_{\phi} \leq 1$ \\
   \hline
   $-\frac{\chi}{M^4}+\left(\frac{\chi}{M^4}\right)^n$ & 
   $\left(\frac{1-w_{\phi}}{1-(2n-1)w_{\phi}}\right)^{\frac{1}{n-1}}$ & 
   $\frac{1+w_{\phi}}{(2n+1)-(2n-1)w_{\phi}}$ & $-1\leq w_{\phi} \leq 1$ for $n>0$ \\
   \hline
   $-\sqrt{1+2\eta\frac{\chi}{M^4}}$ & $-\eta\frac{1+w_{\phi}}{2}$ & $-w_{\phi}$ & $-1\leq w_{\phi} \leq 0$ \\
   \hline
   $\left[2\left(\frac{\chi}{M^4}\right)^n-1\right]^{\frac{1}{2n}}$ & 
   $\left(\frac{1+w_{\phi}}{2}\right)^{\frac{1}{n}}$ & 
   $-\frac{1}{2n-1}w_{\phi}$ & $1-2n\leq w_{\phi}\leq 0$ for $n>\frac{1}{2}$ \\
   & & & $0\leq w_{\phi}\leq 1-2n$ for $n<\frac{1}{2}$ \\
   \hline
   $-\left[1+2\eta\left(\frac{\chi}{M^4}\right)^n\right]^{\frac{1}{2n}}$ & 
   $\left(-\eta\frac{1+w_{\phi}}{2}\right)^{\frac{1}{n}}$ & 
   $-\frac{1}{2n-1}w_{\phi}$ & $1-2n\leq w_{\phi}\leq 0$ for $n>\frac{1}{2}$ \\
   & & & $0\leq w_{\phi}\leq 1-2n$ for $n<\frac{1}{2}$ \\
   \hline
   $\frac{\chi}{M^4}-\sqrt{\frac{\chi}{M^4}}$ & $\frac{1}{(1-w_{\phi})^2}$ & $\frac{1+w_{\phi}}{2}$ & 
   $-1\leq w_{\phi} \leq 1$ \\
   \hline
   $-\left(1-2\frac{\chi}{M^4}\right)^{\beta}$ & $\frac{1+w_{\phi}}{2[1+(1-2\beta)w_{\phi}]}$ & 
   $\frac{\beta w_{\phi}}{\beta-1+(2\beta-1)w_{\phi}}$ & $w_{\phi}\leq -1$ or $w_{\phi}\geq 0$ for $\beta<0$ or 
   $\beta>1$ \\
   & & & $-1\leq w_{\phi}\leq 0$ for $0<\beta<1$ \\
   \hline
   $A_1\sqrt{\frac{\chi}{M^2}}-A_2\left(\frac{\chi}{M^2}\right)^{\alpha}$ & 
   $\left\{\frac{A_2}{A_1}[1-(2\alpha-1)w_{\phi}]\right\}^{\frac{2}{1-2\alpha}}$ & $\frac{1+w_{\phi}}{2\alpha}$ & 
   $-1\leq w_{\phi}\leq 2\alpha-1$ for $\alpha>0$ \\
   & & & $2\alpha-1\leq w_{\phi}\leq -1$ for $\alpha<0$ \\
   \hline
  \end{tabular}
  \label{tab:typeB}
 \end{center}
\end{table}

\subsection{Type C models with constant \texorpdfstring{$w_{\phi}$}{w}}\label{sect:keC}
Type C models resemble phantom models discussed in \autoref{sect:mcsf}. They reduce to phantom models when the 
function $N(\chi)$ is constant, that is, $n=0$ for the power-law choice of $N(\chi)$. Despite the apparent complexity, 
type C models, in contrast to type A and B models, have a general analytical solution when $N(\chi)$ is a power-law. 
Key equations for general $n>0$ and $w_{\phi}(a)$ are
\begin{align}
 V(a) & = -\frac{[1-w(a)]M^{4n}}{[1-(2n-1)w_{\phi}(a)]^n}(n-1)^{n-1}
           \left(\frac{3H_0^2M_{\rm pl}^2\Omega_{\rm de}g(a)}{16\pi}\right)^{1-n}\;,\label{eqn:VphiaC}\\
 \frac{d\phi}{da} & = \left\{\frac{3M_{\rm pl}^2\Omega_{\rm de}[1-(2n-1)w_{\phi}(a)]g(a)}
                      {8\pi(n-1)a^2E^2(a)}\right\}^{\frac{1}{2}}\;,\label{eqn:phiaC}\\
 \alpha & = \frac{1+w_{\phi}}{2n+1-(2n-1)w_{\phi}}\;,\label{eqn:alphaC}
\end{align}
for the evolution of the potential, of the scalar field and of the sound speed, respectively. Note in particular that 
for $n=0$ we recover results for the phantom models. The evolution of the scalar field as a function of the scale 
factor is
\begin{equation}\label{eqn:phiC}
 \frac{\phi-\phi_0}{M_{\rm pl}}=-\frac{2}{3w_{\phi}}\sqrt{\frac{3[1+(1-2n)w_{\phi}]}{8\pi(n-1)}}
 \left[\sinh^{-1}{\left(\sqrt{\frac{\Omega_{\rm de}}{\Omega_{\rm m}}}a^{-\frac{3w_{\phi}}{2}}\right)}
 -\sinh^{-1}{\left(\sqrt{\frac{\Omega_{\rm de}}{\Omega_{\rm m}}}\right)}\right]\;,
\end{equation}
and the corresponding potential is
\begin{equation}\label{eqn:VphiC}
 V(\phi) = k\sinh^{-\frac{2(n-1)(1+w_{\phi})}{w_{\phi}}}{\left[-\frac{3w_{\phi}}{2}
           \sqrt{\frac{8\pi(n-1)}{3[1+(1-2n)w_{\phi}]}}\frac{\phi}{M_{\rm pl}}\right]}\;,
\end{equation}
with the constant $k$
\begin{equation}
 k = \frac{3H_0^2M_{\rm pl}^2\Omega_{\rm de}(1-w_{\phi})}{16\pi(1-n)}
     \left[\frac{3H_0^2M_{\rm pl}^2\Omega_{\rm de}[1-(2n-1)w_{\phi}]}{16\pi(n-1)M^4}\right]^{-n}
     \left(\frac{\Omega_{\rm m}}{\Omega_{\rm de}}\right)^{-\frac{(n-1)(1+w_{\phi})}{w_{\phi}}}\;,
\end{equation}
when we choose as we did in the Quintessence case
\begin{equation}
 \frac{\phi_0}{M_{\rm pl}}=-\frac{2}{3w_{\phi}}\sqrt{\frac{3[1+(1-2n)w_{\phi}]}{8\pi(n-1)}}
 \sinh^{-1}{\left(\sqrt{\frac{\Omega_{\rm de}}{\Omega_{\rm m}}}\right)}\;.
\end{equation}
These expressions are as expected similar to those for Quintessence models.

\begin{figure}
 \centering
 \includegraphics[width=0.3\textwidth,angle=-90]{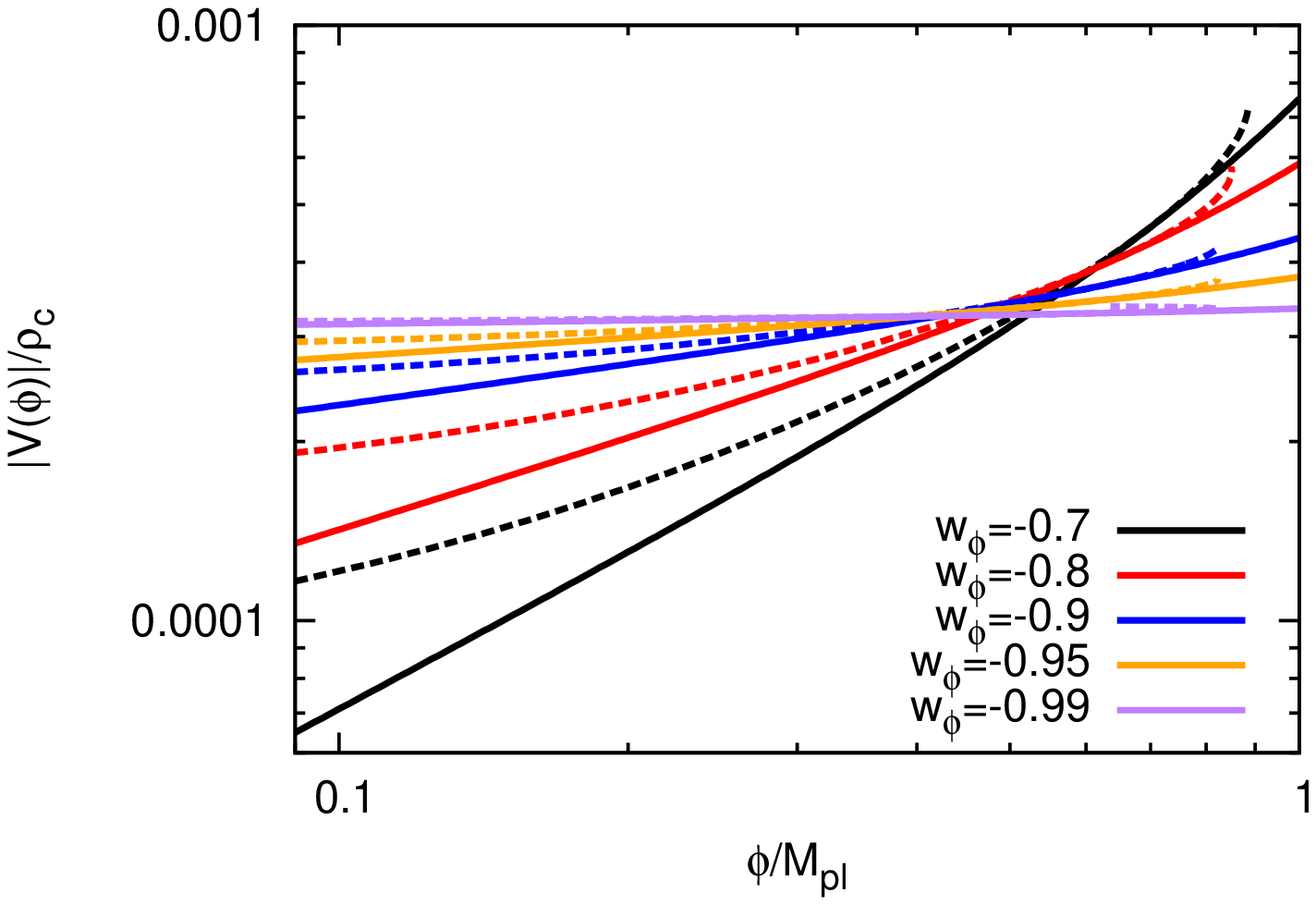}
 \includegraphics[width=0.3\textwidth,angle=-90]{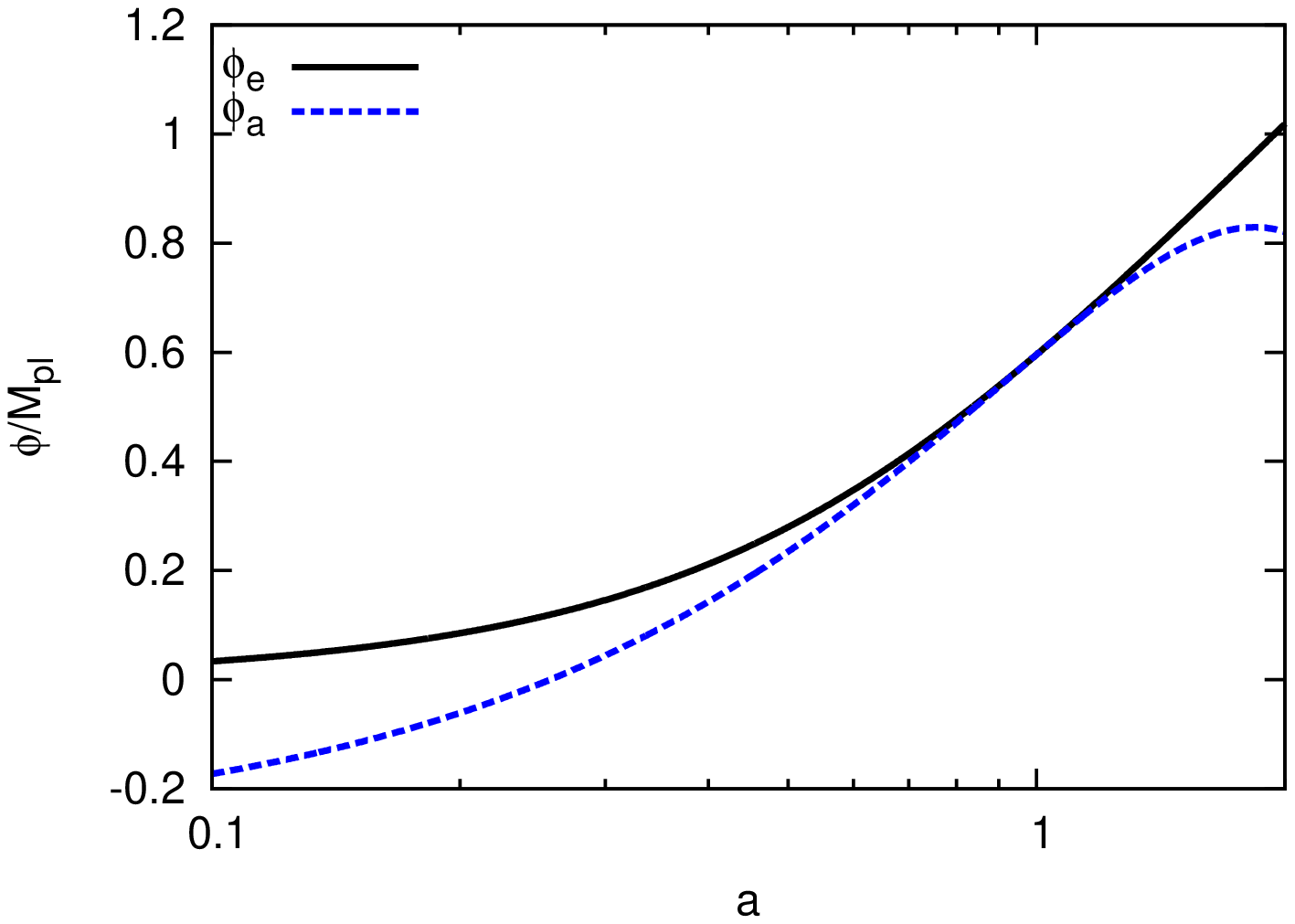}\\
 \includegraphics[width=0.3\textwidth,angle=-90]{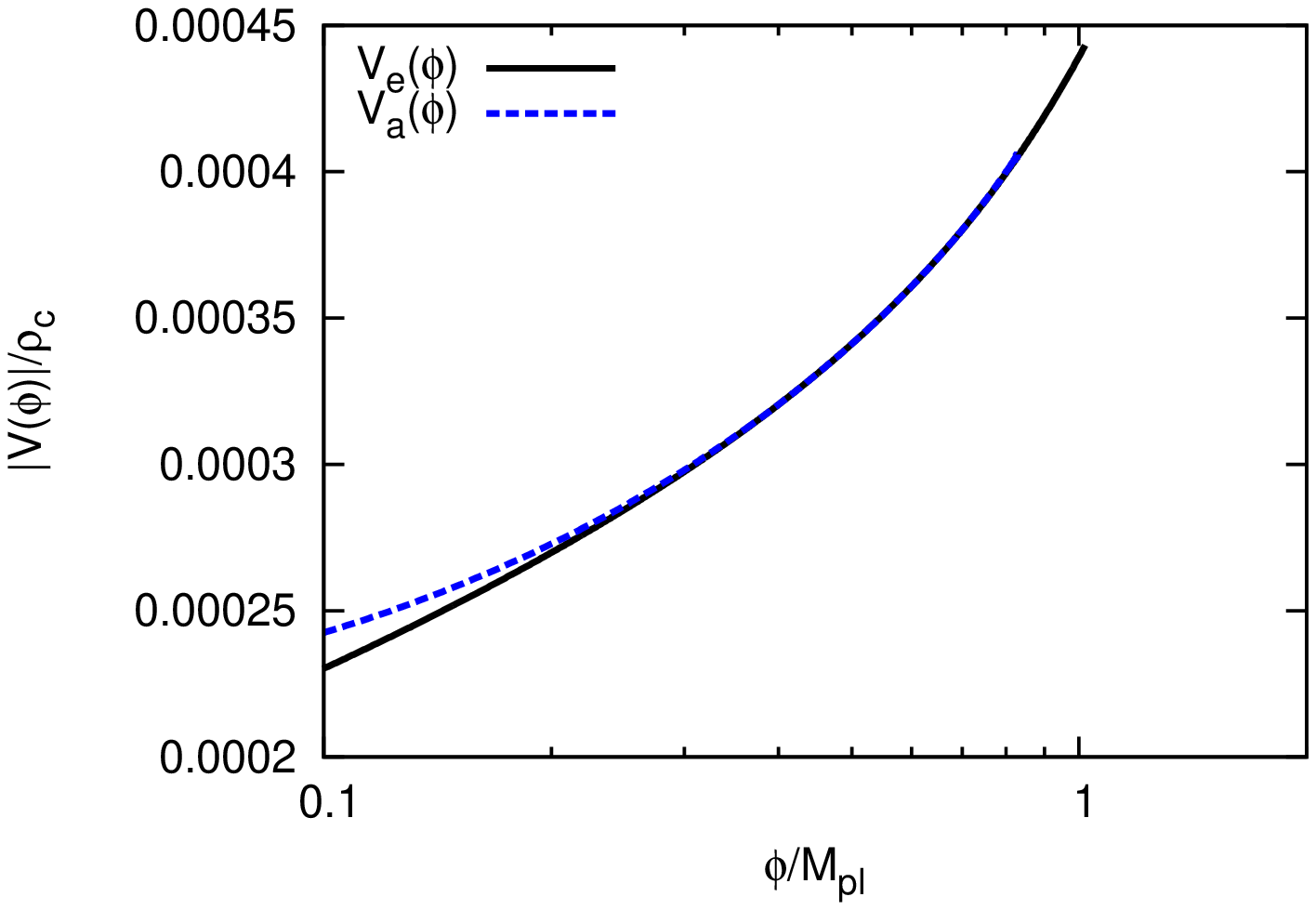}
 \caption{\textit{Top Left panel}: Comparison between the exact solution for constant equation of state for the 
 absolute value of the potential (solid line) and its approximate expression, for $a\approx 1$ (dashed line) for 
 type C models. Line styles and colours are as in \autoref{fig:VphiQn}. 
 \textit{Top right} (\textit{middle}) \textit{panel}: scalar field (absolute value of the potential) for the 
 approximated solution compared with the exact expression in (\ref{eqn:phiC}) [(\ref{eqn:VphiC})] for a model with 
 $w_{\phi}=-0.9$. The subscripts \textit{a} and \textit{e} represent the approximated (blue dashed line) and the exact 
 (black solid line) solutions, respectively.}
 \label{fig:VphiC}
\end{figure}

Given the general form of the potential for type C models, we can expect a similar behaviour to Quintessence models 
for $a\ll1$, $a\approx 1$ and $a\gg 1$. 
In particular, at early times $V_{\rm E}(\phi)\propto\phi^{\frac{2(1-n)(1+w_{\phi})}{w_{\phi}}}$ and at late times 
we recover the usual exponential behaviour $V_{\rm L}(\phi)\propto\exp{(\phi/M_{\rm pl})}$ typical of the minimally 
coupled models.

For $a\approx 1$, the  scalar field is
\begin{equation}\label{eqn:phiCa1}
 \frac{\phi-\phi_0}{M_{\rm pl}}=\sqrt{\frac{3\Omega_{\rm de}[1-(2n-1)w_{\phi}]}{8\pi(n-1)}}
 \left[a-1-\frac{1}{4}\left(2+3\Omega_{\rm m}w_{\phi}\right)(a-1)^2\right]\;,
\end{equation}
and the corresponding potential can be once again written in the general approximated form of (\ref{eqn:Vphi_ABCD}), 
with coefficients:
\begin{equation}\label{eqn:kCcoeff}
 \begin{split}
  A & = \frac{3\Omega_{\rm de}(1-w_{\phi})}{16\pi(1-n)}
        \left[\frac{3H_0^2M_{\rm pl}^2\Omega_{\rm de}[1+(1-2n)w_{\phi}]}{16\pi(n-1)M^4}\right]^{-n}
        \left(\frac{4+3\Omega_{\rm m}w_{\phi}}{2+3\Omega_{\rm m}w_{\phi}}\right)^{-3(1-n)(1+w_{\phi})}\;, \quad 
  B = \frac{4}{(4+3\Omega_{\rm m}w_{\phi})^2}\;,\\
  C & = -4\frac{2+3\Omega_{\rm m}w_{\phi}}{(4+3\Omega_{\rm m}w_{\phi})^2}
         \sqrt{\frac{8\pi(n-1)}{3\Omega_{\rm de}[1+(1-2n)w_{\phi}]}}\;, \qquad \qquad \qquad \qquad \qquad \qquad 
  \qquad \qquad \quad 
  D = -3(1-n)(1+w_{\phi})\;.
 \end{split}
\end{equation}

As with Quintessence, Type A and B models, we show a comparison between the approximation and the exact solutions 
in \autoref{fig:VphiC}. The picture is similar to the previous ones but the range of scale factor where the 
approximation is good is more restricted, $0.8\lesssim a\lesssim 1.2$. 
This is due to the fact that for type C models, the potential has a stronger dependence on the scale factor with 
respect to the other models, given by the $1-n$ power of $g(a)$ in (\ref{eqn:VphiaC}).

\section{Potential for non-constant \texorpdfstring{$w_{\phi}$}{w}}\label{sect:wa}

In the previous sections we have shown that a potential of the form (\ref{eqn:Vphi_ABCD}) is a good approximation to 
that for scalar field models with constant $w_\phi$ - for both minimal and non-minimal kinetic terms - for some choice 
of the parameters $A$, $B$, $C$ and $D$ over a range of the scale factors around $a\approx 1$. 

\begin{figure}
 \centering
 \includegraphics[width=0.3\textwidth,angle=-90]{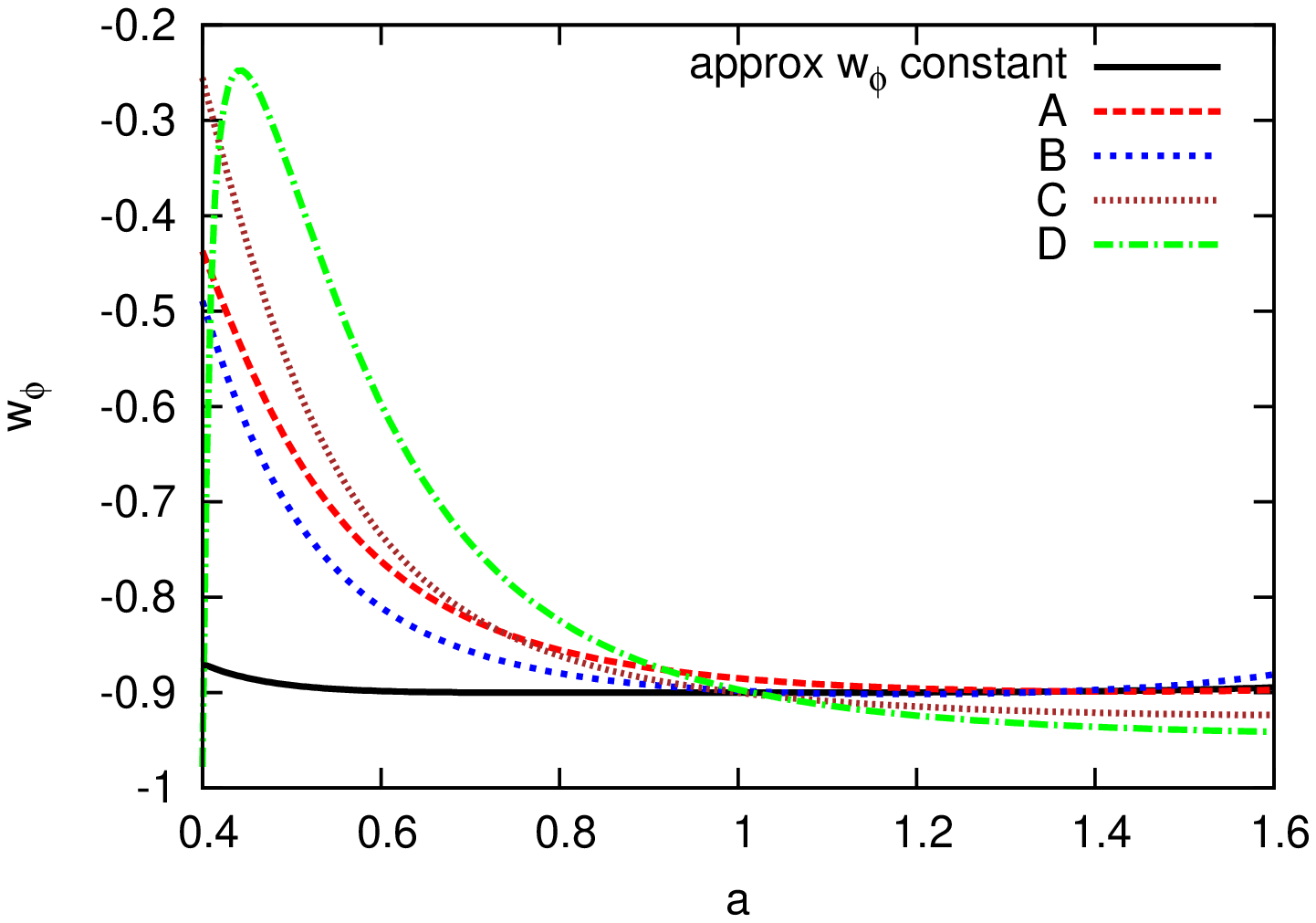}
 \includegraphics[width=0.3\textwidth,angle=-90]{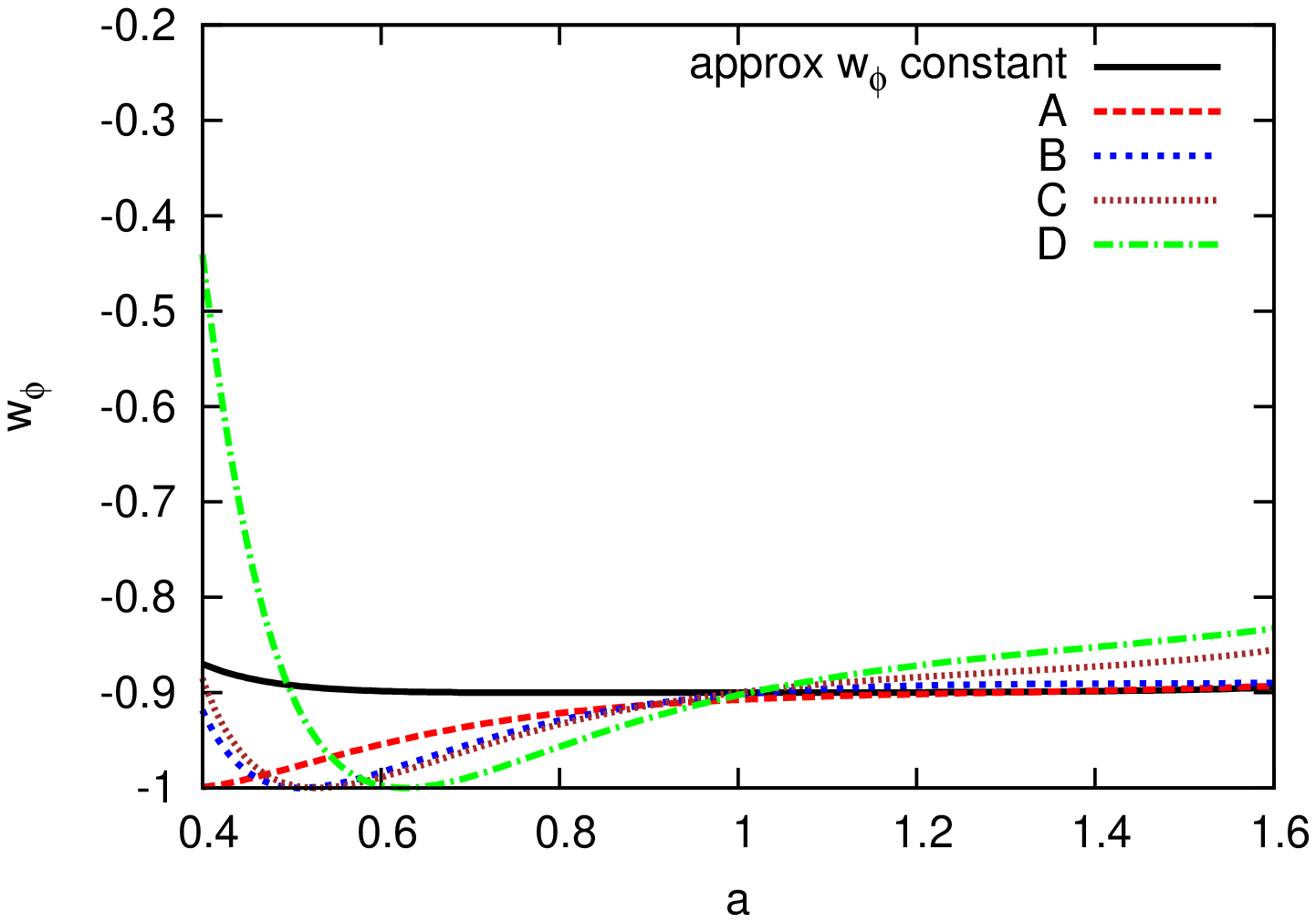}
 \caption{Effect of the variations of the parameters $A$, $B$, $C$ and $D$ on the equation of state $w_{\phi}$. In 
 each panel, three of the coefficients are held constant at their exact value for $w_\phi=-0.9$ while the fourth one 
 is varied by $\pm 30\%$ - left negative and right positive. The black curve shows the equation of state using the 
 exact values of the coefficients as represented in \autoref{fig:VphiQn}. Effects of variations of $A$, $B$, $C$ and 
 $D$ are shown with the red dashed, blue short dashed, brown dotted and green dot-dashed curve, respectively.}
 \label{fig:params}
\end{figure}

In \autoref{fig:params} we have varied the parameters around their values for a specific constant $w_\phi=-0.9$ model 
with a minimal kinetic term. We do this by keeping three of the parameters fixed and vary the fourth one by $\pm30\%$ 
while requiring that $\phi_0$ coincides with the exact solution. We see that a wide range of behaviour of the actual $w(
a)$ can be achieved in these models suggesting that this parametrization of the potential could be used as a proxy for 
a significant range of models, albeit with some restrictions.

We can attempt to generalise the set of coefficients of (\ref{eqn:Vphi_ABCD}) to models with a non-constant equation of 
state. It is not possible to adapt the exact method used for constant $w_\phi$ because usually there is not a general 
expression for $g(a)$. However, we have been able to make some progress by realising that 
$(1+\alpha x)^{\beta}\approx 1+\alpha\beta x$ for $x\ll 1$ and performing an expansion around $\phi\approx \phi_0$ in 
our set up. Expanding (\ref{eqn:Vphi_ABCD}) to first order and matching the coefficients with a similar expansion 
derived from $V(a)$ and $\phi(a)$, we can determine a new set of parameters $A$-$D$. As before, they will depend on the 
background cosmological parameters, $\Omega_{\rm m}$, $w_{\phi}$ and in this case, also on its derivative with respect 
to the scale factor, $w_{\phi}^{\prime}$, evaluated at $a=1$. In the appendix  we report the explicit expression for 
the four coefficients for minimally coupled, type A and type C models, respectively. To understand them note that 
the relation between the scale factor and the scalar field is now, for Quintessence models,
\begin{equation}
 a = 1+\frac{2}{2+3\Omega_{\rm m}w_{\phi}(1)-\frac{w_{\phi}^{\prime}(1)}{1+w_{\phi}(1)}}
     \left\{1-\sqrt{1-\left[2+3\Omega_{\rm m}w_{\phi}(1)-\frac{w_{\phi}^{\prime}(1)}{1+w_{\phi}(1)}\right]
     \sqrt{\frac{8\eta\pi}{3\Omega_{\rm de}[1+w_{\phi}(1)]}}\frac{\phi-\phi_0}{M_{\rm pl}}}\right\}\;.
\end{equation}
Also note that for $w_{\phi}^{\prime}(1)=0$, we recover the result in (\ref{eqn:aphi}) for Quintessence models. 
Similar expressions hold for type A and type C models.

\begin{figure}
 \centering
 \includegraphics[width=0.3\textwidth,angle=-90]{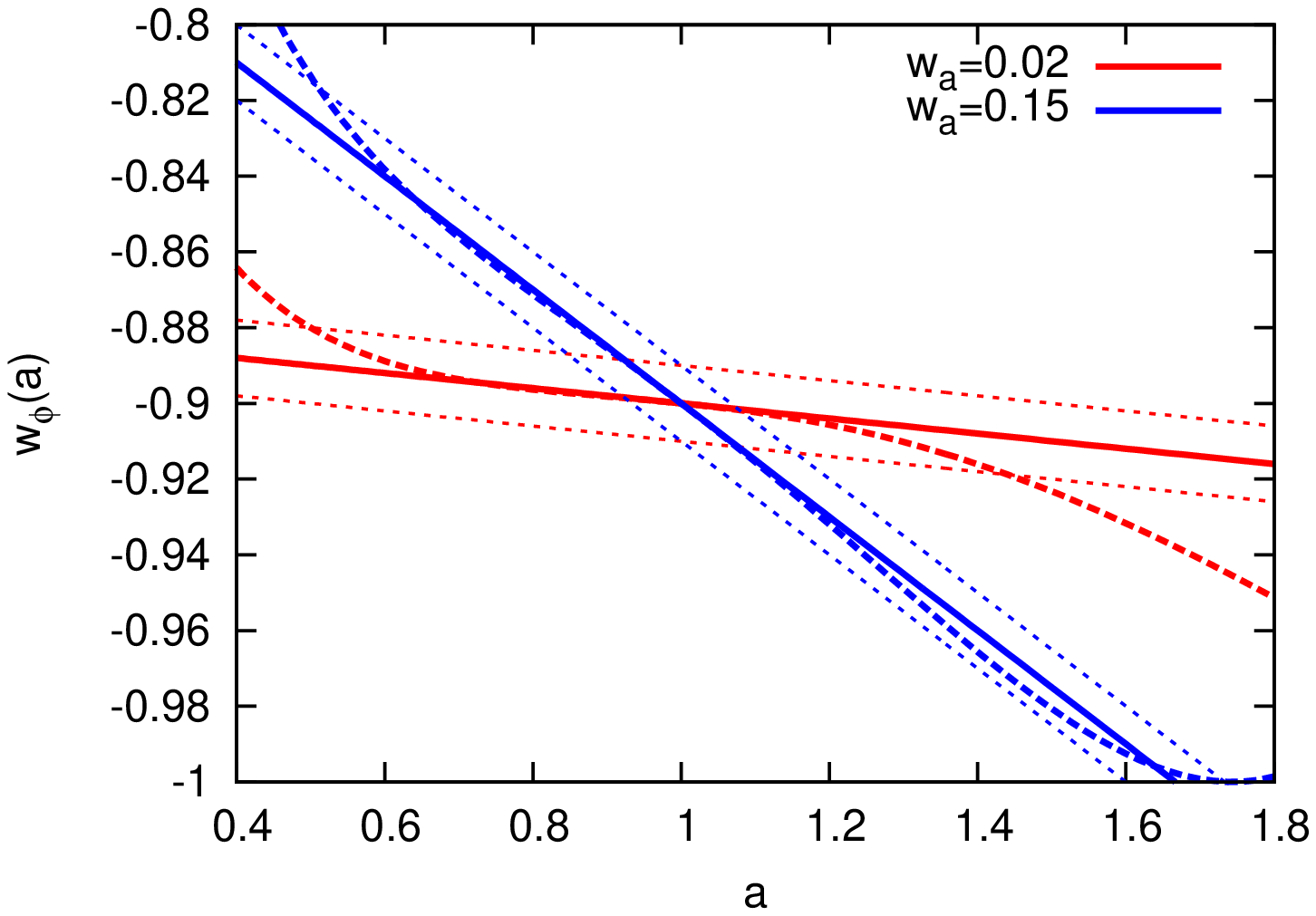}
 \includegraphics[width=0.3\textwidth,angle=-90]{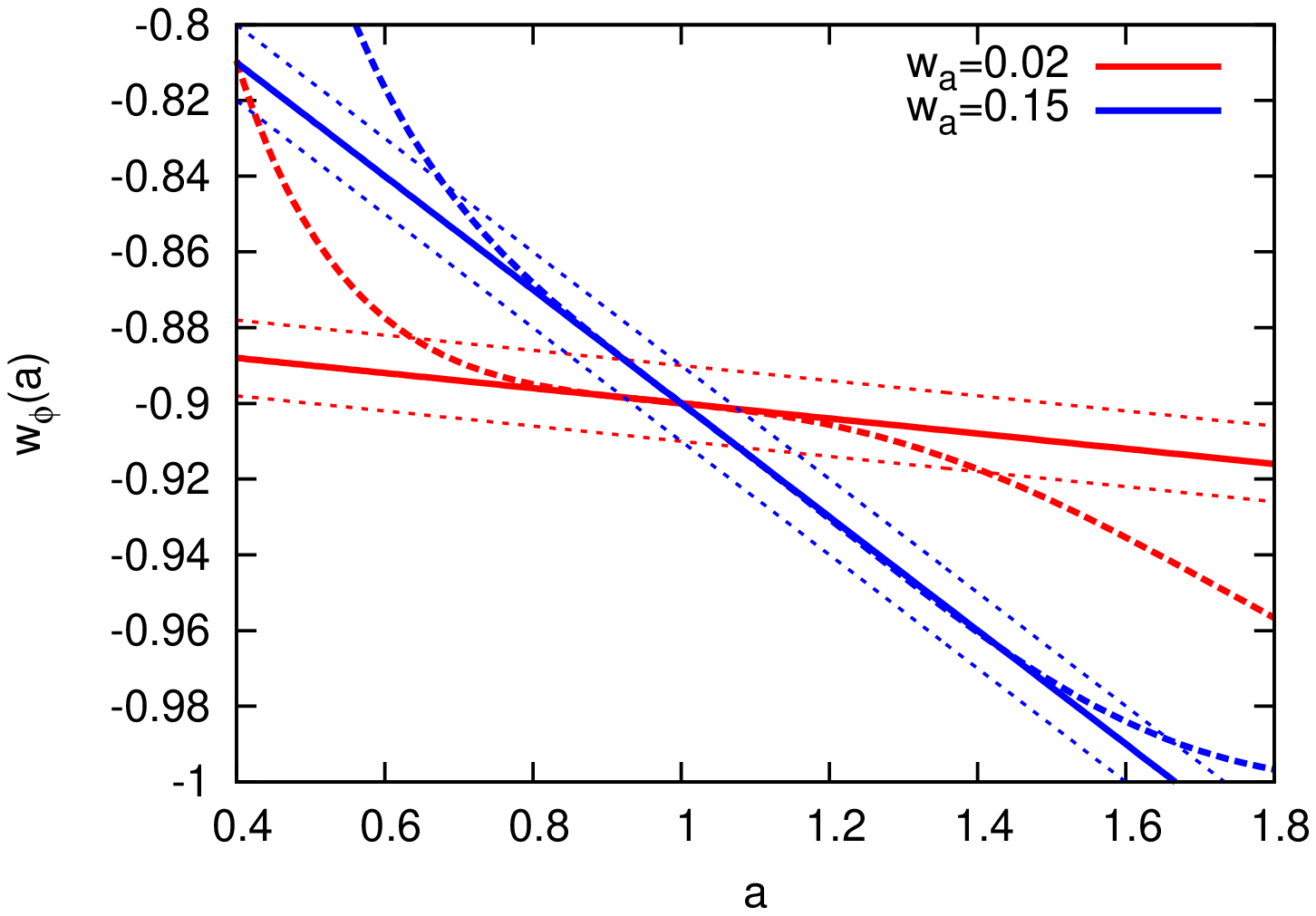}
 \caption{\textit{Left} (\textit{right}) \textit{panel}: Equation of state for the approximated potential of 
 (\ref{eqn:Vphi_ABCD}) for a Quintessence (type A) model described by a CPL equation of state. 
 Red (blue) curve represents a model with $w_\phi(a)=-0.9+0.02(1-a)$ ($w_\phi(a)=-0.9+0.15(1-a)$). Dotted lines 
 represent a 1\% difference from the true equation of state. Solid (dashed) lines show the true (approximated) equation 
 of state.}
\label{fig:Vphiwa_approx}
\end{figure}

To see how well this new parametrization performs we use the Chevallier-Polarski-Linder (CPL) parametrization 
\citep{Chevallier2001,Linder2003},
\begin{equation}
 w_{\phi}(a)=w_0+w_a(1-a)\;,
\end{equation}
where $w_0$ and $w_a$ are constants. In \autoref{fig:Vphiwa_approx} we show the comparison between the true and 
the approximated equation of state evaluated from the potential in (\ref{eqn:Vphi_ABCD}) with the set of 
coefficients given in the appendix for minimally coupled models and type A k-essence models. We use two different 
sets of coefficients ($w_0,w_a$): one with a very gentle slope, $w_\phi(a)=-0.9+0.02(1-a)$, and one with a more 
pronounced variation, $w_\phi(a)=-0.9+0.15(1-a)$. At early times we find a better agreement for models not 
differing too much from a constant equation of state, while at late times, the agreement is better for models with 
$w_a=0.15$. As it can be seen in \autoref{fig:Vphiwa_approx}, this is due to the fact that $\phi^{\prime}$ shows 
fluctuations around the true value. Note that if we limit ourselves to a sub-percent agreement between the true 
and the reconstructed equation of state, then the agreement is much more limited with respect to the case of constant 
$w_{\phi}$. This is because we poorly approximate the function $g(a)$: for a CPL model, it consists of two 
elements: a power-law and an exponential and we only include the power-law. 
When the exponential behaviour dominates, our proposed potential is a less good fit to the true behaviour. 
Note also that the range of agreement is similar for both Quintessence and type A models. Deviations in type A models 
are suppressed with respect to Quintessence models thanks to a higher value of $n$ (2 in the example). From a 
quantitative point of view, for $w_a=0.02$ ($w_a=0.15$), for quintessence models we reach a 1\% agreement for 
$0.5\lesssim a\lesssim 1.4$ ($0.5\lesssim a\lesssim 1.7$), while for type A models we have $0.5\lesssim a\lesssim 1.4$ 
and $0.7\lesssim a\lesssim 1.7$, respectively. 
This is similar to what found before for a model with constant $w_{\phi}=-0.9$. 
One caveat to our approach is that the only knowledge of the evolution of the equation of state is given by its 
value and its time derivative, both evaluated at $a=1$. Therefore nothing is known about its general time evolution and 
as consequence, nothing is known about the functional form of $g(a)$. This implies that our approach would work well 
for models with a monotonic equation of state (and hence a monotonic $g(a)$), but we expect it to fail and not be a 
good representation for the true potential for oscillating dark energy models, 
\cite[see e.g.][for a recent study of their properties and comparison with observations]{Pace2012}.

\section{Conclusions}\label{sect:conclusions}
Scalar fields are an important field of research in cosmology and are one of the most studied candidates used to 
explain and describe the accelerated expansion of the Universe. In this work, we consider two main classes of models: 
minimally coupled models (both quintessence and phantom) and $k$-essence models. For this second class, we specialise 
the Lagrangian to assume three particular functional forms, dubbed type A, type B and type C models. In each case, we 
have shown that specifying the scalar field potential $V(\phi)$, one can determine the evolution of the scalar field 
and the corresponding equation of state $w_{\phi}(a)$.

This is true generally but in order to make it clear, we have assumed the equation of state to be known and we 
calculated explicitly the time evolution of the scalar field and of the potential in some cases. We showed that it is 
possible to obtain an exact analytic solution for minimally coupled and the type C models with constant equation of 
state. This is not possible for more general $k$-essence models or for models with a time-varying $w_{\phi}(a)$, but 
we have solutions for $\phi(a)$ as definite integrals and these can be used to establish the potentials, $V(\phi)$ 
numerically.

We have also derived useful approximate forms of the potential which are valid in different epochs, corresponding to 
the domination of one cosmic fluid. In particular we deduce the form of the potential at early times ($a\ll 1$, 
corresponding to the matter dominated epoch) and at late times ($a\gg 1$, corresponding to the scalar field dominated 
regime), showing that in general the potential is often very well approximated by a power-law.

From an observational point of view, the most important regime to understand the potential is around $a\approx 1$. 
Assuming initially a constant equation of state $w_{\phi}$, we showed that the scalar field potential can be 
approximated by the expression given in (\ref{eqn:Vphi_ABCD}). This expression depends only on four parameters and with 
the appropriate choice of coefficients can cover all the classes of models studied in this work. 
In \autoref{sect:wa} we discussed how this expression might be applied to dynamical dark energy models, by  
appropriately choosing a new set of parameters which reduces to the correct expression in the limit of constant 
$w_{\phi}$. Note that this can not be done for type B models, since our formalism only works for constant equations of 
state.

In some respect our approach is similar to the work of \cite{Sahlen2005}. To derive our expression in 
(\ref{eqn:Vphi_ABCD}), we performed a Taylor expansion of the scalar field evolution, so the same critique could be 
applied: where to stop the series? 
Our approximate potential, for $a\approx 1$ ($\phi\approx\phi_0$) can be expanded in powers of $\phi-\phi_0$ leading 
to the same form of the potential proposed by \cite{Sahlen2005}. In contrast to that work, our proposed potential has 
well motivated coefficients and in the regime of interest it would be possible to map the $V_i$ coefficients of 
(\ref{eqn:Vphi_series}) in terms of our four parameters. For example, at zeroth order, we can write $V_0$ in 
(\ref{eqn:Vphi_series}) as $V_0=A(1-\sqrt{B})H_0^2M_{\rm pl}^2$.

\section*{Acknowledgements}

We would like to thank Stefano Camera and Robert Reischke for useful comments. The work for this article was funded by 
an STFC postdoctoral fellowship.

\section*{Appendix: Coefficients for dynamical dark energy models}

In this section we write explicitly the generalization of the set of coefficients $A$-$D$  for non-constant equations 
of state using the approach discussed in the text.  For minimally coupled models we find 
\begin{equation}
 \begin{split}
  A & = \frac{3\Omega_{\rm de}[1-w_{\phi}(1)]}{16\pi}
        \left\{\frac{4+3\Omega_{\rm m}w_{\phi}(1)-\frac{w_{\phi}^{\prime}(1)}{1+w_{\phi}(1)}}
        {2+3\Omega_{\rm m}w_{\phi}(1)-\frac{w_{\phi}^{\prime}(1)}{1+w_{\phi}(1)}}\right\}
        ^{-\left\{\frac{w_{\phi}^{\prime}(1)}{1-w_{\phi}(1)}+3[1+w_{\phi}(1)]\right\}}\;,\\
  B & = \frac{4}{\left[4+3\Omega_{\rm m}w_{\phi}(1)-\frac{w_{\phi}^{\prime}(1)}{1+w_{\phi}(1)}\right]^2}\;,\\
  C & = -4\frac{2+3\Omega_{\rm m}w_{\phi}(1)-\frac{w_{\phi}^{\prime}(1)}{1+w_{\phi}(1)}}
        {\left[4+3\Omega_{\rm m}w_{\phi}(1)-\frac{w_{\phi}^{\prime}(1)}{1+w_{\phi}(1)}\right]^2}
        \sqrt{\frac{8\eta\pi}{3\Omega_{\rm de}[1+w_{\phi}(1)]}}\;,\\
  D & = -\left\{\frac{w_{\phi}^{\prime}(1)}{1-w_{\phi}(1)}+3[1+w_{\phi}(1)]\right\}\;.
 \end{split}
\end{equation}
For models of type A we find 
\begin{equation}
 \begin{split}
  A & = \frac{3\Omega_{\rm de}[1-(2n-1)w_{\phi}(1)]}{16\pi n}
        \left\{\frac{3+n-3(\Omega_{\rm de}n-1)w_{\phi}(1)-\frac{w_{\phi}^{\prime}(1)}{1+w_{\phi}(1)}}
        {3-n-3(\Omega_{\rm de}n-1)w_{\phi}(1)-\frac{w_{\phi}^{\prime}(1)}{1+w_{\phi}(1)}}\right\}
        ^{-\left\{\frac{(2n-1)w_{\phi}^{\prime}(1)}{1-(2n-1)w_{\phi}(1)}+3[1+w_{\phi}(1)]\right\}}\;,\\
  B & = \frac{4n^2}
        {\left[3+n-3(\Omega_{\rm de}n-1)w_{\phi}(1)-\frac{w_{\phi}^{\prime}(1)}{1+w_{\phi}(1)}\right]^2}\;,\\
  C & = -2\sqrt{2}n
         \frac{3-n-3(\Omega_{\rm de}n-1)w_{\phi}(1)-\frac{w_{\phi}^{\prime}(1)}{1+w_{\phi}(1)}}
         {\left[3+n-3(\Omega_{\rm de}n-1)w_{\phi}(1)-\frac{w_{\phi}^{\prime}(1)}{1+w_{\phi}(1)}\right]^2}
         \frac{H_0M_{\rm pl}}{M^2}
         \left[\frac{16\pi nM^4}{3H_0^2M_{\rm pl}^2\Omega_{\rm de}(1+w_{\phi})}\right]^{\frac{1}{2n}}\;,\\
  D & = -\left\{\frac{(2n-1)w_{\phi}^{\prime}(1)}{1-(2n-1)w_{\phi}(1)}+3[1+w_{\phi}(1)]\right\}\,.
 \end{split}
\end{equation}
For models of type C we find 
\begin{equation}
 \begin{split}
  A & = \frac{3\Omega_{\rm de}(1-w_{\phi})}{16\pi(1-n)}
        \left[\frac{3H_0^2M_{\rm pl}^2\Omega_{\rm de}[1+(1-2n)w_{\phi}]}{16\pi(n-1)M^4}\right]^{-n}\times\\
    &   \left[\frac{4+3\Omega_{\rm m}w_{\phi}-\frac{(1-2n)w_{\phi}^{\prime}(1)}{1+(1-2n)w_{\phi}(1)}}
        {2+3\Omega_{\rm m}w_{\phi}-\frac{(1-2n)w_{\phi}^{\prime}(1)}{1+(1-2n)w_{\phi}(1)}}\right]
        ^{-(1-n)\left\{\frac{[1+2n+(1-2n)w_{\phi}(1)]w_{\phi}^{\prime}(1)}{[1+(1-2n)w_{\phi}(1)][1-w_{\phi}(1)]}+
        3[1+w_{\phi}(1)]\right\}}\;,\\
  B & = \frac{4}
        {\left[4+3\Omega_{\rm m}w_{\phi}-\frac{(1-2n)w_{\phi}^{\prime}(1)}{1+(1-2n)w_{\phi}(1)}\right]^2}\;,\\
  C & = -4\frac{2+3\Omega_{\rm m}w_{\phi}-\frac{(1-2n)w_{\phi}^{\prime}(1)}{1+(1-2n)w_{\phi}(1)}}
        {\left[4+3\Omega_{\rm m}w_{\phi}-\frac{(1-2n)w_{\phi}^{\prime}(1)}{1+(1-2n)w_{\phi}(1)}\right]^2}\;,\\
  D & = -(1-n)\left\{\frac{[1+2n+(1-2n)w_{\phi}(1)]w_{\phi}^{\prime}(1)}{[1+(1-2n)w_{\phi}(1)][1-w_{\phi}(1)]}+
        3[1+w_{\phi}(1)]\right\}\;.
 \end{split}
\end{equation}

\bibliographystyle{apsrev4-1}
\bibliography{ScalarFields.bbl}

\begin{thebibliography}{77}%
\makeatletter
\providecommand \@ifxundefined [1]{%
 \@ifx{#1\undefined}
}%
\providecommand \@ifnum [1]{%
 \ifnum #1\expandafter \@firstoftwo
 \else \expandafter \@secondoftwo
 \fi
}%
\providecommand \@ifx [1]{%
 \ifx #1\expandafter \@firstoftwo
 \else \expandafter \@secondoftwo
 \fi
}%
\providecommand \natexlab [1]{#1}%
\providecommand \enquote  [1]{``#1''}%
\providecommand \bibnamefont  [1]{#1}%
\providecommand \bibfnamefont [1]{#1}%
\providecommand \citenamefont [1]{#1}%
\providecommand \href@noop [0]{\@secondoftwo}%
\providecommand \href [0]{\begingroup \@sanitize@url \@href}%
\providecommand \@href[1]{\@@startlink{#1}\@@href}%
\providecommand \@@href[1]{\endgroup#1\@@endlink}%
\providecommand \@sanitize@url [0]{\catcode `\\12\catcode `\$12\catcode
  `\&12\catcode `\#12\catcode `\^12\catcode `\_12\catcode `\%12\relax}%
\providecommand \@@startlink[1]{}%
\providecommand \@@endlink[0]{}%
\providecommand \url  [0]{\begingroup\@sanitize@url \@url }%
\providecommand \@url [1]{\endgroup\@href {#1}{\urlprefix }}%
\providecommand \urlprefix  [0]{URL }%
\providecommand \Eprint [0]{\href }%
\providecommand \doibase [0]{http://dx.doi.org/}%
\providecommand \selectlanguage [0]{\@gobble}%
\providecommand \bibinfo  [0]{\@secondoftwo}%
\providecommand \bibfield  [0]{\@secondoftwo}%
\providecommand \translation [1]{[#1]}%
\providecommand \BibitemOpen [0]{}%
\providecommand \bibitemStop [0]{}%
\providecommand \bibitemNoStop [0]{.\EOS\space}%
\providecommand \EOS [0]{\spacefactor3000\relax}%
\providecommand \BibitemShut  [1]{\csname bibitem#1\endcsname}%
\let\auto@bib@innerbib\@empty
\bibitem [{\citenamefont {{Planck Collaboration
  XIII}}(2015)}]{Planck2015_XIII}%
  \BibitemOpen
  \bibfield  {author} {\bibinfo {author} {\bibnamefont {{Planck Collaboration
  XIII}}},\ }\href@noop {} {\bibfield  {journal} {\bibinfo  {journal} {ArXiv
  e-prints, 1502.01589}\ } (\bibinfo {year} {2015})},\ \Eprint
  {http://arxiv.org/abs/1502.01589} {arXiv:1502.01589} \BibitemShut {NoStop}%
\bibitem [{\citenamefont {{Planck Collaboration XIV}}(2015)}]{Planck2015_XIV}%
  \BibitemOpen
  \bibfield  {author} {\bibinfo {author} {\bibnamefont {{Planck Collaboration
  XIV}}},\ }\href@noop {} {\bibfield  {journal} {\bibinfo  {journal} {ArXiv
  e-prints, 1502.01590}\ } (\bibinfo {year} {2015})},\ \Eprint
  {http://arxiv.org/abs/1502.01590} {arXiv:1502.01590} \BibitemShut {NoStop}%
\bibitem [{\citenamefont {{Ford}}(1987)}]{Ford1987}%
  \BibitemOpen
  \bibfield  {author} {\bibinfo {author} {\bibfnamefont {L.~H.}\ \bibnamefont
  {{Ford}}},\ }\href {\doibase 10.1103/PhysRevD.35.2339} {\bibfield  {journal}
  {\bibinfo  {journal} {\prd}\ }\textbf {\bibinfo {volume} {35}},\ \bibinfo
  {pages} {2339} (\bibinfo {year} {1987})}\BibitemShut {NoStop}%
\bibitem [{\citenamefont {{Peebles}}\ and\ \citenamefont
  {{Ratra}}(1988)}]{Peebles1988}%
  \BibitemOpen
  \bibfield  {author} {\bibinfo {author} {\bibfnamefont {P.~J.~E.}\
  \bibnamefont {{Peebles}}}\ and\ \bibinfo {author} {\bibfnamefont
  {B.}~\bibnamefont {{Ratra}}},\ }\href {\doibase 10.1086/185100} {\bibfield
  {journal} {\bibinfo  {journal} {\apjl}\ }\textbf {\bibinfo {volume} {325}},\
  \bibinfo {pages} {L17} (\bibinfo {year} {1988})}\BibitemShut {NoStop}%
\bibitem [{\citenamefont {{Ratra}}\ and\ \citenamefont
  {{Peebles}}(1988)}]{Ratra1988a}%
  \BibitemOpen
  \bibfield  {author} {\bibinfo {author} {\bibfnamefont {B.}~\bibnamefont
  {{Ratra}}}\ and\ \bibinfo {author} {\bibfnamefont {P.~J.~E.}\ \bibnamefont
  {{Peebles}}},\ }\href {\doibase 10.1103/PhysRevD.37.3406} {\bibfield
  {journal} {\bibinfo  {journal} {\prd}\ }\textbf {\bibinfo {volume} {37}},\
  \bibinfo {pages} {3406} (\bibinfo {year} {1988})}\BibitemShut {NoStop}%
\bibitem [{\citenamefont {{Wetterich}}(1988)}]{Wetterich1988}%
  \BibitemOpen
  \bibfield  {author} {\bibinfo {author} {\bibfnamefont {C.}~\bibnamefont
  {{Wetterich}}},\ }\href {\doibase 10.1016/0550-3213(88)90193-9} {\bibfield
  {journal} {\bibinfo  {journal} {Nuclear Physics B}\ }\textbf {\bibinfo
  {volume} {302}},\ \bibinfo {pages} {668} (\bibinfo {year}
  {1988})}\BibitemShut {NoStop}%
\bibitem [{\citenamefont {{Caldwell}}\ \emph {et~al.}(1998)\citenamefont
  {{Caldwell}}, \citenamefont {{Dave}},\ and\ \citenamefont
  {{Steinhardt}}}]{Caldwell1998}%
  \BibitemOpen
  \bibfield  {author} {\bibinfo {author} {\bibfnamefont {R.~R.}\ \bibnamefont
  {{Caldwell}}}, \bibinfo {author} {\bibfnamefont {R.}~\bibnamefont {{Dave}}},
  \ and\ \bibinfo {author} {\bibfnamefont {P.~J.}\ \bibnamefont
  {{Steinhardt}}},\ }\href {\doibase 10.1103/PhysRevLett.80.1582} {\bibfield
  {journal} {\bibinfo  {journal} {Physical Review Letters}\ }\textbf {\bibinfo
  {volume} {80}},\ \bibinfo {pages} {1582} (\bibinfo {year} {1998})},\ \Eprint
  {http://arxiv.org/abs/arXiv:astro-ph/9708069} {arXiv:astro-ph/9708069}
  \BibitemShut {NoStop}%
\bibitem [{\citenamefont {{Copeland}}\ \emph {et~al.}(1998)\citenamefont
  {{Copeland}}, \citenamefont {{Liddle}},\ and\ \citenamefont
  {{Wands}}}]{Copeland1998}%
  \BibitemOpen
  \bibfield  {author} {\bibinfo {author} {\bibfnamefont {E.~J.}\ \bibnamefont
  {{Copeland}}}, \bibinfo {author} {\bibfnamefont {A.~R.}\ \bibnamefont
  {{Liddle}}}, \ and\ \bibinfo {author} {\bibfnamefont {D.}~\bibnamefont
  {{Wands}}},\ }\href {\doibase 10.1103/PhysRevD.57.4686} {\bibfield  {journal}
  {\bibinfo  {journal} {\prd}\ }\textbf {\bibinfo {volume} {57}},\ \bibinfo
  {pages} {4686} (\bibinfo {year} {1998})},\ \Eprint
  {http://arxiv.org/abs/gr-qc/9711068} {gr-qc/9711068} \BibitemShut {NoStop}%
\bibitem [{\citenamefont {{Steinhardt}}\ \emph {et~al.}(1999)\citenamefont
  {{Steinhardt}}, \citenamefont {{Wang}},\ and\ \citenamefont
  {{Zlatev}}}]{Steinhardt1999}%
  \BibitemOpen
  \bibfield  {author} {\bibinfo {author} {\bibfnamefont {P.~J.}\ \bibnamefont
  {{Steinhardt}}}, \bibinfo {author} {\bibfnamefont {L.}~\bibnamefont
  {{Wang}}}, \ and\ \bibinfo {author} {\bibfnamefont {I.}~\bibnamefont
  {{Zlatev}}},\ }\href {\doibase 10.1103/PhysRevD.59.123504} {\bibfield
  {journal} {\bibinfo  {journal} {\prd}\ }\textbf {\bibinfo {volume} {59}},\
  \bibinfo {pages} {123504} (\bibinfo {year} {1999})},\ \Eprint
  {http://arxiv.org/abs/arXiv:astro-ph/9812313} {arXiv:astro-ph/9812313}
  \BibitemShut {NoStop}%
\bibitem [{\citenamefont {{Barreiro}}\ \emph {et~al.}(2000)\citenamefont
  {{Barreiro}}, \citenamefont {{Copeland}},\ and\ \citenamefont
  {{Nunes}}}]{Barreiro2000}%
  \BibitemOpen
  \bibfield  {author} {\bibinfo {author} {\bibfnamefont {T.}~\bibnamefont
  {{Barreiro}}}, \bibinfo {author} {\bibfnamefont {E.~J.}\ \bibnamefont
  {{Copeland}}}, \ and\ \bibinfo {author} {\bibfnamefont {N.~J.}\ \bibnamefont
  {{Nunes}}},\ }\href {\doibase 10.1103/PhysRevD.61.127301} {\bibfield
  {journal} {\bibinfo  {journal} {\prd}\ }\textbf {\bibinfo {volume} {61}},\
  \bibinfo {pages} {127301} (\bibinfo {year} {2000})},\ \Eprint
  {http://arxiv.org/abs/arXiv:astro-ph/9910214} {arXiv:astro-ph/9910214}
  \BibitemShut {NoStop}%
\bibitem [{\citenamefont {{Armend{\'a}riz-Pic{\'o}n}}\ \emph
  {et~al.}(1999)\citenamefont {{Armend{\'a}riz-Pic{\'o}n}}, \citenamefont
  {{Damour}},\ and\ \citenamefont {{Mukhanov}}}]{ArmendarizPicon1999}%
  \BibitemOpen
  \bibfield  {author} {\bibinfo {author} {\bibfnamefont {C.}~\bibnamefont
  {{Armend{\'a}riz-Pic{\'o}n}}}, \bibinfo {author} {\bibfnamefont
  {T.}~\bibnamefont {{Damour}}}, \ and\ \bibinfo {author} {\bibfnamefont
  {V.}~\bibnamefont {{Mukhanov}}},\ }\href {\doibase
  10.1016/S0370-2693(99)00603-6} {\bibfield  {journal} {\bibinfo  {journal}
  {Physics Letters B}\ }\textbf {\bibinfo {volume} {458}},\ \bibinfo {pages}
  {209} (\bibinfo {year} {1999})},\ \Eprint
  {http://arxiv.org/abs/hep-th/9904075} {hep-th/9904075} \BibitemShut {NoStop}%
\bibitem [{\citenamefont {{Chiba}}\ \emph {et~al.}(2000)\citenamefont
  {{Chiba}}, \citenamefont {{Okabe}},\ and\ \citenamefont
  {{Yamaguchi}}}]{Chiba2000}%
  \BibitemOpen
  \bibfield  {author} {\bibinfo {author} {\bibfnamefont {T.}~\bibnamefont
  {{Chiba}}}, \bibinfo {author} {\bibfnamefont {T.}~\bibnamefont {{Okabe}}}, \
  and\ \bibinfo {author} {\bibfnamefont {M.}~\bibnamefont {{Yamaguchi}}},\
  }\href {\doibase 10.1103/PhysRevD.62.023511} {\bibfield  {journal} {\bibinfo
  {journal} {\prd}\ }\textbf {\bibinfo {volume} {62}},\ \bibinfo {eid} {023511}
  (\bibinfo {year} {2000})},\ \Eprint {http://arxiv.org/abs/astro-ph/9912463}
  {astro-ph/9912463} \BibitemShut {NoStop}%
\bibitem [{\citenamefont {{Mukhanov}}\ and\ \citenamefont
  {{Vikman}}(2006)}]{Mukhanov2006}%
  \BibitemOpen
  \bibfield  {author} {\bibinfo {author} {\bibfnamefont {V.}~\bibnamefont
  {{Mukhanov}}}\ and\ \bibinfo {author} {\bibfnamefont {A.}~\bibnamefont
  {{Vikman}}},\ }\href {\doibase 10.1088/1475-7516/2006/02/004} {\bibfield
  {journal} {\bibinfo  {journal} {\jcap}\ }\textbf {\bibinfo {volume} {2}},\
  \bibinfo {eid} {004} (\bibinfo {year} {2006})},\ \Eprint
  {http://arxiv.org/abs/astro-ph/0512066} {astro-ph/0512066} \BibitemShut
  {NoStop}%
\bibitem [{\citenamefont {{Chevallier}}\ and\ \citenamefont
  {{Polarski}}(2001)}]{Chevallier2001}%
  \BibitemOpen
  \bibfield  {author} {\bibinfo {author} {\bibfnamefont {M.}~\bibnamefont
  {{Chevallier}}}\ and\ \bibinfo {author} {\bibfnamefont {D.}~\bibnamefont
  {{Polarski}}},\ }\href {\doibase 10.1142/S0218271801000822} {\bibfield
  {journal} {\bibinfo  {journal} {International Journal of Modern Physics D}\
  }\textbf {\bibinfo {volume} {10}},\ \bibinfo {pages} {213} (\bibinfo {year}
  {2001})},\ \Eprint {http://arxiv.org/abs/arXiv:gr-qc/0009008}
  {arXiv:gr-qc/0009008} \BibitemShut {NoStop}%
\bibitem [{\citenamefont {{Linder}}(2003)}]{Linder2003}%
  \BibitemOpen
  \bibfield  {author} {\bibinfo {author} {\bibfnamefont {E.~V.}\ \bibnamefont
  {{Linder}}},\ }\href {\doibase 10.1103/PhysRevLett.90.091301} {\bibfield
  {journal} {\bibinfo  {journal} {Physical Review Letters}\ }\textbf {\bibinfo
  {volume} {90}},\ \bibinfo {pages} {091301} (\bibinfo {year} {2003})},\
  \Eprint {http://arxiv.org/abs/arXiv:astro-ph/0208512}
  {arXiv:astro-ph/0208512} \BibitemShut {NoStop}%
\bibitem [{\citenamefont {{Laureijs}}\ \emph {et~al.}(2011)\citenamefont
  {{Laureijs}}, \citenamefont {{Amiaux}}, \citenamefont {{Arduini}},
  \citenamefont {{Augu{\`e}res}}, \citenamefont {{Brinchmann}}, \citenamefont
  {{Cole}}, \citenamefont {{Cropper}}, \citenamefont {{Dabin}}, \citenamefont
  {{Duvet}}, \citenamefont {{Ealet}},\ and\ \citenamefont
  {et~al.}}]{Laureijs2011}%
  \BibitemOpen
  \bibfield  {author} {\bibinfo {author} {\bibfnamefont {R.}~\bibnamefont
  {{Laureijs}}}, \bibinfo {author} {\bibfnamefont {J.}~\bibnamefont
  {{Amiaux}}}, \bibinfo {author} {\bibfnamefont {S.}~\bibnamefont {{Arduini}}},
  \bibinfo {author} {\bibfnamefont {J.~.}\ \bibnamefont {{Augu{\`e}res}}},
  \bibinfo {author} {\bibfnamefont {J.}~\bibnamefont {{Brinchmann}}}, \bibinfo
  {author} {\bibfnamefont {R.}~\bibnamefont {{Cole}}}, \bibinfo {author}
  {\bibfnamefont {M.}~\bibnamefont {{Cropper}}}, \bibinfo {author}
  {\bibfnamefont {C.}~\bibnamefont {{Dabin}}}, \bibinfo {author} {\bibfnamefont
  {L.}~\bibnamefont {{Duvet}}}, \bibinfo {author} {\bibfnamefont
  {A.}~\bibnamefont {{Ealet}}}, \ and\ \bibinfo {author} {\bibnamefont
  {et~al.}},\ }\href@noop {} {\bibfield  {journal} {\bibinfo  {journal} {ArXiv
  e-prints, 1110.3193}\ } (\bibinfo {year} {2011})},\ \Eprint
  {http://arxiv.org/abs/1110.3193} {arXiv:1110.3193 [astro-ph.CO]} \BibitemShut
  {NoStop}%
\bibitem [{\citenamefont {{Amendola}}\ \emph {et~al.}(2013)\citenamefont
  {{Amendola}}, \citenamefont {{Appleby}}, \citenamefont {{Bacon}},
  \citenamefont {{Baker}}, \citenamefont {{Baldi}}, \citenamefont {{Bartolo}},
  \citenamefont {{Blanchard}}, \citenamefont {{Bonvin}}, \citenamefont
  {{Borgani}},\ and\ \citenamefont {{et~al.}}}]{Amendola2013}%
  \BibitemOpen
  \bibfield  {author} {\bibinfo {author} {\bibfnamefont {L.}~\bibnamefont
  {{Amendola}}}, \bibinfo {author} {\bibfnamefont {S.}~\bibnamefont
  {{Appleby}}}, \bibinfo {author} {\bibfnamefont {D.}~\bibnamefont {{Bacon}}},
  \bibinfo {author} {\bibfnamefont {T.}~\bibnamefont {{Baker}}}, \bibinfo
  {author} {\bibfnamefont {M.}~\bibnamefont {{Baldi}}}, \bibinfo {author}
  {\bibfnamefont {N.}~\bibnamefont {{Bartolo}}}, \bibinfo {author}
  {\bibfnamefont {A.}~\bibnamefont {{Blanchard}}}, \bibinfo {author}
  {\bibfnamefont {C.}~\bibnamefont {{Bonvin}}}, \bibinfo {author}
  {\bibfnamefont {S.}~\bibnamefont {{Borgani}}}, \ and\ \bibinfo {author}
  {\bibnamefont {{et~al.}}},\ }\href {\doibase 10.12942/lrr-2013-6} {\bibfield
  {journal} {\bibinfo  {journal} {Living Reviews in Relativity}\ }\textbf
  {\bibinfo {volume} {16}},\ \bibinfo {pages} {6} (\bibinfo {year} {2013})},\
  \Eprint {http://arxiv.org/abs/1206.1225} {arXiv:1206.1225 [astro-ph.CO]}
  \BibitemShut {NoStop}%
\bibitem [{\citenamefont {{LSST Dark Energy Science
  Collaboration}}(2012)}]{LSST2012}%
  \BibitemOpen
  \bibfield  {author} {\bibinfo {author} {\bibnamefont {{LSST Dark Energy
  Science Collaboration}}},\ }\href@noop {} {\bibfield  {journal} {\bibinfo
  {journal} {ArXiv e-prints}\ } (\bibinfo {year} {2012})},\ \Eprint
  {http://arxiv.org/abs/1211.0310} {arXiv:1211.0310 [astro-ph.CO]} \BibitemShut
  {NoStop}%
\bibitem [{\citenamefont {{Bull}}\ \emph {et~al.}(2015)\citenamefont {{Bull}},
  \citenamefont {{Camera}}, \citenamefont {{Raccanelli}}, \citenamefont
  {{Blake}}, \citenamefont {{Ferreira}}, \citenamefont {{Santos}},\ and\
  \citenamefont {{Schwarz}}}]{Bull2015}%
  \BibitemOpen
  \bibfield  {author} {\bibinfo {author} {\bibfnamefont {P.}~\bibnamefont
  {{Bull}}}, \bibinfo {author} {\bibfnamefont {S.}~\bibnamefont {{Camera}}},
  \bibinfo {author} {\bibfnamefont {A.}~\bibnamefont {{Raccanelli}}}, \bibinfo
  {author} {\bibfnamefont {C.}~\bibnamefont {{Blake}}}, \bibinfo {author}
  {\bibfnamefont {P.}~\bibnamefont {{Ferreira}}}, \bibinfo {author}
  {\bibfnamefont {M.}~\bibnamefont {{Santos}}}, \ and\ \bibinfo {author}
  {\bibfnamefont {D.~J.}\ \bibnamefont {{Schwarz}}},\ }\href@noop {} {\bibfield
   {journal} {\bibinfo  {journal} {Advancing Astrophysics with the Square
  Kilometre Array (AASKA14)}\ ,\ \bibinfo {eid} {24}} (\bibinfo {year}
  {2015})},\ \Eprint {http://arxiv.org/abs/1501.04088} {arXiv:1501.04088}
  \BibitemShut {NoStop}%
\bibitem [{\citenamefont {{Camera}}\ \emph {et~al.}(2015)\citenamefont
  {{Camera}}, \citenamefont {{Raccanelli}}, \citenamefont {{Bull}},
  \citenamefont {{Bertacca}}, \citenamefont {{Chen}}, \citenamefont
  {{Ferreira}}, \citenamefont {{Kunz}}, \citenamefont {{Maartens}},
  \citenamefont {{Mao}}, \citenamefont {{Santos}}, \citenamefont {{Shapiro}},
  \citenamefont {{Viel}},\ and\ \citenamefont {{Xu}}}]{Camera2015}%
  \BibitemOpen
  \bibfield  {author} {\bibinfo {author} {\bibfnamefont {S.}~\bibnamefont
  {{Camera}}}, \bibinfo {author} {\bibfnamefont {A.}~\bibnamefont
  {{Raccanelli}}}, \bibinfo {author} {\bibfnamefont {P.}~\bibnamefont
  {{Bull}}}, \bibinfo {author} {\bibfnamefont {D.}~\bibnamefont {{Bertacca}}},
  \bibinfo {author} {\bibfnamefont {X.}~\bibnamefont {{Chen}}}, \bibinfo
  {author} {\bibfnamefont {P.}~\bibnamefont {{Ferreira}}}, \bibinfo {author}
  {\bibfnamefont {M.}~\bibnamefont {{Kunz}}}, \bibinfo {author} {\bibfnamefont
  {R.}~\bibnamefont {{Maartens}}}, \bibinfo {author} {\bibfnamefont
  {Y.}~\bibnamefont {{Mao}}}, \bibinfo {author} {\bibfnamefont
  {M.}~\bibnamefont {{Santos}}}, \bibinfo {author} {\bibfnamefont {P.~R.}\
  \bibnamefont {{Shapiro}}}, \bibinfo {author} {\bibfnamefont {M.}~\bibnamefont
  {{Viel}}}, \ and\ \bibinfo {author} {\bibfnamefont {Y.}~\bibnamefont
  {{Xu}}},\ }\href@noop {} {\bibfield  {journal} {\bibinfo  {journal}
  {Advancing Astrophysics with the Square Kilometre Array (AASKA14)}\ ,\
  \bibinfo {eid} {25}} (\bibinfo {year} {2015})},\ \Eprint
  {http://arxiv.org/abs/1501.03851} {arXiv:1501.03851} \BibitemShut {NoStop}%
\bibitem [{\citenamefont {{Raccanelli}}\ \emph {et~al.}(2015)\citenamefont
  {{Raccanelli}}, \citenamefont {{Bull}}, \citenamefont {{Camera}},
  \citenamefont {{Blake}}, \citenamefont {{Ferreira}}, \citenamefont
  {{Maartens}}, \citenamefont {{Santos}}, \citenamefont {{Bull}}, \citenamefont
  {{Bacon}}, \citenamefont {{Dor{\'e}}}, \citenamefont {{Ferreira}},
  \citenamefont {{Santos}}, \citenamefont {{Viel}},\ and\ \citenamefont
  {{Zhao}}}]{Raccanelli2015}%
  \BibitemOpen
  \bibfield  {author} {\bibinfo {author} {\bibfnamefont {A.}~\bibnamefont
  {{Raccanelli}}}, \bibinfo {author} {\bibfnamefont {P.}~\bibnamefont
  {{Bull}}}, \bibinfo {author} {\bibfnamefont {S.}~\bibnamefont {{Camera}}},
  \bibinfo {author} {\bibfnamefont {C.}~\bibnamefont {{Blake}}}, \bibinfo
  {author} {\bibfnamefont {P.}~\bibnamefont {{Ferreira}}}, \bibinfo {author}
  {\bibfnamefont {R.}~\bibnamefont {{Maartens}}}, \bibinfo {author}
  {\bibfnamefont {M.}~\bibnamefont {{Santos}}}, \bibinfo {author}
  {\bibfnamefont {P.}~\bibnamefont {{Bull}}}, \bibinfo {author} {\bibfnamefont
  {D.}~\bibnamefont {{Bacon}}}, \bibinfo {author} {\bibfnamefont
  {O.}~\bibnamefont {{Dor{\'e}}}}, \bibinfo {author} {\bibfnamefont
  {P.}~\bibnamefont {{Ferreira}}}, \bibinfo {author} {\bibfnamefont {M.~G.}\
  \bibnamefont {{Santos}}}, \bibinfo {author} {\bibfnamefont {M.}~\bibnamefont
  {{Viel}}}, \ and\ \bibinfo {author} {\bibfnamefont {G.~B.}\ \bibnamefont
  {{Zhao}}},\ }\href@noop {} {\bibfield  {journal} {\bibinfo  {journal}
  {Advancing Astrophysics with the Square Kilometre Array (AASKA14)}\ ,\
  \bibinfo {eid} {31}} (\bibinfo {year} {2015})},\ \Eprint
  {http://arxiv.org/abs/1501.03821} {arXiv:1501.03821} \BibitemShut {NoStop}%
\bibitem [{\citenamefont {{Santos}}\ \emph {et~al.}(2015)\citenamefont
  {{Santos}}, \citenamefont {{Bull}}, \citenamefont {{Alonso}}, \citenamefont
  {{Camera}}, \citenamefont {{Ferreira}}, \citenamefont {{Bernardi}},
  \citenamefont {{Maartens}}, \citenamefont {{Viel}}, \citenamefont
  {{Villaescusa-Navarro}}, \citenamefont {{Abdalla}}, \citenamefont {{Jarvis}},
  \citenamefont {{Metcalf}}, \citenamefont {{Pourtsidou}},\ and\ \citenamefont
  {{Wolz}}}]{Santos2015}%
  \BibitemOpen
  \bibfield  {author} {\bibinfo {author} {\bibfnamefont {M.~G.}\ \bibnamefont
  {{Santos}}}, \bibinfo {author} {\bibfnamefont {P.}~\bibnamefont {{Bull}}},
  \bibinfo {author} {\bibfnamefont {D.}~\bibnamefont {{Alonso}}}, \bibinfo
  {author} {\bibfnamefont {S.}~\bibnamefont {{Camera}}}, \bibinfo {author}
  {\bibfnamefont {P.~G.}\ \bibnamefont {{Ferreira}}}, \bibinfo {author}
  {\bibfnamefont {G.}~\bibnamefont {{Bernardi}}}, \bibinfo {author}
  {\bibfnamefont {R.}~\bibnamefont {{Maartens}}}, \bibinfo {author}
  {\bibfnamefont {M.}~\bibnamefont {{Viel}}}, \bibinfo {author} {\bibfnamefont
  {F.}~\bibnamefont {{Villaescusa-Navarro}}}, \bibinfo {author} {\bibfnamefont
  {F.~B.}\ \bibnamefont {{Abdalla}}}, \bibinfo {author} {\bibfnamefont
  {M.}~\bibnamefont {{Jarvis}}}, \bibinfo {author} {\bibfnamefont {R.~B.}\
  \bibnamefont {{Metcalf}}}, \bibinfo {author} {\bibfnamefont {A.}~\bibnamefont
  {{Pourtsidou}}}, \ and\ \bibinfo {author} {\bibfnamefont {L.}~\bibnamefont
  {{Wolz}}},\ }\href@noop {} {\bibfield  {journal} {\bibinfo  {journal} {ArXiv
  e-prints}\ } (\bibinfo {year} {2015})},\ \Eprint
  {http://arxiv.org/abs/1501.03989} {arXiv:1501.03989} \BibitemShut {NoStop}%
\bibitem [{\citenamefont {{Hannestad}}\ and\ \citenamefont
  {{M{\"o}rtsell}}(2004)}]{Hannestad2004}%
  \BibitemOpen
  \bibfield  {author} {\bibinfo {author} {\bibfnamefont {S.}~\bibnamefont
  {{Hannestad}}}\ and\ \bibinfo {author} {\bibfnamefont {E.}~\bibnamefont
  {{M{\"o}rtsell}}},\ }\href {\doibase 10.1088/1475-7516/2004/09/001}
  {\bibfield  {journal} {\bibinfo  {journal} {\jcap}\ }\textbf {\bibinfo
  {volume} {9}},\ \bibinfo {eid} {001} (\bibinfo {year} {2004})},\ \Eprint
  {http://arxiv.org/abs/astro-ph/0407259} {astro-ph/0407259} \BibitemShut
  {NoStop}%
\bibitem [{\citenamefont {{Jassal}}\ \emph {et~al.}(2005)\citenamefont
  {{Jassal}}, \citenamefont {{Bagla}},\ and\ \citenamefont
  {{Padmanabhan}}}]{Jassal2005}%
  \BibitemOpen
  \bibfield  {author} {\bibinfo {author} {\bibfnamefont {H.~K.}\ \bibnamefont
  {{Jassal}}}, \bibinfo {author} {\bibfnamefont {J.~S.}\ \bibnamefont
  {{Bagla}}}, \ and\ \bibinfo {author} {\bibfnamefont {T.}~\bibnamefont
  {{Padmanabhan}}},\ }\href {\doibase 10.1103/PhysRevD.72.103503} {\bibfield
  {journal} {\bibinfo  {journal} {\prd}\ }\textbf {\bibinfo {volume} {72}},\
  \bibinfo {eid} {103503} (\bibinfo {year} {2005})},\ \Eprint
  {http://arxiv.org/abs/astro-ph/0506748} {astro-ph/0506748} \BibitemShut
  {NoStop}%
\bibitem [{\citenamefont {{Barboza}}\ and\ \citenamefont
  {{Alcaniz}}(2008)}]{Barboza2008}%
  \BibitemOpen
  \bibfield  {author} {\bibinfo {author} {\bibfnamefont {E.~M.}\ \bibnamefont
  {{Barboza}}}\ and\ \bibinfo {author} {\bibfnamefont {J.~S.}\ \bibnamefont
  {{Alcaniz}}},\ }\href {\doibase 10.1016/j.physletb.2008.08.012} {\bibfield
  {journal} {\bibinfo  {journal} {Physics Letters B}\ }\textbf {\bibinfo
  {volume} {666}},\ \bibinfo {pages} {415} (\bibinfo {year} {2008})},\ \Eprint
  {http://arxiv.org/abs/0805.1713} {arXiv:0805.1713} \BibitemShut {NoStop}%
\bibitem [{\citenamefont {{Sahni}}\ \emph {et~al.}(2008)\citenamefont
  {{Sahni}}, \citenamefont {{Shafieloo}},\ and\ \citenamefont
  {{Starobinsky}}}]{Sahni2008}%
  \BibitemOpen
  \bibfield  {author} {\bibinfo {author} {\bibfnamefont {V.}~\bibnamefont
  {{Sahni}}}, \bibinfo {author} {\bibfnamefont {A.}~\bibnamefont
  {{Shafieloo}}}, \ and\ \bibinfo {author} {\bibfnamefont {A.~A.}\ \bibnamefont
  {{Starobinsky}}},\ }\href {\doibase 10.1103/PhysRevD.78.103502} {\bibfield
  {journal} {\bibinfo  {journal} {\prd}\ }\textbf {\bibinfo {volume} {78}},\
  \bibinfo {eid} {103502} (\bibinfo {year} {2008})},\ \Eprint
  {http://arxiv.org/abs/0807.3548} {arXiv:0807.3548} \BibitemShut {NoStop}%
\bibitem [{\citenamefont {{Zunckel}}\ and\ \citenamefont
  {{Clarkson}}(2008)}]{Zunckel2008}%
  \BibitemOpen
  \bibfield  {author} {\bibinfo {author} {\bibfnamefont {C.}~\bibnamefont
  {{Zunckel}}}\ and\ \bibinfo {author} {\bibfnamefont {C.}~\bibnamefont
  {{Clarkson}}},\ }\href {\doibase 10.1103/PhysRevLett.101.181301} {\bibfield
  {journal} {\bibinfo  {journal} {Physical Review Letters}\ }\textbf {\bibinfo
  {volume} {101}},\ \bibinfo {eid} {181301} (\bibinfo {year} {2008})},\ \Eprint
  {http://arxiv.org/abs/0807.4304} {arXiv:0807.4304} \BibitemShut {NoStop}%
\bibitem [{\citenamefont {{Alam}}\ \emph {et~al.}(2003)\citenamefont {{Alam}},
  \citenamefont {{Sahni}}, \citenamefont {{Saini}},\ and\ \citenamefont
  {{Starobinsky}}}]{Alam2003}%
  \BibitemOpen
  \bibfield  {author} {\bibinfo {author} {\bibfnamefont {U.}~\bibnamefont
  {{Alam}}}, \bibinfo {author} {\bibfnamefont {V.}~\bibnamefont {{Sahni}}},
  \bibinfo {author} {\bibfnamefont {T.~D.}\ \bibnamefont {{Saini}}}, \ and\
  \bibinfo {author} {\bibfnamefont {A.~A.}\ \bibnamefont {{Starobinsky}}},\
  }\href {\doibase 10.1046/j.1365-8711.2003.06871.x} {\bibfield  {journal}
  {\bibinfo  {journal} {\mnras}\ }\textbf {\bibinfo {volume} {344}},\ \bibinfo
  {pages} {1057} (\bibinfo {year} {2003})},\ \Eprint
  {http://arxiv.org/abs/astro-ph/0303009} {astro-ph/0303009} \BibitemShut
  {NoStop}%
\bibitem [{\citenamefont {{Sahni}}\ \emph {et~al.}(2003)\citenamefont
  {{Sahni}}, \citenamefont {{Saini}}, \citenamefont {{Starobinsky}},\ and\
  \citenamefont {{Alam}}}]{Sahni2003}%
  \BibitemOpen
  \bibfield  {author} {\bibinfo {author} {\bibfnamefont {V.}~\bibnamefont
  {{Sahni}}}, \bibinfo {author} {\bibfnamefont {T.~D.}\ \bibnamefont
  {{Saini}}}, \bibinfo {author} {\bibfnamefont {A.~A.}\ \bibnamefont
  {{Starobinsky}}}, \ and\ \bibinfo {author} {\bibfnamefont {U.}~\bibnamefont
  {{Alam}}},\ }\href {\doibase 10.1134/1.1574831} {\bibfield  {journal}
  {\bibinfo  {journal} {Soviet Journal of Experimental and Theoretical Physics
  Letters}\ }\textbf {\bibinfo {volume} {77}},\ \bibinfo {pages} {201}
  (\bibinfo {year} {2003})},\ \Eprint {http://arxiv.org/abs/astro-ph/0201498}
  {astro-ph/0201498} \BibitemShut {NoStop}%
\bibitem [{\citenamefont {{Crittenden}}\ \emph {et~al.}(2009)\citenamefont
  {{Crittenden}}, \citenamefont {{Pogosian}},\ and\ \citenamefont
  {{Zhao}}}]{Crittenden2009}%
  \BibitemOpen
  \bibfield  {author} {\bibinfo {author} {\bibfnamefont {R.~G.}\ \bibnamefont
  {{Crittenden}}}, \bibinfo {author} {\bibfnamefont {L.}~\bibnamefont
  {{Pogosian}}}, \ and\ \bibinfo {author} {\bibfnamefont {G.-B.}\ \bibnamefont
  {{Zhao}}},\ }\href {\doibase 10.1088/1475-7516/2009/12/025} {\bibfield
  {journal} {\bibinfo  {journal} {\jcap}\ }\textbf {\bibinfo {volume} {12}},\
  \bibinfo {eid} {025} (\bibinfo {year} {2009})},\ \Eprint
  {http://arxiv.org/abs/astro-ph/0510293} {astro-ph/0510293} \BibitemShut
  {NoStop}%
\bibitem [{\citenamefont {{Sahni}}\ and\ \citenamefont
  {{Starobinsky}}(2006)}]{Sahni2006}%
  \BibitemOpen
  \bibfield  {author} {\bibinfo {author} {\bibfnamefont {V.}~\bibnamefont
  {{Sahni}}}\ and\ \bibinfo {author} {\bibfnamefont {A.}~\bibnamefont
  {{Starobinsky}}},\ }\href {\doibase 10.1142/S0218271806009704} {\bibfield
  {journal} {\bibinfo  {journal} {International Journal of Modern Physics D}\
  }\textbf {\bibinfo {volume} {15}},\ \bibinfo {pages} {2105} (\bibinfo {year}
  {2006})},\ \Eprint {http://arxiv.org/abs/astro-ph/0610026} {astro-ph/0610026}
  \BibitemShut {NoStop}%
\bibitem [{\citenamefont {{Sahl{\'e}n}}\ \emph {et~al.}(2005)\citenamefont
  {{Sahl{\'e}n}}, \citenamefont {{Liddle}},\ and\ \citenamefont
  {{Parkinson}}}]{Sahlen2005}%
  \BibitemOpen
  \bibfield  {author} {\bibinfo {author} {\bibfnamefont {M.}~\bibnamefont
  {{Sahl{\'e}n}}}, \bibinfo {author} {\bibfnamefont {A.~R.}\ \bibnamefont
  {{Liddle}}}, \ and\ \bibinfo {author} {\bibfnamefont {D.}~\bibnamefont
  {{Parkinson}}},\ }\href {\doibase 10.1103/PhysRevD.72.083511} {\bibfield
  {journal} {\bibinfo  {journal} {\prd}\ }\textbf {\bibinfo {volume} {72}},\
  \bibinfo {eid} {083511} (\bibinfo {year} {2005})},\ \Eprint
  {http://arxiv.org/abs/astro-ph/0506696} {astro-ph/0506696} \BibitemShut
  {NoStop}%
\bibitem [{\citenamefont {{Grivell}}\ and\ \citenamefont
  {{Liddle}}(2000)}]{Grivell2000}%
  \BibitemOpen
  \bibfield  {author} {\bibinfo {author} {\bibfnamefont {I.~J.}\ \bibnamefont
  {{Grivell}}}\ and\ \bibinfo {author} {\bibfnamefont {A.~R.}\ \bibnamefont
  {{Liddle}}},\ }\href {\doibase 10.1103/PhysRevD.61.081301} {\bibfield
  {journal} {\bibinfo  {journal} {\prd}\ }\textbf {\bibinfo {volume} {61}},\
  \bibinfo {eid} {081301} (\bibinfo {year} {2000})},\ \Eprint
  {http://arxiv.org/abs/astro-ph/9906327} {astro-ph/9906327} \BibitemShut
  {NoStop}%
\bibitem [{\citenamefont {{Copeland}}\ \emph {et~al.}(1993)\citenamefont
  {{Copeland}}, \citenamefont {{Kolb}}, \citenamefont {{Liddle}},\ and\
  \citenamefont {{Lidsey}}}]{Copeland1993}%
  \BibitemOpen
  \bibfield  {author} {\bibinfo {author} {\bibfnamefont {E.~J.}\ \bibnamefont
  {{Copeland}}}, \bibinfo {author} {\bibfnamefont {E.~W.}\ \bibnamefont
  {{Kolb}}}, \bibinfo {author} {\bibfnamefont {A.~R.}\ \bibnamefont
  {{Liddle}}}, \ and\ \bibinfo {author} {\bibfnamefont {J.~E.}\ \bibnamefont
  {{Lidsey}}},\ }\href {\doibase 10.1103/PhysRevD.48.2529} {\bibfield
  {journal} {\bibinfo  {journal} {\prd}\ }\textbf {\bibinfo {volume} {48}},\
  \bibinfo {pages} {2529} (\bibinfo {year} {1993})},\ \Eprint
  {http://arxiv.org/abs/hep-ph/9303288} {hep-ph/9303288} \BibitemShut {NoStop}%
\bibitem [{\citenamefont {{Daly}}\ and\ \citenamefont
  {{Djorgovski}}(2003)}]{Daly2003}%
  \BibitemOpen
  \bibfield  {author} {\bibinfo {author} {\bibfnamefont {R.~A.}\ \bibnamefont
  {{Daly}}}\ and\ \bibinfo {author} {\bibfnamefont {S.~G.}\ \bibnamefont
  {{Djorgovski}}},\ }\href {\doibase 10.1086/378230} {\bibfield  {journal}
  {\bibinfo  {journal} {\apj}\ }\textbf {\bibinfo {volume} {597}},\ \bibinfo
  {pages} {9} (\bibinfo {year} {2003})},\ \Eprint
  {http://arxiv.org/abs/astro-ph/0305197} {astro-ph/0305197} \BibitemShut
  {NoStop}%
\bibitem [{\citenamefont {{Daly}}\ and\ \citenamefont
  {{Djorgovski}}(2004)}]{Daly2004}%
  \BibitemOpen
  \bibfield  {author} {\bibinfo {author} {\bibfnamefont {R.~A.}\ \bibnamefont
  {{Daly}}}\ and\ \bibinfo {author} {\bibfnamefont {S.~G.}\ \bibnamefont
  {{Djorgovski}}},\ }\href {\doibase 10.1086/422673} {\bibfield  {journal}
  {\bibinfo  {journal} {\apj}\ }\textbf {\bibinfo {volume} {612}},\ \bibinfo
  {pages} {652} (\bibinfo {year} {2004})},\ \Eprint
  {http://arxiv.org/abs/astro-ph/0403664} {astro-ph/0403664} \BibitemShut
  {NoStop}%
\bibitem [{\citenamefont {{Simon}}\ \emph {et~al.}(2005)\citenamefont
  {{Simon}}, \citenamefont {{Verde}},\ and\ \citenamefont
  {{Jimenez}}}]{Simon2005a}%
  \BibitemOpen
  \bibfield  {author} {\bibinfo {author} {\bibfnamefont {J.}~\bibnamefont
  {{Simon}}}, \bibinfo {author} {\bibfnamefont {L.}~\bibnamefont {{Verde}}}, \
  and\ \bibinfo {author} {\bibfnamefont {R.}~\bibnamefont {{Jimenez}}},\ }\href
  {\doibase 10.1103/PhysRevD.71.123001} {\bibfield  {journal} {\bibinfo
  {journal} {\prd}\ }\textbf {\bibinfo {volume} {71}},\ \bibinfo {eid} {123001}
  (\bibinfo {year} {2005})},\ \Eprint {http://arxiv.org/abs/astro-ph/0412269}
  {astro-ph/0412269} \BibitemShut {NoStop}%
\bibitem [{\citenamefont {{Slepian}}\ \emph {et~al.}(2014)\citenamefont
  {{Slepian}}, \citenamefont {{Gott}},\ and\ \citenamefont
  {{Zinn}}}]{Slepian2014}%
  \BibitemOpen
  \bibfield  {author} {\bibinfo {author} {\bibfnamefont {Z.}~\bibnamefont
  {{Slepian}}}, \bibinfo {author} {\bibfnamefont {J.~R.}\ \bibnamefont
  {{Gott}}}, \ and\ \bibinfo {author} {\bibfnamefont {J.}~\bibnamefont
  {{Zinn}}},\ }\href {\doibase 10.1093/mnras/stt2195} {\bibfield  {journal}
  {\bibinfo  {journal} {\mnras}\ }\textbf {\bibinfo {volume} {438}},\ \bibinfo
  {pages} {1948} (\bibinfo {year} {2014})},\ \Eprint
  {http://arxiv.org/abs/1301.4611} {arXiv:1301.4611} \BibitemShut {NoStop}%
\bibitem [{\citenamefont {{Crittenden}}\ \emph {et~al.}(2007)\citenamefont
  {{Crittenden}}, \citenamefont {{Majerotto}},\ and\ \citenamefont
  {{Piazza}}}]{Crittenden2007}%
  \BibitemOpen
  \bibfield  {author} {\bibinfo {author} {\bibfnamefont {R.}~\bibnamefont
  {{Crittenden}}}, \bibinfo {author} {\bibfnamefont {E.}~\bibnamefont
  {{Majerotto}}}, \ and\ \bibinfo {author} {\bibfnamefont {F.}~\bibnamefont
  {{Piazza}}},\ }\href {\doibase 10.1103/PhysRevLett.98.251301} {\bibfield
  {journal} {\bibinfo  {journal} {Physical Review Letters}\ }\textbf {\bibinfo
  {volume} {98}},\ \bibinfo {eid} {251301} (\bibinfo {year} {2007})},\ \Eprint
  {http://arxiv.org/abs/astro-ph/0702003} {astro-ph/0702003} \BibitemShut
  {NoStop}%
\bibitem [{\citenamefont {{Chiba}}(2009{\natexlab{a}})}]{Chiba2009a}%
  \BibitemOpen
  \bibfield  {author} {\bibinfo {author} {\bibfnamefont {T.}~\bibnamefont
  {{Chiba}}},\ }\href {\doibase 10.1103/PhysRevD.79.083517} {\bibfield
  {journal} {\bibinfo  {journal} {\prd}\ }\textbf {\bibinfo {volume} {79}},\
  \bibinfo {eid} {083517} (\bibinfo {year} {2009}{\natexlab{a}})},\ \Eprint
  {http://arxiv.org/abs/0902.4037} {arXiv:0902.4037} \BibitemShut {NoStop}%
\bibitem [{\citenamefont {{Chiba}}(2009{\natexlab{b}})}]{Chiba2009b}%
  \BibitemOpen
  \bibfield  {author} {\bibinfo {author} {\bibfnamefont {T.}~\bibnamefont
  {{Chiba}}},\ }\href {\doibase 10.1103/PhysRevD.80.109902} {\bibfield
  {journal} {\bibinfo  {journal} {\prd}\ }\textbf {\bibinfo {volume} {80}},\
  \bibinfo {eid} {109902} (\bibinfo {year} {2009}{\natexlab{b}})}\BibitemShut
  {NoStop}%
\bibitem [{\citenamefont {{Chiba}}\ \emph {et~al.}(2009)\citenamefont
  {{Chiba}}, \citenamefont {{Dutta}},\ and\ \citenamefont
  {{Scherrer}}}]{Chiba2009c}%
  \BibitemOpen
  \bibfield  {author} {\bibinfo {author} {\bibfnamefont {T.}~\bibnamefont
  {{Chiba}}}, \bibinfo {author} {\bibfnamefont {S.}~\bibnamefont {{Dutta}}}, \
  and\ \bibinfo {author} {\bibfnamefont {R.~J.}\ \bibnamefont {{Scherrer}}},\
  }\href {\doibase 10.1103/PhysRevD.80.043517} {\bibfield  {journal} {\bibinfo
  {journal} {\prd}\ }\textbf {\bibinfo {volume} {80}},\ \bibinfo {eid} {043517}
  (\bibinfo {year} {2009})},\ \Eprint {http://arxiv.org/abs/0906.0628}
  {arXiv:0906.0628 [astro-ph.CO]} \BibitemShut {NoStop}%
\bibitem [{\citenamefont {{Caldwell}}(2002)}]{Caldwell2002}%
  \BibitemOpen
  \bibfield  {author} {\bibinfo {author} {\bibfnamefont {R.~R.}\ \bibnamefont
  {{Caldwell}}},\ }\href {\doibase 10.1016/S0370-2693(02)02589-3} {\bibfield
  {journal} {\bibinfo  {journal} {Physics Letters B}\ }\textbf {\bibinfo
  {volume} {545}},\ \bibinfo {pages} {23} (\bibinfo {year} {2002})},\ \Eprint
  {http://arxiv.org/abs/arXiv:astro-ph/9908168} {arXiv:astro-ph/9908168}
  \BibitemShut {NoStop}%
\bibitem [{\citenamefont {{Ellis}}\ and\ \citenamefont
  {{Madsen}}(1991)}]{Ellis1991}%
  \BibitemOpen
  \bibfield  {author} {\bibinfo {author} {\bibfnamefont {G.~F.~R.}\
  \bibnamefont {{Ellis}}}\ and\ \bibinfo {author} {\bibfnamefont {M.~S.}\
  \bibnamefont {{Madsen}}},\ }\href {\doibase 10.1088/0264-9381/8/4/012}
  {\bibfield  {journal} {\bibinfo  {journal} {Classical and Quantum Gravity}\
  }\textbf {\bibinfo {volume} {8}},\ \bibinfo {pages} {667} (\bibinfo {year}
  {1991})}\BibitemShut {NoStop}%
\bibitem [{\citenamefont {{Sahni}}\ and\ \citenamefont
  {{Starobinsky}}(2000)}]{Sahni2000}%
  \BibitemOpen
  \bibfield  {author} {\bibinfo {author} {\bibfnamefont {V.}~\bibnamefont
  {{Sahni}}}\ and\ \bibinfo {author} {\bibfnamefont {A.}~\bibnamefont
  {{Starobinsky}}},\ }\href {\doibase 10.1142/S0218271800000542} {\bibfield
  {journal} {\bibinfo  {journal} {International Journal of Modern Physics D}\
  }\textbf {\bibinfo {volume} {9}},\ \bibinfo {pages} {373} (\bibinfo {year}
  {2000})},\ \Eprint {http://arxiv.org/abs/astro-ph/9904398} {astro-ph/9904398}
  \BibitemShut {NoStop}%
\bibitem [{\citenamefont {{Paliathanasis}}\ \emph {et~al.}(2015)\citenamefont
  {{Paliathanasis}}, \citenamefont {{Tsamparlis}}, \citenamefont
  {{Basilakos}},\ and\ \citenamefont {{Barrow}}}]{Paliathanasis2015}%
  \BibitemOpen
  \bibfield  {author} {\bibinfo {author} {\bibfnamefont {A.}~\bibnamefont
  {{Paliathanasis}}}, \bibinfo {author} {\bibfnamefont {M.}~\bibnamefont
  {{Tsamparlis}}}, \bibinfo {author} {\bibfnamefont {S.}~\bibnamefont
  {{Basilakos}}}, \ and\ \bibinfo {author} {\bibfnamefont {J.~D.}\ \bibnamefont
  {{Barrow}}},\ }\href {\doibase 10.1103/PhysRevD.91.123535} {\bibfield
  {journal} {\bibinfo  {journal} {\prd}\ }\textbf {\bibinfo {volume} {91}},\
  \bibinfo {eid} {123535} (\bibinfo {year} {2015})},\ \Eprint
  {http://arxiv.org/abs/1503.05750} {arXiv:1503.05750 [gr-qc]} \BibitemShut
  {NoStop}%
\bibitem [{\citenamefont {{Armendariz-Picon}}\ \emph
  {et~al.}(2000)\citenamefont {{Armendariz-Picon}}, \citenamefont
  {{Mukhanov}},\ and\ \citenamefont {{Steinhardt}}}]{ArmendarizPicon2000}%
  \BibitemOpen
  \bibfield  {author} {\bibinfo {author} {\bibfnamefont {C.}~\bibnamefont
  {{Armendariz-Picon}}}, \bibinfo {author} {\bibfnamefont {V.}~\bibnamefont
  {{Mukhanov}}}, \ and\ \bibinfo {author} {\bibfnamefont {P.~J.}\ \bibnamefont
  {{Steinhardt}}},\ }\href {\doibase 10.1103/PhysRevLett.85.4438} {\bibfield
  {journal} {\bibinfo  {journal} {Physical Review Letters}\ }\textbf {\bibinfo
  {volume} {85}},\ \bibinfo {pages} {4438} (\bibinfo {year} {2000})},\ \Eprint
  {http://arxiv.org/abs/astro-ph/0004134} {astro-ph/0004134} \BibitemShut
  {NoStop}%
\bibitem [{\citenamefont {{Armendariz-Picon}}\ \emph
  {et~al.}(2001)\citenamefont {{Armendariz-Picon}}, \citenamefont
  {{Mukhanov}},\ and\ \citenamefont {{Steinhardt}}}]{ArmendarizPicon2001}%
  \BibitemOpen
  \bibfield  {author} {\bibinfo {author} {\bibfnamefont {C.}~\bibnamefont
  {{Armendariz-Picon}}}, \bibinfo {author} {\bibfnamefont {V.}~\bibnamefont
  {{Mukhanov}}}, \ and\ \bibinfo {author} {\bibfnamefont {P.~J.}\ \bibnamefont
  {{Steinhardt}}},\ }\href {\doibase 10.1103/PhysRevD.63.103510} {\bibfield
  {journal} {\bibinfo  {journal} {\prd}\ }\textbf {\bibinfo {volume} {63}},\
  \bibinfo {eid} {103510} (\bibinfo {year} {2001})},\ \Eprint
  {http://arxiv.org/abs/arXiv:astro-ph/0006373} {arXiv:astro-ph/0006373}
  \BibitemShut {NoStop}%
\bibitem [{\citenamefont {{Padmanabhan}}(2002)}]{Padmanabhan2002}%
  \BibitemOpen
  \bibfield  {author} {\bibinfo {author} {\bibfnamefont {T.}~\bibnamefont
  {{Padmanabhan}}},\ }\href {\doibase 10.1103/PhysRevD.66.021301} {\bibfield
  {journal} {\bibinfo  {journal} {\prd}\ }\textbf {\bibinfo {volume} {66}},\
  \bibinfo {eid} {021301} (\bibinfo {year} {2002})},\ \Eprint
  {http://arxiv.org/abs/hep-th/0204150} {hep-th/0204150} \BibitemShut {NoStop}%
\bibitem [{\citenamefont {{Chimento}}(2004)}]{Chimento2004}%
  \BibitemOpen
  \bibfield  {author} {\bibinfo {author} {\bibfnamefont {L.~P.}\ \bibnamefont
  {{Chimento}}},\ }\href {\doibase 10.1103/PhysRevD.69.123517} {\bibfield
  {journal} {\bibinfo  {journal} {\prd}\ }\textbf {\bibinfo {volume} {69}},\
  \bibinfo {eid} {123517} (\bibinfo {year} {2004})},\ \Eprint
  {http://arxiv.org/abs/astro-ph/0311613} {astro-ph/0311613} \BibitemShut
  {NoStop}%
\bibitem [{\citenamefont {{Garriga}}\ and\ \citenamefont
  {{Mukhanov}}(1999)}]{Garriga1999}%
  \BibitemOpen
  \bibfield  {author} {\bibinfo {author} {\bibfnamefont {J.}~\bibnamefont
  {{Garriga}}}\ and\ \bibinfo {author} {\bibfnamefont {V.~F.}\ \bibnamefont
  {{Mukhanov}}},\ }\href {\doibase 10.1016/S0370-2693(99)00602-4} {\bibfield
  {journal} {\bibinfo  {journal} {Physics Letters B}\ }\textbf {\bibinfo
  {volume} {458}},\ \bibinfo {pages} {219} (\bibinfo {year} {1999})},\ \Eprint
  {http://arxiv.org/abs/hep-th/9904176} {hep-th/9904176} \BibitemShut {NoStop}%
\bibitem [{\citenamefont {{Copeland}}\ \emph {et~al.}(2005)\citenamefont
  {{Copeland}}, \citenamefont {{Garousi}}, \citenamefont {{Sami}},\ and\
  \citenamefont {{Tsujikawa}}}]{Copeland2005}%
  \BibitemOpen
  \bibfield  {author} {\bibinfo {author} {\bibfnamefont {E.~J.}\ \bibnamefont
  {{Copeland}}}, \bibinfo {author} {\bibfnamefont {M.~R.}\ \bibnamefont
  {{Garousi}}}, \bibinfo {author} {\bibfnamefont {M.}~\bibnamefont {{Sami}}}, \
  and\ \bibinfo {author} {\bibfnamefont {S.}~\bibnamefont {{Tsujikawa}}},\
  }\href {\doibase 10.1103/PhysRevD.71.043003} {\bibfield  {journal} {\bibinfo
  {journal} {\prd}\ }\textbf {\bibinfo {volume} {71}},\ \bibinfo {eid} {043003}
  (\bibinfo {year} {2005})},\ \Eprint {http://arxiv.org/abs/hep-th/0411192}
  {hep-th/0411192} \BibitemShut {NoStop}%
\bibitem [{\citenamefont {{Rendall}}(2006)}]{Rendall2006}%
  \BibitemOpen
  \bibfield  {author} {\bibinfo {author} {\bibfnamefont {A.~D.}\ \bibnamefont
  {{Rendall}}},\ }\href {\doibase 10.1088/0264-9381/23/5/008} {\bibfield
  {journal} {\bibinfo  {journal} {Classical and Quantum Gravity}\ }\textbf
  {\bibinfo {volume} {23}},\ \bibinfo {pages} {1557} (\bibinfo {year}
  {2006})},\ \Eprint {http://arxiv.org/abs/gr-qc/0511158} {gr-qc/0511158}
  \BibitemShut {NoStop}%
\bibitem [{\citenamefont {{De-Santiago}}\ \emph {et~al.}(2013)\citenamefont
  {{De-Santiago}}, \citenamefont {{Cervantes-Cota}},\ and\ \citenamefont
  {{Wands}}}]{DeSantiago2013}%
  \BibitemOpen
  \bibfield  {author} {\bibinfo {author} {\bibfnamefont {J.}~\bibnamefont
  {{De-Santiago}}}, \bibinfo {author} {\bibfnamefont {J.~L.}\ \bibnamefont
  {{Cervantes-Cota}}}, \ and\ \bibinfo {author} {\bibfnamefont
  {D.}~\bibnamefont {{Wands}}},\ }\href {\doibase 10.1103/PhysRevD.87.023502}
  {\bibfield  {journal} {\bibinfo  {journal} {\prd}\ }\textbf {\bibinfo
  {volume} {87}},\ \bibinfo {eid} {023502} (\bibinfo {year} {2013})},\ \Eprint
  {http://arxiv.org/abs/1204.3631} {arXiv:1204.3631 [gr-qc]} \BibitemShut
  {NoStop}%
\bibitem [{\citenamefont {{De-Santiago}}\ and\ \citenamefont
  {{Cervantes-Cota}}(2014)}]{DeSantiago2014}%
  \BibitemOpen
  \bibfield  {author} {\bibinfo {author} {\bibfnamefont {J.}~\bibnamefont
  {{De-Santiago}}}\ and\ \bibinfo {author} {\bibfnamefont {J.~L.}\ \bibnamefont
  {{Cervantes-Cota}}},\ }\href {\doibase 10.1088/1742-6596/485/1/012017}
  {\bibfield  {journal} {\bibinfo  {journal} {Journal of Physics Conference
  Series}\ }\textbf {\bibinfo {volume} {485}},\ \bibinfo {eid} {012017}
  (\bibinfo {year} {2014})},\ \Eprint {http://arxiv.org/abs/1404.0946}
  {arXiv:1404.0946 [gr-qc]} \BibitemShut {NoStop}%
\bibitem [{\citenamefont {{Gasperini}}\ and\ \citenamefont
  {{Veneziano}}(2003)}]{Gasperini2003}%
  \BibitemOpen
  \bibfield  {author} {\bibinfo {author} {\bibfnamefont {M.}~\bibnamefont
  {{Gasperini}}}\ and\ \bibinfo {author} {\bibfnamefont {G.}~\bibnamefont
  {{Veneziano}}},\ }\href {\doibase 10.1016/S0370-1573(02)00389-7} {\bibfield
  {journal} {\bibinfo  {journal} {\physrep}\ }\textbf {\bibinfo {volume}
  {373}},\ \bibinfo {pages} {1} (\bibinfo {year} {2003})},\ \Eprint
  {http://arxiv.org/abs/hep-th/0207130} {hep-th/0207130} \BibitemShut {NoStop}%
\bibitem [{\citenamefont {{Abramo}}\ and\ \citenamefont
  {{Finelli}}(2003)}]{Abramo2003}%
  \BibitemOpen
  \bibfield  {author} {\bibinfo {author} {\bibfnamefont {L.~R.}\ \bibnamefont
  {{Abramo}}}\ and\ \bibinfo {author} {\bibfnamefont {F.}~\bibnamefont
  {{Finelli}}},\ }\href {\doibase 10.1016/j.physletb.2003.09.065} {\bibfield
  {journal} {\bibinfo  {journal} {Physics Letters B}\ }\textbf {\bibinfo
  {volume} {575}},\ \bibinfo {pages} {165} (\bibinfo {year} {2003})},\ \Eprint
  {http://arxiv.org/abs/astro-ph/0307208} {astro-ph/0307208} \BibitemShut
  {NoStop}%
\bibitem [{\citenamefont {{Aguirregabiria}}\ and\ \citenamefont
  {{Lazkoz}}(2004)}]{Aguirregabiria2004}%
  \BibitemOpen
  \bibfield  {author} {\bibinfo {author} {\bibfnamefont {J.~M.}\ \bibnamefont
  {{Aguirregabiria}}}\ and\ \bibinfo {author} {\bibfnamefont {R.}~\bibnamefont
  {{Lazkoz}}},\ }\href {\doibase 10.1103/PhysRevD.69.123502} {\bibfield
  {journal} {\bibinfo  {journal} {\prd}\ }\textbf {\bibinfo {volume} {69}},\
  \bibinfo {eid} {123502} (\bibinfo {year} {2004})},\ \Eprint
  {http://arxiv.org/abs/hep-th/0402190} {hep-th/0402190} \BibitemShut {NoStop}%
\bibitem [{\citenamefont {{Hamed}}\ \emph {et~al.}(2004)\citenamefont
  {{Hamed}}, \citenamefont {{Cheng}}, \citenamefont {{Luty}},\ and\
  \citenamefont {{Mukohyama}}}]{Hamed2004}%
  \BibitemOpen
  \bibfield  {author} {\bibinfo {author} {\bibfnamefont {N.~A.}\ \bibnamefont
  {{Hamed}}}, \bibinfo {author} {\bibfnamefont {H.~S.}\ \bibnamefont
  {{Cheng}}}, \bibinfo {author} {\bibfnamefont {M.~A.}\ \bibnamefont {{Luty}}},
  \ and\ \bibinfo {author} {\bibfnamefont {S.}~\bibnamefont {{Mukohyama}}},\
  }\href {\doibase 10.1088/1126-6708/2004/05/074} {\bibfield  {journal}
  {\bibinfo  {journal} {Journal of High Energy Physics}\ }\textbf {\bibinfo
  {volume} {5}},\ \bibinfo {eid} {074} (\bibinfo {year} {2004})},\ \Eprint
  {http://arxiv.org/abs/hep-th/0312099} {hep-th/0312099} \BibitemShut {NoStop}%
\bibitem [{\citenamefont {{Piazza}}\ and\ \citenamefont
  {{Tsujikawa}}(2004)}]{Piazza2004}%
  \BibitemOpen
  \bibfield  {author} {\bibinfo {author} {\bibfnamefont {F.}~\bibnamefont
  {{Piazza}}}\ and\ \bibinfo {author} {\bibfnamefont {S.}~\bibnamefont
  {{Tsujikawa}}},\ }\href {\doibase 10.1088/1475-7516/2004/07/004} {\bibfield
  {journal} {\bibinfo  {journal} {\jcap}\ }\textbf {\bibinfo {volume} {7}},\
  \bibinfo {eid} {004} (\bibinfo {year} {2004})},\ \Eprint
  {http://arxiv.org/abs/hep-th/0405054} {hep-th/0405054} \BibitemShut {NoStop}%
\bibitem [{\citenamefont {{Alishahiha}}\ \emph {et~al.}(2004)\citenamefont
  {{Alishahiha}}, \citenamefont {{Silverstein}},\ and\ \citenamefont
  {{Tong}}}]{Alishahiha2004}%
  \BibitemOpen
  \bibfield  {author} {\bibinfo {author} {\bibfnamefont {M.}~\bibnamefont
  {{Alishahiha}}}, \bibinfo {author} {\bibfnamefont {E.}~\bibnamefont
  {{Silverstein}}}, \ and\ \bibinfo {author} {\bibfnamefont {D.}~\bibnamefont
  {{Tong}}},\ }\href {\doibase 10.1103/PhysRevD.70.123505} {\bibfield
  {journal} {\bibinfo  {journal} {\prd}\ }\textbf {\bibinfo {volume} {70}},\
  \bibinfo {eid} {123505} (\bibinfo {year} {2004})},\ \Eprint
  {http://arxiv.org/abs/hep-th/0404084} {hep-th/0404084} \BibitemShut {NoStop}%
\bibitem [{\citenamefont {{Silverstein}}\ and\ \citenamefont
  {{Tong}}(2004)}]{Silverstein2004}%
  \BibitemOpen
  \bibfield  {author} {\bibinfo {author} {\bibfnamefont {E.}~\bibnamefont
  {{Silverstein}}}\ and\ \bibinfo {author} {\bibfnamefont {D.}~\bibnamefont
  {{Tong}}},\ }\href {\doibase 10.1103/PhysRevD.70.103505} {\bibfield
  {journal} {\bibinfo  {journal} {\prd}\ }\textbf {\bibinfo {volume} {70}},\
  \bibinfo {eid} {103505} (\bibinfo {year} {2004})},\ \Eprint
  {http://arxiv.org/abs/hep-th/0310221} {hep-th/0310221} \BibitemShut {NoStop}%
\bibitem [{\citenamefont {{Guo}}\ and\ \citenamefont {{Ohta}}(2008)}]{Guo2008}%
  \BibitemOpen
  \bibfield  {author} {\bibinfo {author} {\bibfnamefont {Z.-K.}\ \bibnamefont
  {{Guo}}}\ and\ \bibinfo {author} {\bibfnamefont {N.}~\bibnamefont {{Ohta}}},\
  }\href {\doibase 10.1088/1475-7516/2008/04/035} {\bibfield  {journal}
  {\bibinfo  {journal} {\jcap}\ }\textbf {\bibinfo {volume} {0804}},\ \bibinfo
  {pages} {035} (\bibinfo {year} {2008})},\ \Eprint
  {http://arxiv.org/abs/hep-th/0803.1013} {hep-th/0803.1013} \BibitemShut
  {NoStop}%
\bibitem [{\citenamefont {{Bose}}\ and\ \citenamefont
  {{Majumdar}}(2009)}]{Bose2009}%
  \BibitemOpen
  \bibfield  {author} {\bibinfo {author} {\bibfnamefont {N.}~\bibnamefont
  {{Bose}}}\ and\ \bibinfo {author} {\bibfnamefont {A.~S.}\ \bibnamefont
  {{Majumdar}}},\ }\href {\doibase 10.1103/PhysRevD.80.103508} {\bibfield
  {journal} {\bibinfo  {journal} {\prd}\ }\textbf {\bibinfo {volume} {80}},\
  \bibinfo {eid} {103508} (\bibinfo {year} {2009})},\ \Eprint
  {http://arxiv.org/abs/0907.2330} {arXiv:0907.2330} \BibitemShut {NoStop}%
\bibitem [{\citenamefont {{De-Santiago}}\ and\ \citenamefont
  {{Cervantes-Cota}}(2011)}]{DeSantiago2011}%
  \BibitemOpen
  \bibfield  {author} {\bibinfo {author} {\bibfnamefont {J.}~\bibnamefont
  {{De-Santiago}}}\ and\ \bibinfo {author} {\bibfnamefont {J.~L.}\ \bibnamefont
  {{Cervantes-Cota}}},\ }\href {\doibase 10.1103/PhysRevD.83.063502} {\bibfield
   {journal} {\bibinfo  {journal} {\prd}\ }\textbf {\bibinfo {volume} {83}},\
  \bibinfo {eid} {063502} (\bibinfo {year} {2011})},\ \Eprint
  {http://arxiv.org/abs/1102.1777} {arXiv:1102.1777 [astro-ph.CO]} \BibitemShut
  {NoStop}%
\bibitem [{\citenamefont {{Sharif}}\ \emph {et~al.}(2012)\citenamefont
  {{Sharif}}, \citenamefont {{Yesmakhanova}}, \citenamefont {{Rani}},\ and\
  \citenamefont {{Myrzakulov}}}]{Sharif2012}%
  \BibitemOpen
  \bibfield  {author} {\bibinfo {author} {\bibfnamefont {M.}~\bibnamefont
  {{Sharif}}}, \bibinfo {author} {\bibfnamefont {K.}~\bibnamefont
  {{Yesmakhanova}}}, \bibinfo {author} {\bibfnamefont {S.}~\bibnamefont
  {{Rani}}}, \ and\ \bibinfo {author} {\bibfnamefont {R.}~\bibnamefont
  {{Myrzakulov}}},\ }\href {\doibase 10.1140/epjc/s10052-012-2067-1} {\bibfield
   {journal} {\bibinfo  {journal} {European Physical Journal C}\ }\textbf
  {\bibinfo {volume} {72}},\ \bibinfo {eid} {2067} (\bibinfo {year} {2012})},\
  \Eprint {http://arxiv.org/abs/1204.2181} {arXiv:1204.2181 [physics.gen-ph]}
  \BibitemShut {NoStop}%
\bibitem [{\citenamefont {{Scherrer}}(2004)}]{Scherrer2004}%
  \BibitemOpen
  \bibfield  {author} {\bibinfo {author} {\bibfnamefont {R.~J.}\ \bibnamefont
  {{Scherrer}}},\ }\href {\doibase 10.1103/PhysRevLett.93.011301} {\bibfield
  {journal} {\bibinfo  {journal} {Physical Review Letters}\ }\textbf {\bibinfo
  {volume} {93}},\ \bibinfo {eid} {011301} (\bibinfo {year} {2004})},\ \Eprint
  {http://arxiv.org/abs/astro-ph/0402316} {astro-ph/0402316} \BibitemShut
  {NoStop}%
\bibitem [{\citenamefont {{Sen}}(2002{\natexlab{a}})}]{Sen2002}%
  \BibitemOpen
  \bibfield  {author} {\bibinfo {author} {\bibfnamefont {A.}~\bibnamefont
  {{Sen}}},\ }\href {\doibase 10.1142/S0217732302008071} {\bibfield  {journal}
  {\bibinfo  {journal} {Modern Physics Letters A}\ }\textbf {\bibinfo {volume}
  {17}},\ \bibinfo {pages} {1797} (\bibinfo {year} {2002}{\natexlab{a}})},\
  \Eprint {http://arxiv.org/abs/hep-th/0204143} {hep-th/0204143} \BibitemShut
  {NoStop}%
\bibitem [{\citenamefont {{Sen}}(2002{\natexlab{b}})}]{Sen2002a}%
  \BibitemOpen
  \bibfield  {author} {\bibinfo {author} {\bibfnamefont {A.}~\bibnamefont
  {{Sen}}},\ }\href {\doibase 10.1088/1126-6708/2002/04/048} {\bibfield
  {journal} {\bibinfo  {journal} {Journal of High Energy Physics}\ }\textbf
  {\bibinfo {volume} {4}},\ \bibinfo {eid} {048} (\bibinfo {year}
  {2002}{\natexlab{b}})},\ \Eprint {http://arxiv.org/abs/hep-th/0203211}
  {hep-th/0203211} \BibitemShut {NoStop}%
\bibitem [{\citenamefont {{Sen}}(2005)}]{Sen2005}%
  \BibitemOpen
  \bibfield  {author} {\bibinfo {author} {\bibfnamefont {A.}~\bibnamefont
  {{Sen}}},\ }\href {\doibase 10.1142/S0217751X0502519X} {\bibfield  {journal}
  {\bibinfo  {journal} {International Journal of Modern Physics A}\ }\textbf
  {\bibinfo {volume} {20}},\ \bibinfo {pages} {5513} (\bibinfo {year}
  {2005})},\ \Eprint {http://arxiv.org/abs/hep-th/0410103} {hep-th/0410103}
  \BibitemShut {NoStop}%
\bibitem [{\citenamefont {{Yang}}\ and\ \citenamefont
  {{Zhang}}(2008)}]{Yang2008}%
  \BibitemOpen
  \bibfield  {author} {\bibinfo {author} {\bibfnamefont {R.-J.}\ \bibnamefont
  {{Yang}}}\ and\ \bibinfo {author} {\bibfnamefont {S.-N.}\ \bibnamefont
  {{Zhang}}},\ }\href {\doibase 10.1088/0256-307X/25/1/092} {\bibfield
  {journal} {\bibinfo  {journal} {Chinese Physics Letters}\ }\textbf {\bibinfo
  {volume} {25}},\ \bibinfo {pages} {344} (\bibinfo {year} {2008})}\BibitemShut
  {NoStop}%
\bibitem [{\citenamefont {{Yang}}\ and\ \citenamefont {{Qi}}(2012)}]{Yang2012}%
  \BibitemOpen
  \bibfield  {author} {\bibinfo {author} {\bibfnamefont {R.}~\bibnamefont
  {{Yang}}}\ and\ \bibinfo {author} {\bibfnamefont {J.}~\bibnamefont {{Qi}}},\
  }\href {\doibase 10.1140/epjc/s10052-012-2095-x} {\bibfield  {journal}
  {\bibinfo  {journal} {European Physical Journal C}\ }\textbf {\bibinfo
  {volume} {72}},\ \bibinfo {eid} {2095} (\bibinfo {year} {2012})},\ \Eprint
  {http://arxiv.org/abs/1205.5968} {arXiv:1205.5968 [gr-qc]} \BibitemShut
  {NoStop}%
\bibitem [{\citenamefont {{Yang}}\ \emph {et~al.}(2015)\citenamefont {{Yang}},
  \citenamefont {{Chen}}, \citenamefont {{Li}},\ and\ \citenamefont
  {{Qi}}}]{Yang2015}%
  \BibitemOpen
  \bibfield  {author} {\bibinfo {author} {\bibfnamefont {R.}~\bibnamefont
  {{Yang}}}, \bibinfo {author} {\bibfnamefont {B.}~\bibnamefont {{Chen}}},
  \bibinfo {author} {\bibfnamefont {J.}~\bibnamefont {{Li}}}, \ and\ \bibinfo
  {author} {\bibfnamefont {J.}~\bibnamefont {{Qi}}},\ }\href {\doibase
  10.1007/s10509-014-2218-y} {\bibfield  {journal} {\bibinfo  {journal}
  {\apss}\ }\textbf {\bibinfo {volume} {356}},\ \bibinfo {pages} {399}
  (\bibinfo {year} {2015})},\ \Eprint {http://arxiv.org/abs/1311.5307}
  {arXiv:1311.5307 [gr-qc]} \BibitemShut {NoStop}%
\bibitem [{\citenamefont {{Mamon}}\ and\ \citenamefont
  {{Das}}(2015)}]{Mamon2015}%
  \BibitemOpen
  \bibfield  {author} {\bibinfo {author} {\bibfnamefont {A.~A.}\ \bibnamefont
  {{Mamon}}}\ and\ \bibinfo {author} {\bibfnamefont {S.}~\bibnamefont
  {{Das}}},\ }\href {\doibase 10.1140/epjc/s10052-015-3467-9} {\bibfield
  {journal} {\bibinfo  {journal} {European Physical Journal C}\ }\textbf
  {\bibinfo {volume} {75}},\ \bibinfo {eid} {244} (\bibinfo {year} {2015})},\
  \Eprint {http://arxiv.org/abs/1503.06280} {arXiv:1503.06280 [gr-qc]}
  \BibitemShut {NoStop}%
\bibitem [{\citenamefont {{Ossoulian}}\ \emph {et~al.}(2016)\citenamefont
  {{Ossoulian}}, \citenamefont {{Golanbari}}, \citenamefont {{Sheikhahmadi}},\
  and\ \citenamefont {{Saaidi}}}]{Ossoulian2016}%
  \BibitemOpen
  \bibfield  {author} {\bibinfo {author} {\bibfnamefont {Z.}~\bibnamefont
  {{Ossoulian}}}, \bibinfo {author} {\bibfnamefont {T.}~\bibnamefont
  {{Golanbari}}}, \bibinfo {author} {\bibfnamefont {H.}~\bibnamefont
  {{Sheikhahmadi}}}, \ and\ \bibinfo {author} {\bibfnamefont {K.}~\bibnamefont
  {{Saaidi}}},\ }\href {\doibase 10.1155/2016/3047461} {\bibfield  {journal}
  {\bibinfo  {journal} {Adv. High Energy Phys.}\ }\textbf {\bibinfo {volume}
  {2016}},\ \bibinfo {pages} {3047461} (\bibinfo {year} {2016})},\ \Eprint
  {http://arxiv.org/abs/1512.05571} {arXiv:1512.05571 [gr-qc]} \BibitemShut
  {NoStop}%
\bibitem [{\citenamefont {{Battye}}\ \emph {et~al.}(2015)\citenamefont
  {{Battye}}, \citenamefont {{Moss}},\ and\ \citenamefont
  {{Pearson}}}]{Battye2015}%
  \BibitemOpen
  \bibfield  {author} {\bibinfo {author} {\bibfnamefont {R.~A.}\ \bibnamefont
  {{Battye}}}, \bibinfo {author} {\bibfnamefont {A.}~\bibnamefont {{Moss}}}, \
  and\ \bibinfo {author} {\bibfnamefont {J.~A.}\ \bibnamefont {{Pearson}}},\
  }\href {\doibase 10.1088/1475-7516/2015/04/048} {\bibfield  {journal}
  {\bibinfo  {journal} {\jcap}\ }\textbf {\bibinfo {volume} {4}},\ \bibinfo
  {eid} {048} (\bibinfo {year} {2015})},\ \Eprint
  {http://arxiv.org/abs/1409.4650} {arXiv:1409.4650} \BibitemShut {NoStop}%
\bibitem [{\citenamefont {{Pace}}\ \emph {et~al.}(2012)\citenamefont {{Pace}},
  \citenamefont {{Fedeli}}, \citenamefont {{Moscardini}},\ and\ \citenamefont
  {{Bartelmann}}}]{Pace2012}%
  \BibitemOpen
  \bibfield  {author} {\bibinfo {author} {\bibfnamefont {F.}~\bibnamefont
  {{Pace}}}, \bibinfo {author} {\bibfnamefont {C.}~\bibnamefont {{Fedeli}}},
  \bibinfo {author} {\bibfnamefont {L.}~\bibnamefont {{Moscardini}}}, \ and\
  \bibinfo {author} {\bibfnamefont {M.}~\bibnamefont {{Bartelmann}}},\ }\href
  {\doibase 10.1111/j.1365-2966.2012.20692.x} {\bibfield  {journal} {\bibinfo
  {journal} {\mnras}\ }\textbf {\bibinfo {volume} {422}},\ \bibinfo {pages}
  {1186} (\bibinfo {year} {2012})},\ \Eprint {http://arxiv.org/abs/1111.1556}
  {arXiv:1111.1556 [astro-ph.CO]} \BibitemShut {NoStop}%
\end{thebibliography}%

\label{lastpage}

\end{document}